\documentclass[twocolumn,10pt]{article}
\usepackage[top=1in, bottom=1in, left=.65in, right=.65in]{geometry}
\usepackage[utf8]{inputenc}
\usepackage[T1]{fontenc}
\usepackage{listings}
\usepackage{setspace}
\usepackage{etoolbox}
\usepackage{xcolor}
\usepackage{courier}

\usepackage{graphicx}
\usepackage{cprotect}
\usepackage{xspace}
\usepackage{colortbl}
\usepackage[affil-it]{authblk} 
\usepackage[symbol]{footmisc}
\graphicspath{ {./figures/} }
\usepackage{mathpazo}
\patchcmd{\maketitle}{\@fnsymbol}{\@alph}{}{}  

\usepackage[numbers,sort&compress]{natbib}
\usepackage{tikz}
\usetikzlibrary{arrows,calc,positioning,shadows,shapes,shapes.geometric,arrows.meta}
\usepackage{pgfplots}
\usepgfplotslibrary{groupplots}
\usepackage{adjustbox}
\usepackage{stfloats}
\usepackage{pifont}
\usepackage{booktabs}
\usepackage{array}
\usepackage{amsmath}
\usepackage{siunitx}
\usepackage{wrapfig}

\usepackage[nohyperlinks,nolist]{acronym}

\usepackage[title]{appendix}

\definecolor{bg}{rgb}{0.97, 0.97, 0.97}

\definecolor{keywordcolor}{rgb}{0.0, 0.0, 0.5}      
\definecolor{stringcolor}{rgb}{0.5, 0.0, 0.0}       
\definecolor{commentcolor}{rgb}{0.5, 0.5, 0.5}      

\title{\textbf{\huge{The HydroGym Reinforcement Learning Platform \\ for Fluid Dynamics}}\vspace{-.1in}}
\author[1,2]{Christian Lagemann$^*$}
\author[1,2]{Sajeda Mokbel}
\author[3]{Miro Gondrum}
\author[4]{Mario Rüttgers}
\author[5]{Yuning Wang}
\author[6]{Pol Suárez}
\author[7]{Ludger Paehler}
\author[7]{Deniz A. Bezgin}
\author[7]{Aaron B. Buhendwa}
\author[1,2]{Jared L. Callaham}
\author[1,2]{Samuel Ahnert}
\author[1,2]{Nicholas Zolman}
\author[3]{Xiao Shao}
\author[8]{Jean-Christophe Loiseau}
\author[7,9]{Nikolaus A. Adams}
\author[3,10]{Matthias Meinke}
\author[3,10]{Wolfgang Schr{\"o}der}
\author[11, 12]{Kai Lagemann}
\author[1,2]{Esther Lagemann}
\author[5,6]{Ricardo Vinuesa$^*$}
\author[1,2]{Steven L. Brunton$^*$\vspace{-.1in}}
{\small
\affil[1]{\small Department of Mechanical Engineering, University of Washington, Seattle, WA 98195, United States}
\affil[2]{\small AI Institute in Dynamic Systems, University of Washington, Seattle, WA 98195, United States}
\affil[3]{\small Chair of Fluid Mechanics and Institute of Aerodynamics, RWTH Aachen University, Aachen, Germany}
\affil[4]{\small Data-Driven Fluid Engineering Laboratory, Inha University, Incheon, South Korea}
\affil[5]{\small Department of Aerospace Engineering, University of Michigan, Ann Arbor, MI, United States}
\affil[6]{\small FLOW, Engineering Mechanics, KTH Royal Institute of Technology, Stockholm, Sweden}
\affil[7]{\small School of Engineering and Design, Technical University of Munich, Munich, Germany}
\affil[8]{\small Arts et Métiers Institute of Technology, CNAM, DynFluid, HESAM Université, Paris, France}
\affil[9]{\small Munich Institute of Integrated Materials, Energy and Process Engineering, Technical University of Munich, Munich, Germany}
\affil[10]{\small JARA Center for Simulation and Data Science, RWTH Aachen University, Aachen, Germany}
\affil[11]{\small MediaTek Research, London, United Kingdom}
\affil[12]{\small Statistics and Machine Learning, German Center for Neurodegenerative Diseases, Bonn, Germany}

\affil[ ]{\vspace{10pt}}   
\affil[*]{\small Corresponding authors: christian.lagemann@rwth-aachen.de, rvinuesa@umich.edu, sbrunton@uw.edu}
}
\date{}

\usepackage{url}
\usepackage{placeins}

\usepackage{microtype}
\usepackage{graphicx}
\usepackage{booktabs} 
\usepackage{hyperref}
\usepackage{natbib}
\usepackage{amsmath,amssymb,amsfonts,amsthm}
\usepackage{mathrsfs}
\usepackage{xcolor}
\usepackage{mathtools}
\usepackage{dsfont}
\usepackage{hyperref}
\usepackage{bm}
\usepackage{nicefrac}
\usepackage{wrapfig}
\usepackage{lipsum}
\usepackage{enumerate}
\usepackage{outlines}
\usepackage{amsmath,algorithm,float,tabularx,amsfonts}

\usepackage[utf8]{inputenc}
\usepackage{algorithm,algcompatible,amssymb,amsmath}

\algnewcommand\algorithmicto{\textbf{to}}
\algnewcommand\RETURN{\State \textbf{return} }

\usepackage{algpseudocode}
\makeatletter
\newcommand{\multiline}[1]{%
    \begin{tabularx}{\dimexpr\linewidth-\ALG@thistlm}[t]{@{}X@{}}
        #1
    \end{tabularx}
}

\newcounter{assumption}%
\renewcommand{\theassumption}{\arabic{assumption}}

\numberwithin{lemma}{section} 
\numberwithin{theorem}{section} 
\numberwithin{corollary}{section} 
\numberwithin{proposition}{section} 
\numberwithin{definition}{section} 
\numberwithin{example}{section} 
\numberwithin{question}{section} 

\usepackage{multirow}
\usepackage{caption}
\usepackage{subcaption} 
\usepackage[font={scriptsize},labelfont={bf}]{caption}
\usepackage{xspace}
\usepackage{cprotect}

\newcommand{\keywords}[1]
{\centerline{
  \small	
  \textbf{\textit{Keywords: }} #1
}
}

\usepackage{bibunits}
\begin{document}

\begin{bibunit}[unsrt]
    \twocolumn[
    \begin{@twocolumnfalse} 
        \maketitle
        \vspace{-30pt}
        \begin{abstract}
           Modeling and controlling fluids is critical for several fields of science and engineering, including transportation, energy, and medicine.  Effective flow control can lead to lift increase, drag reduction, mixing enhancement, and noise attenuation, which may unlock new technologies.  However, controlling a fluid faces significant challenges, including high-dimensional, nonlinear, and multiscale interactions in space and time. While reinforcement learning (RL) has recently shown great success in robotics and protein folding through shared benchmark platforms, fluid dynamics has remained resistant: each controller is typically tuned to one geometry and operating point, making progress difficult to accumulate, transfer, and compare.
           To address these challenges, we introduce HydroGym, a solver-independent RL platform for flow control research, and show that standardized community infrastructure unlocks transferable control intelligence across flow regimes rather than isolated case-specific solutions. HydroGym provides 61+ validated flow control environments spanning canonical laminar to complex turbulent flows with systematic Reynolds number ($Re$) progressions up to $Re=4\times 10^5$ and Mach number variations for 2D and 3D flows. The platform enables diverse computational backends, including finite-volume, spectral-element, finite-element, lattice-Boltzmann, and fully differentiable solvers, which improve efficiency through gradient-enhanced optimization. Comprehensive evaluation reveals that RL agents consistently discover robust control principles, such as boundary layer manipulation, acoustic feedback disruption, and wake reorganization, leading to drag reductions exceeding 90\% across canonical configurations. Critically, we demonstrate zero-shot transfer, where agents trained exclusively on a simplified HydroGym channel flow achieve 38\% friction-drag reduction on an unseen three-dimensional wing section at $Re_c=200{,}000$. This zero-shot transfer reduces exploration costs by four orders of magnitude compared to direct on-wing optimization, establishing a viable pathway for RL in real-world fluid systems that would otherwise remain intractable. These findings suggest RL agents are able to uncover essential physics, rather than configuration-specific patterns, with profound implications for generalizable control strategies.  The highly extensible and scalable HydroGym platform provides community infrastructure for researchers in fluid dynamics, machine learning, and control to advance grand challenges in science and technology through flow control.  
        \end{abstract}

        \vspace{15pt}

    \end{@twocolumnfalse}
]

\begin{figure*}[!t]
    \centering
    \includegraphics[width=0.999\textwidth,trim={7.25cm 1.25cm 6.0cm 1.0cm},clip]{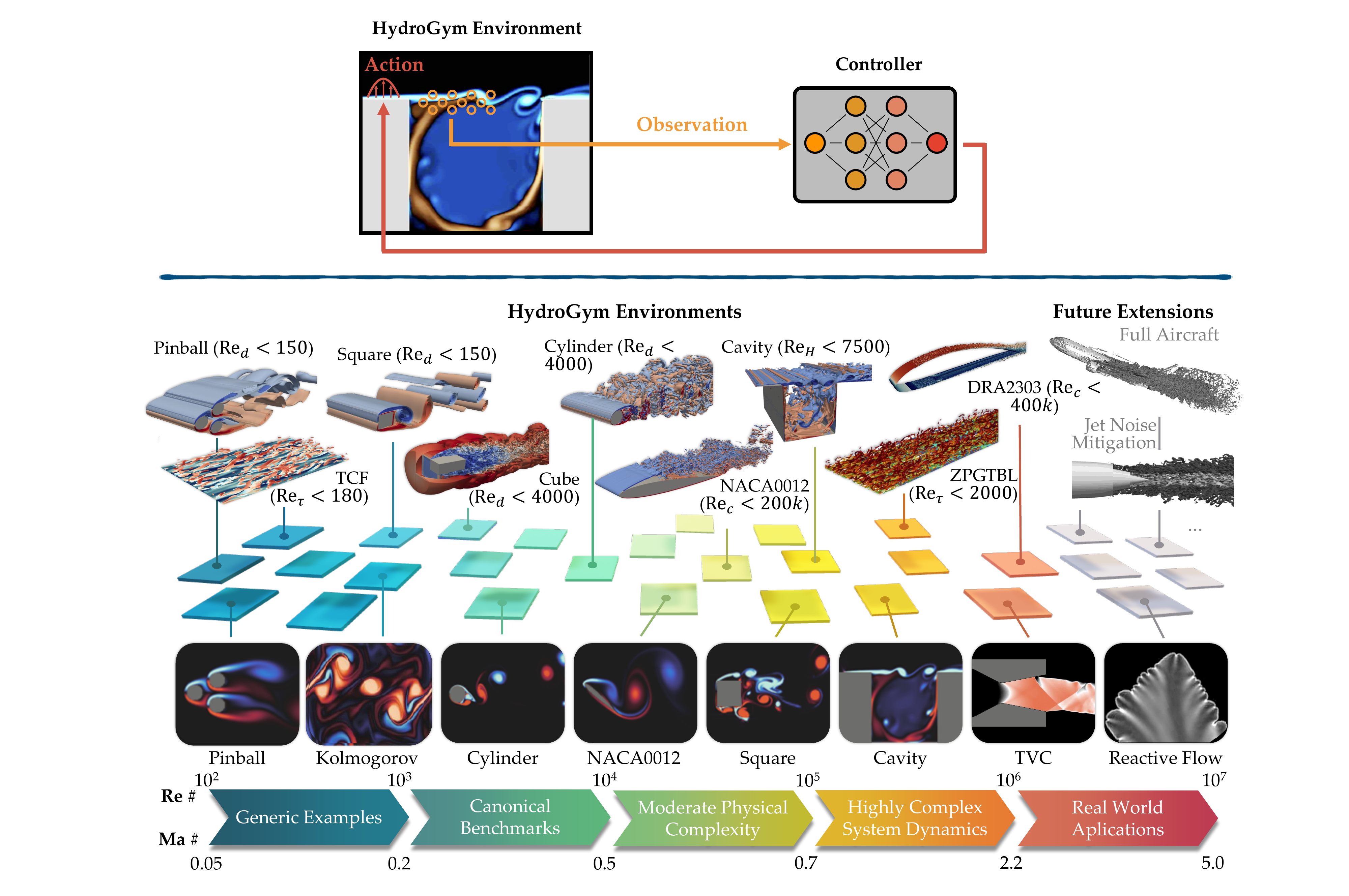}
    \caption{\textbf{The HydroGym framework and environment roadmap.} (Top) The RL paradigm applied to active flow control, framing the fluid solver as an environment that exchanges states and actions with an RL controller. (Bottom) A visual roadmap of HydroGym’s rich collection of flow environments, illustrating the growing physical and computational complexity offered by the framework. The suite spans from computationally inexpensive, 2D canonical problems designed for rapid prototyping (left) to large-scale, high-Reynolds-number turbulent flows utilizing up to 1.6 billion cells (right). By spanning this vast range of Reynolds ($Re$) and Mach ($Ma$) numbers, HydroGym enables researchers to systematically develop, test, and scale flow control algorithms toward real-world applications.}
    \label{figure1}
\end{figure*}

\label{intro}

Fluid flow control represents a critical challenge in several trillion-dollar industries including transportation (e.g., aerospace, automotive, shipping), energy production (e.g., wind farms, gas turbines, pipelines), manufacturing (e.g., spray processes, additive manufacturing, thin films and coatings), and medicine (e.g., circulatory, respiratory, glymphatic system). 
Advanced control systems could dramatically improve energy efficiency globally, for example reducing fuel consumption in the aviation industry by up to 15\% through active drag reduction~\cite{marusic2021energy, mateling2023spanwise, ricco2021review, font2025deep, lagemann2024extending}, or increasing wind farm output by 5\% through coordinated control~\cite{zhao2020cooperative,dalili2009review,chamorro2013drag}.  \\

Despite decades of progress in traditional control approaches, the nonlinear, multiscale nature of turbulence results in control problems that are high-dimensional, non-convex, and computationally intractable using conventional methods~\cite{gao2024generative, brunton2015annual,brunton2020machine,koumoutsakos2024roads}. Recent breakthroughs in RL have demonstrated transformative capabilities across increasingly complex scientific applications, with AlphaFold revolutionizing protein structure prediction~\cite{jumper2021highly} and DeepMind achieving real-time control of fusion reactor plasmas~\cite{degrave2022magnetic}. These successes share a common enabler: benchmark environments that democratize research access and accelerate algorithmic innovation. \\

However, fluid dynamics has remained largely resistant to similar breakthroughs due to fundamental computational and accessibility barriers and a lack of comprehensive benchmark platforms for systematic algorithmic comparison and reproducible research. 
Training effective Reinforcement Learning (RL) agents typically requires thousands or millions of interactions with the environment, and each evaluation in a fluid environment requires expensive computational fluid dynamics (CFD) simulations~\cite{kochkov2021machine, suarez2025flow, bae2022scientific}.  
This bottleneck is magnified by the inherent sample inefficiency of model-free RL~\cite{zolman2024sindy,vinyals2019grandmaster}, creating severe computational limitations for large-scale flow control problems~\cite{chatzimanolakis2024learning, fan2020reinforcement, bhola2023multi}.  
Further, flow control research is currently caught in a ``specificity trap" where controllers trained at specific flow conditions may fail to generalize to different regimes. 
In much of today’s literature, each controller is carefully tuned to a single geometry, a specific operating point, and a unique numerical setup. 
Under these conditions, algorithmic progress and rigorous comparisons are nearly impossible, and generalization of control strategies across physical regimes remains a major challenge. \\

We address these fundamental challenges in HydroGym, a reinforcement learning platform for flow control research. 
HydroGym establishes much needed community infrastructure for common environments, interfaces, and evaluation of algorithms, allowing competing methods to be tested fairly and extended to increasingly difficult regimes.  
Our solver-independent architecture provides sixty-one validated flow control environments spanning canonical laminar flows to advanced three-dimensional turbulent scenarios and Reynolds number progressions across flow regimes. 
The platform's solver diversity helps bridge the disparate scales of fluid research, integrating established Python-based finite-element options for accessible prototyping, highly parallel GPU-accelerated lattice-Boltzmann and finite-volume solvers for large-scale flows, and fully differentiable solvers to facilitate gradient-aware learning.

\begin{figure*}[t!]
    \centering
    \includegraphics[width=\textwidth,trim={0.0cm 0.0cm 0.0cm 0.0cm},clip]{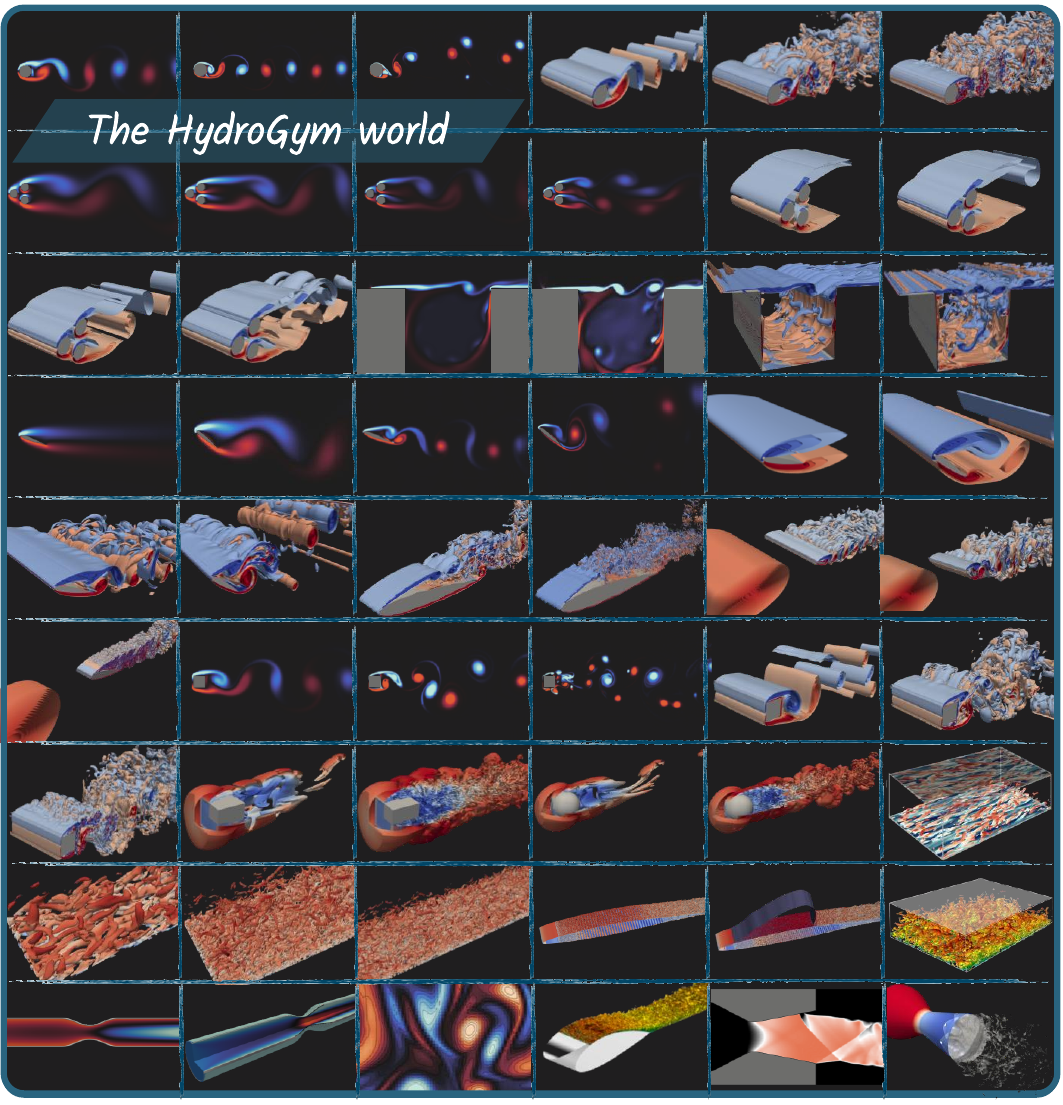}
    \caption{\textbf{The HydroGym Environment Suite.} The HydroGym platform provides standardized access to over 61 benchmarked fluid dynamics environments through a unified Gymnasium-compatible interface. The comprehensive environment gallery showcases diverse flow configurations spanning two-dimensional and three-dimensional scenarios including cylinder flows, cavity flows, airfoils, channel flows, and turbulent cases across multiple Reynolds number regimes. This standardized approach enables consistent benchmarking and comparison of reinforcement learning algorithms across a vast spectrum of active flow control problems.}
    \label{ExtDataFigure1}
\end{figure*}

Figures~\ref{figure1}\&~\ref{ExtDataFigure1} show the progression of HydroGym, from canonical environments of low-to-medium complexity to highly challenging environments with complex three-dimensional geometries and fully turbulent flow physics at high Reynolds and Mach numbers. The HydroGym platform exhibits several important flow classes, including incompressible and compressible high-Reynolds number wall turbulence and airfoil environments for realistic deployment studies; shock-interaction regimes that isolate the physics of convective delays and nonlocal coupling; divergent nozzle configurations for strongly coupled compressible flows; and multi-agent settings for high-dimensional actuation (see Tables \ref{ExtDataTable1} \& \ref{ExtDataTable2} for environment details including $Re$ number, degrees of freedom, control objective, etc.). Together, HydroGym establishes a community infrastructure to learn transferable physical control principles that generalize across geometries and flow regimes, towards increasingly relevant real-world systems.

\begin{figure*}[t!]
    \centering
    \includegraphics[width=\textwidth,trim={9.0cm 11.0cm 9.0cm 5.5cm},clip]{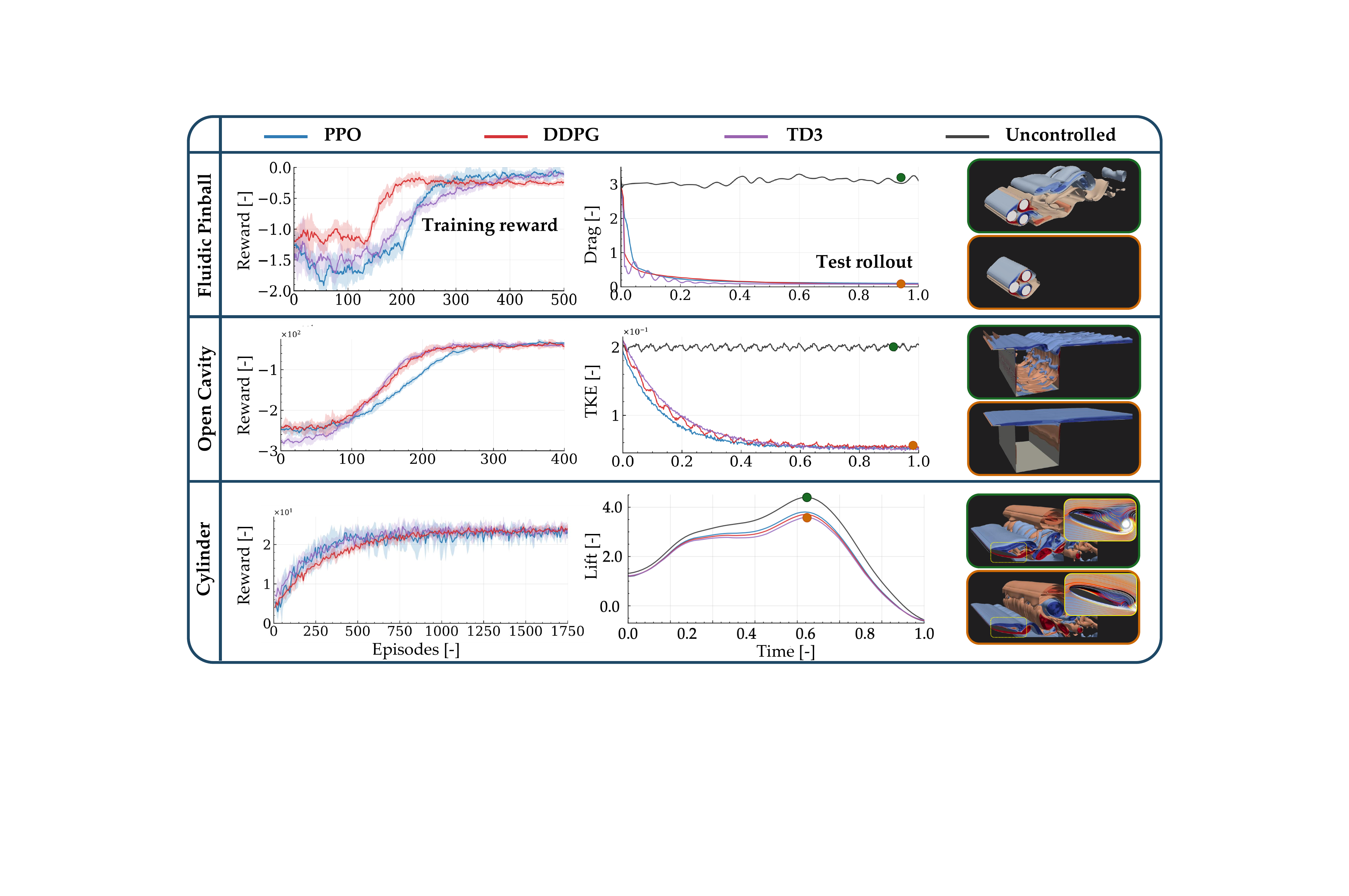}
    \vspace{-.225in}
    \caption{\textbf{Reinforcement learning performance across three-dimensional flow control environments.} Training curves (left) show convergence behavior for PPO, DDPG, and TD3 algorithms across three benchmark scenarios including a fluidic pinball ($Re=150$), an open cavity ($Re=7,500$), and a gust-airfoil interaction ($Re=1,000$) environment. Test rollouts (center) demonstrate control effectiveness, revealing the agents' ability to manage complex three-dimensional phenomena including spanwise instabilities, feedback loop mechanisms, and turbulent structures (TKE denotes turbulent kinetic energy). Flow visualizations (right) highlight the sophisticated wake manipulation and flow control achieved through learned policies.}
    \label{figure2}
\end{figure*}

\begin{figure*}[t!]
    \centering
    \includegraphics[width=\textwidth,trim={0.0cm 20.0cm 0.0cm 0.0cm},clip]{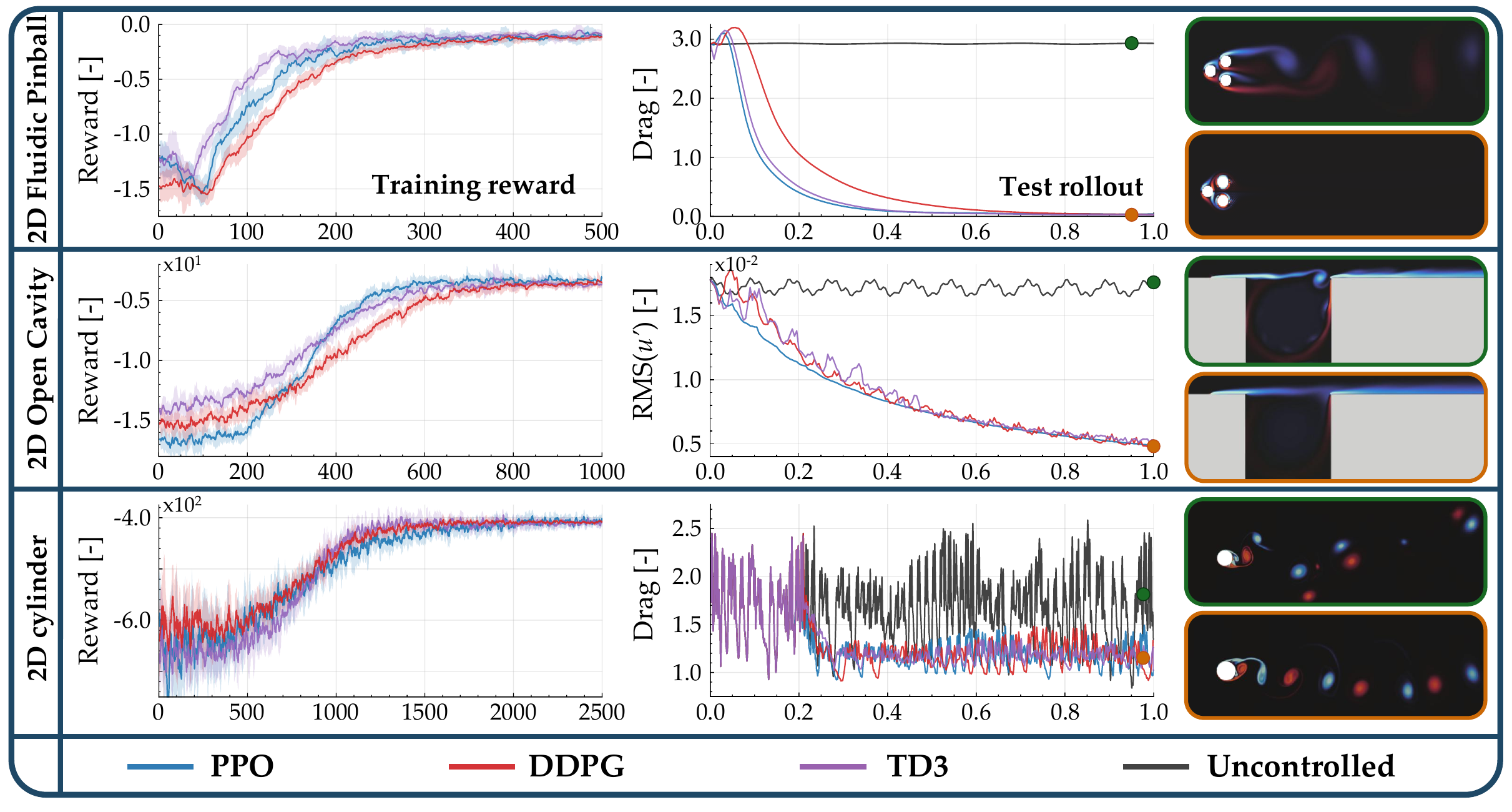}
   \caption{\textbf{Reinforcement learning performance across two-dimensional flow control environments.} Training curves (left) show convergence behavior for PPO, DDPG, and TD3 algorithms across three benchmark scenarios including a fluidic pinball ($Re=100$), an open cavity ($Re=4,200$), and a cylinder ($Re=3,900$) environment. Test rollouts (center) demonstrate control effectiveness, revealing the agents' ability to manage canonical two-dimensional phenomena including vortex shedding and Kelvin-Helmholtz instabilities. Flow visualizations (right) highlight the sophisticated wake manipulation and flow control achieved through learned policies.}
    \label{ExtDataFigure2}
\end{figure*}

\paragraph{Systematic flow control demonstrations.}
We discuss baseline performance metrics by considering four benchmark configurations using industry-standard RL algorithms: Proximal Policy Optimization (PPO)~\cite{schulman2017proximal}, Deep Deterministic Policy Gradient (DDPG)~\cite{lillicrap2015continuous}, and Twin Delayed Deep Deterministic Policy Gradient (TD3)~\cite{fujimoto2018addressing}. In total, training of all validated baselines required over 150,000 GPU hours (further details in Section 4 of the Supplementary Information). Remarkably, the RL agents autonomously discover and exploit complex physical mechanisms such as boundary layer energization, feedback loop disruption, and wake management through interaction alone. They successfully transform detrimental, separated flows into controlled, organized states while minimizing actuation, validating HydroGym's utility for both fundamental physics research and practical control design. Results for presented 2D environments are provided in Figure~\ref{ExtDataFigure2}, while the 3D demonstrations are detailed in Figure~\ref{figure2}.\\

\noindent\textit{Fluidic pinball control.} Three cylinders in an equilateral formation create complex wake interactions, demonstrating coordinated multi-actuator control with fundamentally different dynamics across Reynolds numbers. At $Re=100$ (2D), the uncontrolled system exhibits natural flow asymmetry with irregular pressure oscillations. RL agents discover coordinated rotation strategies achieving ${\sim}90\%$ drag reduction through counterrotating rear cylinders that inject momentum into the wake centerline, preventing shear layer interaction and suppressing vortex formation (Figure~\ref{ExtDataFigure2}). Increasing the Reynolds number to $Re=150$, the flow undergoes a system bifurcation and becomes highly chaotic with irregular, unpredictable vortex shedding. The agents introduce an asymmetric forcing strategy on the front cylinder alongside counter-rotating rear cylinders. As a result, the rotation-induced Magnus forces create favorable pressure gradients, vectoring separated shear layers inward while energizing boundary layers. This transforms irregular, asymmetric shedding into a steady, symmetric wake with reduced velocity deficits and enhanced pressure recovery (Figure~\ref{figure2}). Overall, this example demonstrates how targeted boundary layer manipulation can suppress complex dynamics in multi-body fluid systems.\\

\noindent\textit{Open cavity flow control.} Cavity flows exhibit complex dynamics dominated by shear layer instabilities and acoustic feedback, requiring control strategies that address multiple coupled timescales. At $Re=4,200$, the 2D flow sits precisely at the threshold where small perturbations trigger transition from steady recirculating flow to unsteady vortex roll-up. The shear layer exhibits weakly growing instabilities that manifest as low-amplitude pressure oscillations. RL agents discover that targeted momentum injection at the upstream cavity edge can exploit this marginal stability. The optimal policy employs controlled blowing to thicken the separating shear layer and reduce velocity gradients, suppressing Kelvin-Helmholtz instability growth before their amplification into coherent vortical
\newpage

structures. It exemplifies how small control inputs at near-critical conditions produce disproportionately large stabilization effects.
In 3D at $Re=7,500$, secondary flow structures emerge with well-developed acoustic feedback mechanisms. The shear layer develops coherent vortical structures that convect downstream and impinge on the cavity edge, generating pressure waves that propagate upstream to excite the separation point. This resonant feedback establishes dominant cavity tones and complex three-dimensional features including spanwise instabilities and vortex stretching. As shown in Figure~\ref{figure2}, RL agents discover sophisticated control policies using coordinated blowing and suction with temporal modulation. High-frequency, low-amplitude blowing at the upstream edge thickens the boundary layer and introduces controlled perturbations that interfere destructively with Kelvin-Helmholtz instabilities, while strategically placed suction near the downstream edge weakens the acoustic source by reducing vortex-edge interactions. The coordinated actuation achieves near-complete suppression of dominant cavity tones while controlling three-dimensional secondary flows, significantly reducing recirculation zone complexity and unsteady loading.\\

\FloatBarrier

\noindent\textit{Circular cylinder control at $Re=3,900$.} Moving to isolated bluff-body dynamics, the circular cylinder at $Re=3,900$ (2D) features profound shear layer instabilities, irregular shedding frequencies, and a chaotic wake structure. Operating under a zero-net-mass-flux (ZNMF) constraint, where opposing jets maintain equal and opposite mass flow rates, the controller implements boundary layer manipulation that effectively streamlines the cylinder without adding or removing mass.
The learned policy exploits asymmetric pressure distributions through coordinated suction and injection. Specifically, it pairs suction on one surface (removing low-momentum fluid to delay separation) with blowing on the opposite surface to energize the boundary layer through direct momentum injection. This creates a time-varying "virtual geometry" that actively manages wake formation. This localized boundary layer manipulation successfully transforms the chaotic wake into coherent, predictable vortical patterns that shed with improved regularity and symmetry (Figure~\ref{ExtDataFigure2}). The separation points shift downstream, narrowing the momentum deficit structure and converting random turbulent fluctuations into ordered momentum transport.\\

\noindent\textit{Transverse gust mitigation at $Re=1,000$.} Finally, the gust mitigation environment presents a fundamentally different challenge requiring real-time adaptation. We consider a NACA0012 airfoil at a post-stall angle of attack ($\alpha=20^\circ$) under transverse gusts with a gust ratio $G=u_g/u_\infty=2$ (where $u_g$ and $u_\infty$ are the gust and freestream velocities, respectively). In this 3D environment, rapid variations in the effective angle of attack cause severe load fluctuations and complex vortex-gust interactions. The RL agent utilizes three leading-edge jet actuators to directly inject momentum into the separated boundary layer. This actuation creates a more compact, energetic separation bubble and generates primary vortices with enhanced circulation strength and spatial coherence. By synchronizing the controlled vortex shedding with the natural flow response to the incoming gust, this dynamic circulation control mechanism prevents premature vortex breakdown and destructive interference. The resulting wake exhibits a reduced velocity deficit and decreased turbulence intensity, allowing the airfoil to maintain stable aerodynamic performance with approximately 20\% lower load oscillations despite extreme transverse perturbations (Figure~\ref{figure2}).

\paragraph{Scaling Complexity - Differentiable environments, multi-agent RL, and transfer learning capabilities.}
While conventional reinforcement learning has proven effective for flow control, emerging RL paradigms offer fundamentally different approaches to optimization and learning efficiency. HydroGym's advanced capabilities demonstrate how differentiable reinforcement learning exploits gradient information through entire simulation trajectories, enabling more efficient policy optimization, while multi-agent frameworks address the spatial complexity of three-dimensional flows through distributed actuator coordination. Additionally, transfer learning capabilities enable knowledge reuse across Reynolds numbers, geometries, and dimensions, substantially reducing computational requirements for new control problems. Together, these approaches provide direct access to sensitivity information, enable scalable solutions for complex 3D control scenarios, and unlock generalizable control strategies.\\

\begin{figure*}[t!]
\vspace{-.25cm}
\centering
    \includegraphics[width=.95\textwidth,trim={10cm 9.0cm 10cm 5.5cm},clip]{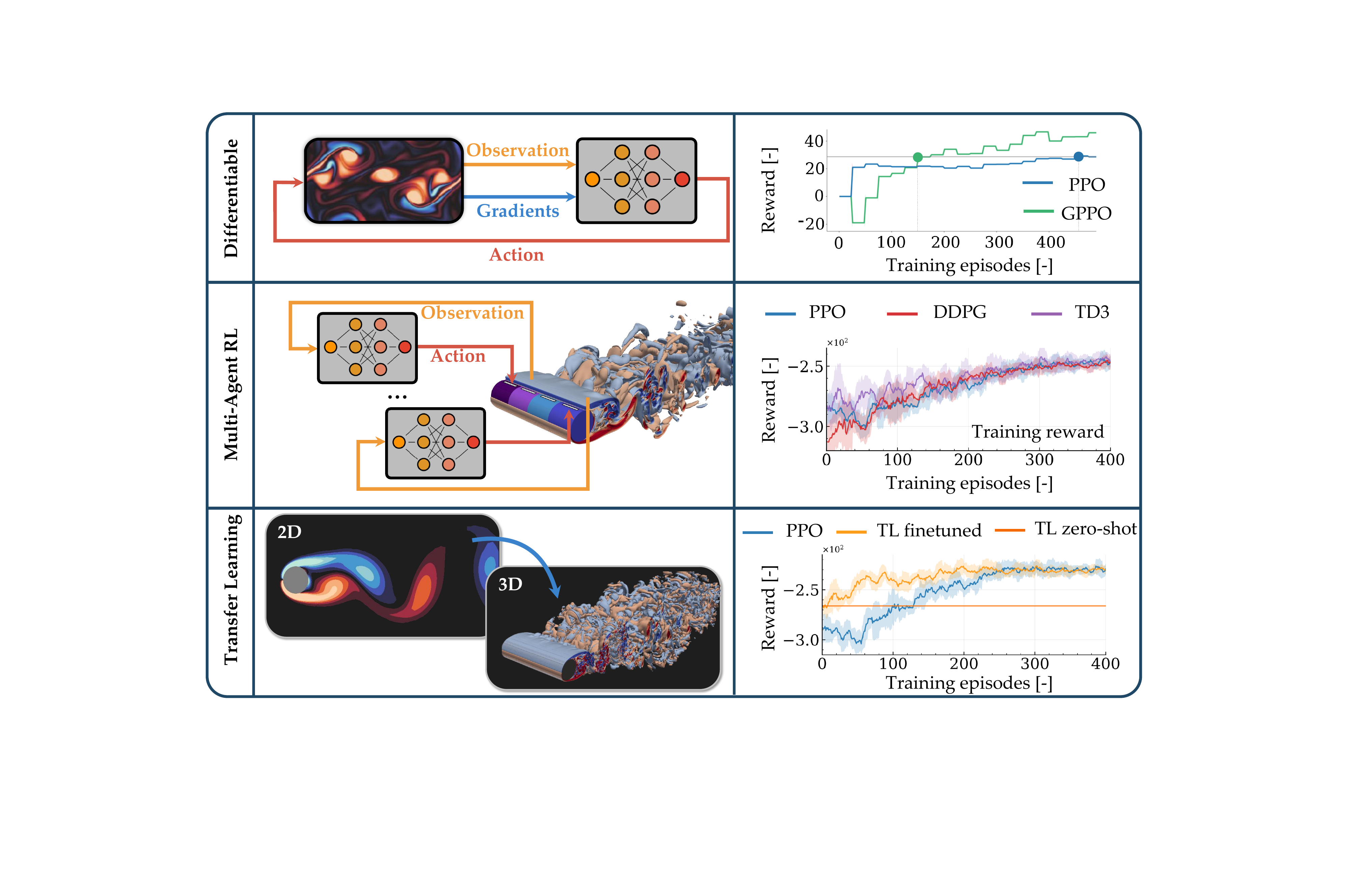}
    \vspace{-.1in}
    \caption{\textbf{Hydrogym's advanced reinforcement learning capabilities.} (Top) Gradient-enhanced policy optimization through differentiable environments substantially improves sample efficiency and the overall reward compared to gradient-free methods. JAX-based solvers enable automatic differentiation through complete simulation trajectories, providing exact policy gradients. Training curves compare GPPO against standard PPO for the Kolmogorov flow environment, illustrating a $65\%$ training sample reduction for GPPO alongside a substantially improved final reward. (Center) Multi-agent reinforcement learning for spatially distributed control. The HydroGym framework decomposes a 3D cylinder environment at $Re=3,900$ into multiple cooperative agents, each managing local actuator pairs while sharing gradient information and experience buffers. This distributed approach scales efficiently with actuator count and enables parallel exploration of high-dimensional control spaces. Training performance and test rollouts demonstrate effective learning and substantial drag reduction through coordinated spanwise actuation. (Bottom) Transfer learning scenarios demonstrate the knowledge transfer and policy generalization across dimensional and Reynolds number scaling. Finetuned policies consistently achieve faster convergence, requiring approximately half the training episodes of baseline methods.}
    \label{figure3}
\end{figure*}

\noindent\textit{Gradient-enhanced policy optimization through differentiable physics.} In contrast to traditional CFD solvers, modern machine learning (ML) libraries enable physics solvers to compute sensitivities of objective functions with respect to control parameters and actuator placement leveraging automatic differentiation. This capability allows policy optimization using sensitivity information, yielding improved performance and sample efficiency~\cite{son2023gippo, xu2022accelerated}. HydroGym provides a variety of differentiable environments including 2D Kolmogorov flows, 3D turbulent channel flows and shock vector control scenarios for flows in 2D and 3D nozzles with complex shock interactions and system dynamics. Here, we briefly demonstrate the advantages of differentiable environments for the chaotic Kolmogorov flow~\cite{Blanchard2019} characterized by extreme energy dissipation events from nonlinear energy transfer across scales. The control task is to leverage these events to enhance mixing through adaptive forcing wavenumbers. We compared standard PPO against gradient-enhanced PPO (GPPO), where agents controlled the amplitudes of the forcing terms. In GPPO, the loss function backpropagates through both the policy and environment, via an analytical gradient term (see supplementary information for details). GPPO reduced training iterations by at least 65\% while producing more efficient controllers with lower action amplitudes than PPO (Figure~\ref{figure3}). These results demonstrate that incorporating gradient information into the optimization process substantially improves sample efficiency and yields more efficient policies. As differentiable simulation and control methods mature, more sophisticated gradient-based approaches may deliver further gains in both efficiency and performance. \\

\begin{figure*}[t!]
\vspace{-.15in}
    \centering
    \includegraphics[width=.98\textwidth,trim={5.0cm 6.9cm 20.5cm 1.5cm},clip]{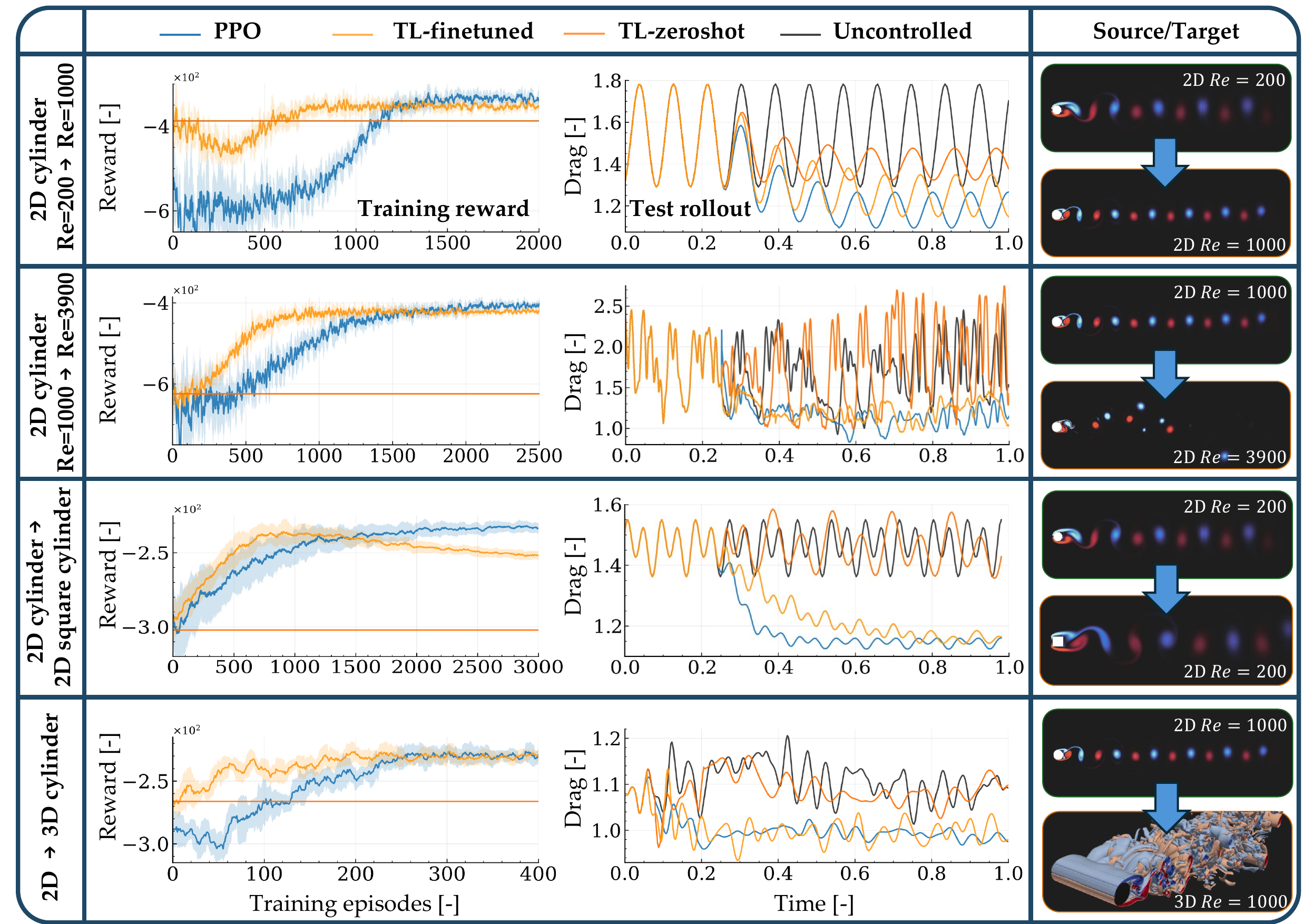}
    \caption{\textbf{Transfer learning capabilities across flow configurations and conditions.} Four transfer scenarios demonstrate knowledge generalization: Reynolds number scaling ($Re=200 \rightarrow 1,000$, $Re=1000 \rightarrow 3,900$), geometric transfer (circular to square cylinder), and dimensional scaling (2D to 3D cylinder). Training curves compare standard PPO training with transfer learning (finetuned and zeroshot). Finetuned policies consistently achieve faster convergence, requiring approximately half the training episodes of baseline methods. Test rollouts confirm that improved training efficiency translates to effective control performance, with drag reduction across all transfer scenarios.}
    \label{ExtDataFigure3}
\end{figure*}

\noindent\textit{Multi-agent coordination for spatially distributed 3D cylinder control.} The HydroGym benchmark platform further supports multi-agent reinforcement learning (MARL) in a variety of environments. For a 3D cylinder at $Re=3,900$, for instance, the MARL framework exploits the inherent spatial locality of flow control by decomposing the global control problem into multiple local pseudo-environments~\cite{suarez2025active}, each managing a pair of ZNMF jets distributed spanwise along the cylinder surface. Unlike~\cite{suarez2025active}, our approach employs shared gradient updates and a unified replay buffer across all agents, enhancing computational efficiency while maintaining the cooperative learning dynamics essential for coordinated actuation. As shown in Figure~\ref{figure3}, this architecture enables the discovery of efficient control strategies that utilize diverse frequency bandwidths and adaptive spanwise coordination, resulting in drag reductions of approximately $8\%$, similar to~\cite{suarez2025active}.
The agents learn to manipulate the shear layer instabilities, effectively enlarging the recirculation bubble and reshaping the wake into a more aerodynamically efficient teardrop-like configuration. The cooperative control strategy attenuates Reynolds stress fluctuations while shifting peak turbulent activity downstream, resulting in reduced pressure drag through pressure recovery at the rear surface. This distributed actuation offers a scalable path for active flow control in complex 3D flows while maintaining computational tractability by exploiting local invariances in the flow. The shared learning architecture allows agents to benefit from collective experience leading to more robust and generalizable control policies. Similar MARL successes in robotic swarm coordination, autonomous vehicle networks, and multi-robot manipulation demonstrate the broad applicability of this approach for distributed control.\\

\noindent\textit{Transfer Learning Capabilities.}
A fundamental question motivating the core HydroGym idea is whether discovered mechanisms represent configuration-specific solutions or more generalizable approaches. To address this, we systematically investigate how control knowledge transfers across Reynolds numbers, geometries, and dimensions, showing that HydroGym enables efficient transfer learning to reduce computational requirements and improve controller generalization.  

When transferring control policies across a wide range of Reynolds numbers, fine-tuning achieves optimal performance in approximately $50\%$ fewer episodes than training from scratch (Figure~\ref{ExtDataFigure3}, rows 1 \& 2). Throughout this range, drag reduction effectiveness is either maintained or improved. This indicates that flow control strategies capture scale-invariant physics, particularly pressure gradient manipulation and momentum injection timing relative to vortex formation cycles.

Beyond Reynolds number invariance, the policies demonstrate geometric transferability. Control strategies for vortex shedding suppression learned on circular cylinders effectively manage vortex shedding in square geometries. This supports the notion that policies learn mathematical structures governing pattern formation and symmetry breaking that persist across boundary conditions, rather than geometry-specific control (Figure~\ref{ExtDataFigure3}, row 3). 

The most stringent test is dimensional transfer from 2D to 3D, which confirms universality across spatial scales. Despite the increased complexity of three-dimensional flows, including spanwise coupling effects and additional instability modes, pre-trained 2D control strategies successfully delay separation and maintain effective drag reduction in 3D applications while reducing computational costs by $\approx 50-60\%$ (Figure~\ref{figure3} and Figure~\ref{ExtDataFigure3} - row 4). This indicates that boundary layer manipulation strategies transcend spatial dimensionality through conserved scaling laws and similarity parameters, with 2D training capturing the dominant flow physics relevant for 3D control.

The transfer mechanism itself provides insights into learning. While zero-shot performance varies across scenarios, the subsequent fine-tuning process consistently benefits from pre-trained initialization, even when immediate transfer appears limited. This pattern suggests that learned policies encode fundamental flow control relationships that may not manifest immediately in performance metrics but provide essential structure for accelerated learning, indicating that deep RL discovers mathematical invariants governing scaled flow problems.

\begin{figure*}[t!]
\vspace{-.15in}
\centering
    \includegraphics[width=.9\textwidth,trim={0.0cm 0.0cm 0.0cm 0.0cm},clip]{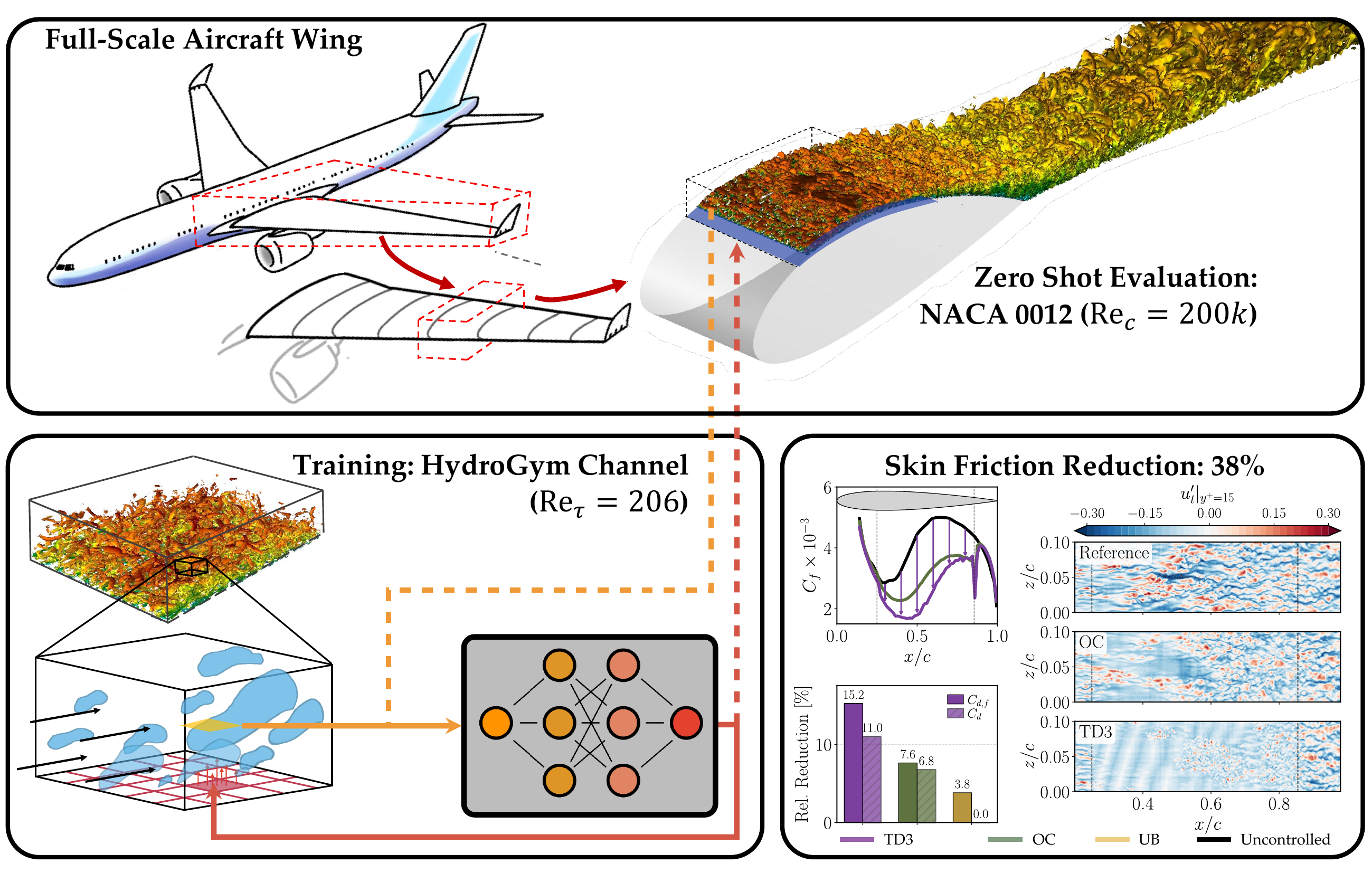}
    \caption{\textbf{Physics-guided Transfer Learning.} Multi-agent reinforcement learning controllers (TD3) trained exclusively in a HydroGym turbulent channel flow surrogate environment at $Re_\tau=206$ are deployed zero-shot to a NACA0012 airfoil at $Re_c = 200{,}000$ without additional training. Spatial distribution of the skin-friction coefficient $c_f$ shows substantial skin friction reduction of approximately 38\% in the controlled region and a total drag reduction of $11\%$ for the NACA0012 configuration. This outperforms gold-standard control techniques including opposition control (OC) and uniform blowing (UB) substantially. Instantaneous wall actuation patterns reveal organized boundary layer manipulation that attenuates high-speed streaks. This physics-guided transfer validates the potential for foundation-model-like generalization in flow control, where canonical training environments yield reusable control priors applicable across geometric configurations.}
    \label{figure4}
\end{figure*}

\paragraph{Physics-Guided Zero-Shot Transfer to a High-Reynolds Number Turbulent Wing.}
Building on these demonstrated spatial and geometric transfer capabilities, we next push the limits of generalizability to a substantially more challenging problem: transferring policies to industrially relevant, high-Reynolds-number environments. Directly training reinforcement learning policies on complex, three-dimensional wings at high Reynolds numbers is computationally prohibitive. To overcome this, we hypothesize that computationally tractable canonical environments can serve as surrogates for discovering transferable control priors, breaking the specificity trap that currently dominates flow control research.

To demonstrate this hypothesis, we employ a surrogate training paradigm that exploits partial invariances in wall-bounded turbulence. Specifically, we train multi-agent RL controllers exclusively on a canonical HydroGym turbulent channel flow at $Re_\tau = 206$. While channel flows lack the streamwise-varying pressure gradients, surface curvature, and inhomogeneous boundary layer development of airfoil flows, they reproduce the inner-layer turbulence scaling and streak dynamics that dominate near-wall momentum transport. Within these channel surrogates, controllers learn distributed actuation strategies that target near-wall flow structures via high-speed streak attenuation and coherent vortex manipulation.

After training, we then deploy these channel-trained controllers zero-shot to a three-dimensional NACA0012 wing section at $Re_c = 200,000$. Despite the radical difference in geometry, pressure gradients, and Reynolds number, the results are striking. The learned mechanisms prove to encode relationships that remain effective across training-deployment mismatch, achieving a 38\% local skin-friction drag reduction (see Figure~\ref{figure4}). Importantly, this zero-shot result substantially outperforms current gold-standard controllers, such as opposition control, and validates a core premise of the HydroGym framework: foundational control strategies discovered in simple environments can successfully generalize to complex aerodynamic configurations.

The computational advantage is substantial: channel surrogate training requires two orders of magnitude fewer grid points than wing simulations, while the additional homogeneous streamwise direction permits aggressive parallelization. Accounting for both spatial resolution and temporal dynamics, this yields an exploration efficiency gain exceeding $10^4$ in computational time. Even with periodic validation on the full wing geometry, total wall-clock time is reduced by approximately a factor of 60 relative to direct on-wing training. This efficiency enables systematic policy optimization including hyperparameter searches, multi-agent coordination strategies, and actuation pattern exploration that would be prohibitively expensive at wing-scale Reynolds numbers.

\section*{Discussion}
\label{discussion}
HydroGym addresses the fundamental bottlenecks preventing reinforcement learning from transforming fluid dynamics control: the computational intractability of training effective policies on expensive CFD simulations, and the steep domain knowledge required to implement these policies and simulations. Our platform's solver-independent architecture and comprehensive benchmark suite demonstrate that standardized environments can accelerate algorithmic development in fluid control just as ImageNet~\cite{deng2009imagenet} transformed computer vision and MuJoCo~\cite{todorov2012mujoco} revolutionized robotics research. 
HydroGym provides an extensive suite of rigorously validated flow environments with comprehensive RL benchmarks.  These environments include 2D and 3D flows across a range of geometries and Reynolds numbers, exhibiting a diversity of control challenges.  Further, the HydroGym framework supports advanced RL features, such as differentiable environments, transfer learning, and multi-agent RL. 

Our physics-guided zero-shot transfer validates a core hypothesis of the HydroGym framework. By successfully deploying controllers trained on simplified surrogate environments to complex, three-dimensional wing sections, we demonstrate that deep reinforcement learning may avoid overfitting to specific geometries, instead capturing invariant physical mechanisms such as near-wall momentum transport. This surrogate approach bypasses the computational bottlenecks of direct high-Reynolds-number training, unlocking control discovery for industrial scales that previously exceeded computational feasibility. This establishes a concrete pathway toward foundation-model-like generalization in fluid dynamics. Moving forward, extending this paradigm to span varying Reynolds numbers, pressure gradients, and geometric complexities will map the boundaries of transferability and accelerate development, from turbomachinery to vehicle aerodynamics. This exemplifies HydroGym's core mission: moving the field from isolated case studies toward accumulating, reproducible progress on flow control as a community challenge.
\paragraph{Limitations and Future Directions.}
While HydroGym addresses fundamental barriers in applying reinforcement learning in fluid control, several current limitations define the present scope and motivate ongoing developments.\\

\noindent\textit{Data sources and physical assumptions.} All current HydroGym environments employ direct numerical simulation (DNS) with fully resolved turbulent structures, providing validated, benchmarked ground truth without turbulence modeling assumptions or closure relations. This approach ensures that learned control strategies exploit genuine physical mechanisms rather than artifacts of turbulence models. However, DNS are computationally intensive, particularly at higher Reynolds numbers. The platform currently spans Reynolds numbers up to $\mathrm{Re}_\tau=2{,}200$ for internal flows and $\mathrm{Re}_c=400{,}000$ for external aerodynamics---regimes sufficient for fundamental research and algorithm development, but below the $\mathrm{Re}\sim10^7$--$10^8$ characteristic of many industrial applications. Critically, this limitation reflects training-phase computational constraints rather than solver capabilities; our production solvers routinely handle $\mathcal{O}(10^9)$ degree-of-freedom simulations on full aircraft geometries. However, embedding such simulations in RL training loops would require enormous compute resources (500+ GPUs and 50+ TB of memory per training run), rendering systematic algorithmic development impractical. Importantly, our demonstrated physics-guided zero-shot transfer validates that computationally tractable canonical environments can serve as surrogates for discovering reusable control priors applicable to industrially relevant configurations, potentially bridging the Reynolds number gap.\\

\noindent\textit{Flow regime coverage.} Current environments emphasize canonical flows and near-equilibrium turbulent boundary layers, reflecting established validation benchmarks. Non-equilibrium flows characterized by strong pressure gradients, flow separation, reattachment, and transient dynamics present additional complexity. We are actively developing environments that incorporate adverse and favorable pressure gradients using body forces. Moreover, future environment releases will include an extended range for multi-physics coupling, including aero-acoustic scenarios (jet noise mitigation, trailing-edge noise suppression), reactive flows in combustion systems, and multiphase dynamics.\\

\noindent\textit{Sample efficiency and reduced-order models.} Despite advances through differentiable environments and multi-agent architectures, effective policy learning still requires thousands of environment interactions, creating computational bottlenecks for complex 3D turbulent flows. While our differentiable environments accelerate learning using gradient information, highly turbulent 3D flows still face fundamental numerical challenges in maintaining stable automatic differentiation through chaotic dynamics. Reduced-order models (ROMs) offer a promising avenue for dramatic efficiency gains~\cite{maulik2020probabilistic,Duraisamy2019arfm, liu2021unsteady,semeraro2011feedback}. Recent work demonstrating that sparse learning methods can approximate flow dynamics and reward functions with up to 100$\times$ fewer high-fidelity interactions~\cite{zolman2024sindy} suggests a hybrid training paradigm: construct fast surrogate environments from limited DNS data, perform extensive policy optimization in the surrogate, then transfer and refine on high-fidelity simulations. ROM integration introduces additional considerations, particularly regarding energy exchange between truncated and retained modes in non-equilibrium regimes, which must be carefully addressed in future implementations.\\

\noindent\textit{Data accessibility and storage constraints.} The platform integrates trajectory and flow field uploads to Hugging Face repositories, enabling offline reinforcement learning, surrogate model training, and community data sharing. However, data volumes pose practical limits: for our largest simulations with $\mathcal{O}(10^9)$ degrees of freedom, individual flow fields require 75~GB of storage, meaning a single 200-step test trajectory generates 15~TB of data. This precludes comprehensive training dataset dissemination but enables selective high-value data sharing and compressed state representations.\\

\noindent\textit{Algorithmic development priorities.} Beyond expanding environment coverage, several algorithmic directions promise future impact. Deep coupling between RL agents and turbulence modeling---where agents learn closure parameters for Reynolds-Averaged Navier--Stokes (RANS) equations, subgrid-scale models, or wall functions on-the-fly during simulation~\cite{bae2022scientific}---could extend the scope. Advancing differentiable physics capabilities to handle fully turbulent flows would dramatically improve sample efficiency and enable gradient-based optimization at scales currently inaccessible~\cite{list2025differentiability,koehler2024apebench}. Surrogate models with uncertainty estimates may guide strategic high-fidelity sampling. Incorporating causal learning could enhance policy robustness and interpretability by identifying invariant physical mechanisms~\cite{cremades2024identifying, lozano2023machine, lagemann2023invariance,martinez2024decomposing, lagemann2023deep, cranmer2020discovering}, transforming black-box controllers into interpretable cause-effect chains. Hierarchical multi-agent architectures may enable control across multiple spatial and temporal scales, from local boundary layer manipulation to global patterns. \\

Overall, the results presented here point toward a broader architectural vision: a foundation model for flow control, pretrained across the diversity of HydroGym environments, and capable of zero-shot or few-shot adaptation to new configurations. The analogy to large language models is instructive as their generalization emerges not from any single algorithmic breakthrough, but from the availability of a standardized, large-scale, and diverse pretraining corpus. Fluid flow control has lacked an equivalent substrate: a unified collection of physically validated, computationally tractable flow control environments spanning geometries, Reynolds numbers, and flow regimes, with consistent interfaces for policy training and evaluation. HydroGym is designed to be precisely that substrate. The transferability results we demonstrate are a strong indicator that such pretraining is meaningful, where control priors learned in canonical flows encode reusable physical structure rather than environment-specific artifacts. Realizing this vision will require systematically mapping the boundaries of transferability across flow families, developing architectures that condition on physical context, and curating pretraining corpora that balance diversity with physical coherence. 
Eventually, HydroGym agents will be deployed in experimental flow control scenarios, further extending and testing their limits.  
These are tractable research questions, and HydroGym provides the benchmarks and infrastructure to pursue them rigorously.


\section*{Methods}
The HydroGym platform formulates flow control problems as discrete-time Markov Decision Processes (MDPs), defined by the tuple $(\mathcal{S}, \mathcal{A}, \mathcal{P}, \mathcal{R})$. At each control step $t$, the agent receives an observation state $s_t \in \mathcal{S}$ representing flow measurements, executes an action $a_t \in \mathcal{A}$ representing actuator inputs, and receives a scalar reward $r_t \in \mathcal{R}$ quantifying the control objective (e.g., drag reduction or flow stabilization). The transition probability $\mathcal{P}$ is governed implicitly by the underlying Navier-Stokes equations integrated by the fluid solver.

To bridge the gap between machine learning and fluid dynamics, HydroGym utilizes the Farama Gymnasium interface~\cite{towers2024gymnasium}, abstracting the complexity of the CFD execution. This abstraction is illustrated by the straightforward initialization and interaction pattern for an open cavity control example:
\begin{lstlisting}
# Swap one line to switch backend solvers
import hydrogym.firedrake as hgym
# import hydrogym.maia as hgym
# import hydrogym.nek as hgym
# import hydrogym.jax as hgym
# import hydrogym.jaxfluids as hgym

# initialize environment
env = hgym.FlowEnv("Cavity_2D_Re7500", **kwargs)

# reset environment
obs, info = env.reset()

# interact with environment
for i in range(num_interactions):
    obs, reward, terminated, truncated, info = env.step(action)
\end{lstlisting}
This design enables control and ML engineers to focus on algorithmic development without CFD expertise, while providing fluids researchers access to state-of-the-art RL techniques. The solver-independent architecture ensures that advances in CFD can be seamlessly integrated into the platform without disrupting existing workflows.

A critical element of the MDP formulation in HydroGym is the temporal coupling between the physical solver and the RL agent. Because turbulent flows evolve continuously while RL agents typically assume discrete actions, control inputs are updated at fixed intervals $\Delta t_{ctrl}$, which correspond to multiple physical CFD time steps designed to capture the characteristic timescale of the target instabilities (e.g., vortex shedding or shear-layer oscillation). 

To prevent numerical instabilities arising from discontinuous boundary conditions, HydroGym enforces temporal smoothing on all control actions. The instantaneous actuation magnitude $A(t)$ is smoothly interpolated between the previous action $A^{old}$ and the new target action $A^{new}$ over a predefined number of sub-iterations using a hyperbolic tangent or exponential ramp function. Furthermore, to standardize learning across disparate physical regimes, all observation features and action bounds are normalized to the $[-1, 1]$ range within the environment wrapper. 

\paragraph{Computational Fluid Dynamics Backends.}
To accommodate diverse computational scales ranging from rapid prototyping to 
large-scale high-performance computing (HPC) campaigns, HydroGym implements a 
solver-independent architecture supporting multiple backends. All computational 
environments have been thoroughly validated against established benchmarks to ensure 
physical correctness and high-fidelity resolution of the underlying fluid dynamics. 
Comprehensive validation studies and grid convergence analyses for each solver and 
flow configuration are detailed in the Supplementary Information 
(Sections~3 and~5).

\textit{Lattice-Boltzmann Solver (m-AIA LB).}
For the majority of weakly compressible, large-scale two- and three-dimensional DNS, HydroGym utilizes the LB method embedded within the m-AIA solver framework~\cite{maia}, which has been continuously 
developed at RWTH Aachen University for over two decades. The solver evolves the 
discrete particle distribution function $f_i$ on a D3Q27 velocity lattice via a 
two-step collision--streaming update. At low-to-moderate Reynolds numbers, a 
Bhatnagar--Gross--Krook (BGK) collision operator is 
employed, relaxing distributions towards a local Maxwellian equilibrium with a single 
relaxation frequency $\omega_{\mathrm{BGK}}$ linked to the kinematic viscosity. For 
enhanced stability at higher Reynolds numbers a cumulant-based collision 
operator is used instead, performing relaxations in a cumulant space 
with individual rates $\omega_\alpha$~\cite{geier2015cumulant}. Local grid refinement 
follows the methodology proposed by Eitel-Amor et al.~\cite{EitelAmor2013}, adjusting $\omega_{\mathrm{BGK}}$ 
across hierarchical grid levels to preserve a constant kinematic viscosity throughout 
the domain. No-slip boundaries are represented via a second-order interpolated bounce-back 
scheme. A hybrid parallelization strategy built on the Message Passing Interface (MPI) and a shared memory model either based on Open Multi-Processing (OpenMP) or on the parallel algorithms in the C\texttt{++} standard provided by the NVIDIA HPC SDK and AMD ROCm  HIPSTDPAR backends allows for a hardware-agnostic GPU implementation.

\textit{Finite-Volume Solver (m-AIA FV).}
Beyond the LBM, m-AIA includes a compressible Navier--Stokes solver based on a FV method targeting DNS and large-eddy simulation (LES) of wall-bounded turbulent flows at higher Mach 
numbers. The compressible Navier--Stokes equations are cast in an arbitrary 
Lagrangian--Eulerian (ALE) formulation and discretized on structured, body-conforming 
curvilinear grids with a cell-centered scheme, which enables anisotropic mesh 
refinement in the wall-normal direction suited for flat-plate and airfoil 
configurations. Inviscid fluxes are evaluated with the advection upstream splitting 
method (AUSM) combined with Monotonic Upwind Scheme for Conservation Laws (MUSCL) reconstruction 
for second-order spatial accuracy, while viscous fluxes are discretized using a 
modified cell-vertex (MCV) approach. Temporal integration employs a five-stage, 
low-storage Runge--Kutta scheme; when the grid deforms under surface actuation, the 
geometric conservation law (GCL) is enforced at every 
stage. Subgrid-scale effects in under-resolved LES are handled implicitly via a 
monotonically integrated LES (MILES) strategy, 
exploiting the inherent numerical dissipation of the upwind-biased AUSM scheme. The 
FV code shares the same MPI/OpenMP/GPU parallelization strategy as the LB solver. 
For RL integration, the solver exposes both synthetic jet actuation and a traveling 
transversal surface-wave boundary condition~\cite{albers2024,lagemann2024impact,mateling2022analysis,mateling2023spanwise,mateling2020detection}, 
where the wave amplitude $A$, wavelength $\lambda$, and phase speed $c$ serve as 
time-varying control parameters updated by the agent at each control step via a 
$C^1$-continuous cosine cross-fade transition to ensure numerical stability.

\textit{Spectral-Element Solver (Nek5000).}
For incompressible high-Reynolds-number configurations requiring higher-order accuracy, 
HydroGym provides a Nek5000~\cite{nek5000} backend employing the spectral-element 
method (SEM)~\cite{demoura_semadvantage_2024}. The velocity and pressure fields are 
approximated by high-order Lagrange interpolants on hexahedral elements following the 
$\mathbb{P}_N\mathbb{P}_{N-2}$ formulation~\cite{maday_spectral_1989}: velocity is 
collocated on $N^3$ Gauss--Lobatto--Legendre (GLL) points per element, while pressure 
resides on a staggered grid of $(N-2)^3$ Gauss--Legendre (GL) points. Temporal 
integration combines a third-order explicit extrapolation scheme (EXT3) for the 
nonlinear convective terms with a third-order implicit backward differentiation scheme 
(BDF3) for viscous contributions; aliasing errors are mitigated by overintegration 
with a $3/2$ oversampling factor in each spatial direction. Flow actuation is 
prescribed as a time-dependent wall-normal Dirichlet velocity $v_n$ with the spatial 
mean removed to satisfy the ZNMF constraint. Parallelization is 
achieved through hybrid OpenMPI distributed- and shared-memory, and 
coupling with RL agents is implemented via dynamic MPI-based communication. 

\textit{Differentiable Incompressible Solver (JAX).}
To enable gradient-enhanced reinforcement learning~\cite{son2023gippo}, HydroGym 
provides natively differentiable solvers written in JAX~\cite{jax2018github} for the 
incompressible Navier--Stokes equations. The Kolmogorov flow environment employs a 
pseudo-spectral method with $2/3$ de-aliasing on a doubly periodic domain, advancing 
the vorticity--streamfunction formulation with a fourth-order Runge--Kutta scheme; 
the three-dimensional turbulent channel flow uses a fully differentiable 
finite-difference DNS solver in which spatial derivatives are formed with sparse 
differentiation matrices and the pressure Poisson equation is solved via a 
biconjugate gradient stabilized (BiCGStab) method. Both environments implement a 
Gymnax interface~\cite{gymnax2022github} and inherit RL frameworks from 
PureJaxRL~\cite{purejaxrl2023github}, enabling a fully synchronous training pipeline 
compatible with JAX's automatic differentiation. Exact analytical gradients of the 
objective with respect to control parameters are obtained via \texttt{jax.grad} 
through the entire simulation trajectory. Parallelization across multiple random seeds 
is achieved using \texttt{vmap} for batch processing, and just-in-time compilation 
via \texttt{jit} optimizes execution on GPU hardware.

\textit{Differentiable Compressible Solver (JAX-Fluids).}
For gradient-based flow control in compressible and multiphase regimes, HydroGym 
integrates the JAX-Fluids solver~\cite{bezgin2023jax,bezgin2025jax}, a 
fully-differentiable, high-order CFD code written entirely in Python using the JAX 
library. The compressible Navier--Stokes equations are solved on structured Cartesian 
grids with arbitrary one-dimensional mesh stretching; convective fluxes are evaluated 
with high-order shock-capturing fifth-order weighted essentially non-oscillatory (WENO5-Z) or sixth-order targeted ENO (TENO6-A) reconstruction paired with an approximate Harten-Lax-van Leer-Contact (HLLC) Riemann solver, diffusive fluxes with high-order central differences, and time integration with explicit total variation diminishing (TVD) Runge--Kutta schemes. Multiphase flows are 
supported through both a sharp-interface level-set method and a five-equation 
diffuse-interface model. Positivity-preserving limiters are used to ensure robust integration in the presence of strong shocks or large density ratios.
The level-set implementation in JAX-Fluids additionally functions as an immersed boundary method (IBM), enabling flow simulations around complex geometries. The IBM is particularly well suited for active flow control problems, as control actions (e.g., blowing or suction) can be naturally incorporated through interface exchange terms. By adhering strictly to JAX's functional 
programming model, the entire simulation pipeline is end-to-end differentiable. Exact 
gradients of arbitrary scalar objectives, e.g., time-averaged drag or RL reward, with 
respect to control inputs or neural network weights are backpropagated through the 
full temporal trajectory via \texttt{jax.grad} or \texttt{value\_and\_grad}, with 
gradient checkpointing (\texttt{jax.checkpoint}) mitigating memory overhead for long 
rollouts. For large-scale HPC deployment, JAX-Fluids employs homogeneous domain 
decomposition across multiple Accelerated Linear Algebra (XLA) devices via \texttt{jax.pmap}, with inter-block 
halo exchanges performed exclusively through \texttt{jax.lax.ppermute} to keep the 
automatic differentiation graph intact across distributed, multi-node clusters.

\textit{Finite-Element Solver (Firedrake).}
For maximum code transparency and rapid prototyping of two-dimensional control 
problems, HydroGym provides a Firedrake~\cite{rathgeber2016firedrake} backend built 
on the Portable, Extensible Toolkit for Scientific Computation (PETSc), which offers automatic code generation for variational problems. The 
incompressible Navier--Stokes equations are discretized with Taylor--Hood elements 
(second-order continuous Galerkin for velocity, first-order for pressure) to ensure 
inf--sup stability, and time integration employs fully implicit schemes with automatic 
solver-parameter selection based on Reynolds number and grid resolution. The 
implementation follows a three-tier modular architecture separating physical problem 
definition (\texttt{PDEBase}), time-stepping (\texttt{TransientSolver}), and RL 
interfacing (\texttt{FlowEnv}), the latter implementing the Farama Foundation 
Gymnasium interface and translating between RL concepts and CFD operations. 
Parallelization leverages Firedrake's distributed mesh capabilities and PETSc's 
parallel sparse linear algebra routines.

\paragraph{Reinforcement Learning and Multi-Agent Infrastructure.}
Rather than heavily tuning hyperparameters for each configuration, which would hinder generalizability, HydroGym relies on robust observation and action normalization to ensure standard off-the-shelf RL algorithms serve as strong baselines. The platform integrates seamlessly with StableBaselines3~\cite{raffin2021stable}, TorchRL~\cite{bou2023torchrl}, and CleanRL~\cite{huang2022cleanrl} for model-free continuous control and PureJaxRL for natively differentiable environments. For standard model-free environments, performance of PPO, DDPG, and TD3 have been studied exemplarily. 

\textit{Gradient-Enhanced PPO.} For differentiable environments, HydroGym implements GPPO. Standard PPO relies on likelihood ratio methods or generalized advantage estimation (GAE) to estimate policy gradients. In GPPO, the analytical gradient of the reward trajectory with respect to the policy parameters, $\nabla_\theta \mathbb{E}_{\tau \sim \pi_\theta}[\sum R(s_t, a_t)]$, is computed exactly by backpropagating through the deterministic fluid dynamics solver. This analytical gradient is incorporated into PPO’s clipped surrogate objective, drastically reducing variance and improving sample efficiency.

\textit{Multi-Agent Reinforcement Learning.} To tackle the dimensionality barrier of spatially distributed 3D flows (such as 3D cylinders, channel and airfoil flows), HydroGym implements a decentralized MARL architecture. Global control domains are partitioned into locally invariant pseudo-environments. Multiple agents operate on these local partitions and may share a common control policy $\pi(a|s)$. To ensure unbiased data distributions in the shared replay buffer, overlapping mesh nodes are excluded from the pseudo-environment definitions, while identical actuation is enforced at boundary interfaces to preserve control continuity. Communication between the parallelized CFD domains and the centralized RL policy is managed efficiently via an MPI-based interface. 
Further details regarding the reinforcement learning agents are provided in the Supplementary Information (Section 4).

\paragraph{Environment Setup for discussed results.}
While the configurations discussed in the main text are briefly outlined below, comprehensive details concerning the physical characteristics, exact boundary conditions, observation/action space normalizations, and specific reward formulations for all environments are provided in the Supplementary Information (Section 5).

\textit{Circular Cylinder ($Re=3,900$).} The subcritical 3D cylinder flow represents a highly chaotic wake. The computational domains extend $[51.2 D \times 48 D]$ in 2D and $32D \times 16D \times 4D$ with periodic spanwise boundaries in 3D. Actuation is applied via ZNMF synthetic jets positioned at the top and bottom of the cylinder. The reward function targets drag minimization while penalizing lift oscillations: $r = -|C_D| - \omega |C_L|$. For MARL configurations, the span is decomposed into independent pseudo-environments, each controlling local jet pairs.

\textit{Fluidic Pinball ($Re=100 - 150$).} The pinball environment features three circular cylinders in an equilateral triangle layout, exposing agents to multiple bifurcation regimes, including symmetry-breaking pitchfork bifurcations and chaotic dynamics. The control mechanism consists of the independent surface rotation of all three cylinders. The reward formulation targets collective drag reduction: $r = -\sum_{i=1}^3 |C_{D,i}| - \omega \sum_{i=1}^3 |C_{L,i}|$, where the scaling factor $\omega$ restricts policies from exploiting asymmetric lift generation.

\textit{Open Cavity Flow ($Re=4,200 - 7,500$).} The cavity flow targets the stabilization of complex feedback loops and Kelvin-Helmholtz instabilities at the shear layer spanning the cavity opening. The reward function penalizes deviations of observed flow quantities from a target reference state: $r = -\sum_{i} \left(\frac{\text{obs}_i - \bar{o}_i}{\sigma_i}\right)^2$, where $\text{obs}_i$ are pressure and velocity measurements at shear-layer probes, $\bar{o}_i$ is the target mean approximated over 1,000 instability cycles, and $\sigma_i$ provides normalization scaling. Actuation is performed using localized jet actuators at the upstream cavity edge and, optionally, inside the cavity, providing multi-point flow manipulation to disrupt resonant interactions.

\textit{Transverse Gust Mitigation (NACA0012, $Re=1,000$).} This environment simulates extreme aerodynamic conditions using a highly disturbed inflow. A 1-cosine transverse gust with a gust ratio $G=2.0$ interacts with an airfoil at a high angle of attack ($\alpha=20^\circ$), threatening dynamic stall and severe load fluctuations. Control is achieved via three independent jet actuators distributed along the leading edge, each covering $3\%$ of the chord length. The reward function is designed to minimize gust-induced force variance while preserving the baseline aerodynamic efficiency: $r = -|C_L(t) - \overline{C}_{L,ref}| - \omega |C_D(t) - \overline{C}_{D,ref}|$.

\paragraph{Physics-Guided Transfer Learning and Zero-Shot Deployment.}
To bridge the gap between computationally tractable RL training and prohibitive industrial-scale CFD, HydroGym utilizes a physics-guided zero-shot transfer protocol. This method was deployed to control the suction-side turbulent boundary layer (TBL) of a NACA0012 wing section at a chord-based Reynolds number of $Re_c = 200,000$.

\textit{Surrogate Construction and Pre-training.} Direct RL exploration on the high-resolution wing is computationally prohibitive. Instead, our working hypothesis argues that the target control region on the wing (ranging from $x/c = 0.25$ to $0.86$) can be systematically partitioned into smaller chordwise blocks based on the local Clauser pressure-gradient parameter $\beta$, with a surrogate environment constructed for each. For the symmetric NACA0012 configuration at a $0^\circ$ angle of attack evaluated here, the moderate adverse pressure gradient across the control region naturally yields a single, continuous control block. Consequently, a single surrogate turbulent channel flow (TCF) is generated. Furthermore, the near-wall spatial resolution of the TCF matches the wing simulations in viscous units ($\Delta x^+, \Delta y^+, \Delta z^+$), ensuring that the effective footprint of the controllers remains geometrically consistent. 

Agents are trained exclusively within these TCF surrogates using the MARL framework (TD3 algorithm). The observation space relies strictly on near-wall metrics accessible in real-world scenarios: the wall-tangential ($u'$) and wall-normal ($v'$) velocity fluctuations sampled at a sensing plane of $y^+ = 15$. The action $a$ dictates the wall-normal blowing/suction velocity, strictly bounded by the local friction velocity ($-u_\tau \leq a \leq u_\tau$) and subjected to a ZNMF constraint across the control region. The reward function seeks to minimize the relative wall-shear stress, $r = 1 - \tau_w^{ctrl}/\tau_w^{ref}$.

\textit{Zero-Shot Deployment.} Following training in the idealized TCF surrogates, the optimized policies are deployed directly onto the corresponding partitioned blocks of the 3D NACA wing sections. No additional on-wing training or fine-tuning is performed. To map the policies appropriately, the agent’s observations are standardized using the spatially varying viscous scales of the wing. Because the viscous time unit $t^*$ evolves along the wing chord, the update frequency of the deployed actions is scaled by the local streamwise-averaged friction velocity $\langle u_\tau \rangle_x$ of each respective block. By learning generalized responses to near-wall streaks rather than overfitting to the macroscopic geometry of the channel, the RL policies successfully attenuate high-speed streaks on the wing, yielding substantial reductions in skin-friction drag at a fraction of the computational cost ($\sim 10^4$ magnitude reduction in required core-hours compared to direct on-wing training). Baseline comparison methods include uniform blowing and opposition control \cite{wang2025opposition}.

\section*{Data Availability}
The flow-field trajectories and datasets underlying the benchmark results reported in this work are available through the HydroGym Hugging Face repositories at \url{https://huggingface.co/datasets/dynamicslab/HydroGym-environments} . Owing to their volume, the largest direct-numerical-simulation datasets (individual flow fields up to 75\,GB; full test trajectories up to 15\,TB) cannot be deposited in their entirety; representative and compressed subsets sufficient to reproduce the reported results are provided, and complete datasets are available from the corresponding authors on reasonable request. Source data for the figures are provided with this paper.

\section*{Code Availability}
The HydroGym platform is publicly available at \url{https://dynamicslab.github.io/hydrogym/}. Pre-built docker containers are provided at \url{https://hub.docker.com/repositories/clagemann}.

\section*{Acknowledgements}
This work was supported by the National Science Foundation AI Institute in Dynamic Systems grant number 2112085 and The Boeing Company (CL, SM, SA, EL, NZ, SLB). SLB acknowledges support from the Army Research Office (W911NF-19-1-0045). CL acknowledges support from the German Research Foundation within the Walter Benjamin fellowship LA 5508/1-1. EL acknowledges support from the German Research Foundation within the Walter Benjamin fellowship MA 10764/1-1. MR acknowledges support from the German Research Foundation within the Walter Benjamin fellowship RU 2771/1-1. MM and WS acknowledge support from the German Research Foundation in the project SCHR309/68-1,2. NAA acknowledges funding through ERC Advanced Grant Project No. 101094463 (GENUFASD). RV acknowledges financial support from the University of Michigan and the Euopean Research Council (ERC) under Grant Agreement No. 2021-CoG-101043998 (DEEPCONTROL). The views and opinions expressed are those of the authors only and do not necessarily reflect those of the European Union or the European Research Council; neither the European Union nor the granting authority can be held responsible for them. The authors acknowledge the Gauss Centre for Supercomputing e.V. for providing computing time on the GCS Supercomputers.

\clearpage

\renewcommand{\arraystretch}{1.3}
\begin{table*}[!ht]
\centering
\begin{adjustbox}{width=\textwidth}
\begin{tabular}{l||c|c|c|c|c}
    \toprule
    \rowcolor{gray!20} \multicolumn{6}{c}{\large\textbf{2D ENVIRONMENTS}} \\
    \midrule
    \textbf{Environments} & \textbf{Re \#} & \textbf{Computational} & \textbf{Degrees of } & \textbf{Computational} & \textbf{Objective} \\
    & \textbf{range} & \textbf{domain} & \textbf{freedom} & \textbf{method} & \\
    \midrule \midrule
    \multirow{4}{*}{Cylinder} & $Re_D=100$ & $[51.2 D \times 48 D]$ & 287,096 & m-AIA | Firedrake & \multirow{4}{*}{\begin{tabular}{c} drag reduction \\ wake stabilization \end{tabular}} \\
    & $Re_D=200$ & $[51.2 D \times 48 D]$ & 287,096 & m-AIA | Firedrake & \\
    & $Re_D=1,000$ & $[51.2 D \times 48 D]$ & 831,456 & m-AIA | Firedrake & \\
    & $Re_D=3,900$ & $[51.2 D \times 48 D]$ & 934,732 & m-AIA | Firedrake & \\\hline
    \multirow{4}{*}{Pinball} & $Re_D=30$ & $[102.4 D \times 80 D]$ & 340,422 & m-AIA | Firedrake & \multirow{4}{*}{drag reduction} \\
    & $Re_D=75$ & $[102.4 D \times 80 D]$ & 340,422 & m-AIA | Firedrake & \\
    & $Re_D=100$ & $[102.4 D \times 80 D]$ & 340,422 & m-AIA | Firedrake & \\
    & $Re_D=150$ & $[102.4 D \times 80 D]$ & 340,422 & m-AIA | Firedrake & \\\hline
    \multirow{2}{*}{Cavity} & $Re_H=4,200$ & $[4.5 H \times 1.5 H]$ & 213,940 & m-AIA & \multirow{2}{*}{\begin{tabular}{c} shear layer \\ stabilization \end{tabular}} \\
    & $Re_H=7,500$ & $[4.5 H \times 1.5 H]$ & 213,940 & m-AIA | Firedrake & \\\hline
    \multirow{3}{*}{Square Cylinder} & $Re_H=200$ & $[51.2 D \times 48 D]$ & 245,288 & m-AIA & \multirow{3}{*}{\begin{tabular}{c} drag reduction \\ wake stabilization \end{tabular}} \\
    & $Re_H=1,000$ & $[51.2 D \times 48 D]$ & 411,168 & m-AIA & \\
    & $Re_H=3,900$ & $[51.2 D \times 48 D]$ & 490,030 & m-AIA & \\\hline
    \multirow{4}{*}{NACA 0012} & $Re_c=100 \mid \alpha=20^\circ$ & $[32 c \times 12 c]$ & 232,823 & m-AIA & \multirow{4}{*}{\begin{tabular}{c} gust mitigation \\ wake stabilization \end{tabular}} \\
    & $Re_c=100 \mid \alpha=40^\circ$ & $[32 c \times 12 c]$ & 232,860 & m-AIA & \\
    & $Re_c=1,000 \mid \alpha=20^\circ$ &$[32 c \times 12 c]$ & 232,823 & m-AIA & \\
    & $Re_c=1,000 \mid \alpha=40^\circ$ & $[32 c \times 12 c]$ & 232,860 & m-AIA & \\\hline
    \multirow{1}{*}{Stenotic Pipe} & $Re_D=100$ & $[16D \times D]$ & 25,000 & m-AIA & thermal regulation \\\hline
    \midrule
    \multirow{2}{*}{\shortstack[l]{Kolmogorov \\ (diff)}} & \multirow{2}{*}{$Re=40$--$500$} & \multirow{2}{*}{$[2\pi \times 2\pi]$} & \multirow{2}{*}{4,096} & \multirow{2}{*}{JAX} & \multirow{2}{*}{\begin{tabular}{c} extreme event mitigation \\ increase mixing \end{tabular}} \\
    & & & & & \\\hline
    \multirow{2}{*}{\shortstack[l]{Single Divergent Nozzle  \\ (diff)}} & \multirow{2}{*}{$Re=\infty \mid p_0 / p_\infty=4.6$} & \multirow{2}{*}{$[250 D \times 150D]$} & \multirow{2}{*}{450,000} & \multirow{2}{*}{JAX-Fluids} & \multirow{2}{*}{shock vector control} \\
    & & & & & \\ \bottomrule
\end{tabular}
\end{adjustbox}
\caption{Overview of HydroGym's 2D flow control environments available for reinforcement learning benchmarking. The table catalogs 2D flow configurations spanning bluff bodies (cylinder, square cylinder, pinball), cavity flows, airfoils (NACA 0012), internal flows (stenotic pipe), and differentiable physics environments (Kolmogorov flow and single divergent nozzle), across a range of Reynolds numbers ($Re$ - based on the diameter $D$, cavity height $H$, or the chord length $c$) and angles of attack ($\alpha$). Differentiable environments (bottom section) support gradient-based policy optimization.}
\label{ExtDataTable1}
\end{table*}

\clearpage

\renewcommand{\arraystretch}{1.3}
\begin{table*}[!ht]
\centering
\begin{adjustbox}{width=\textwidth}
\begin{tabular}{l||c|c|c|c|c}
    \toprule
    \rowcolor{gray!20} \multicolumn{6}{c}{\large\textbf{3D ENVIRONMENTS}} \\
    \midrule
    \textbf{Environments} & \textbf{Re \#} & \textbf{Computational} & \textbf{Degrees of } & \textbf{Computational} & \textbf{Objective} \\
    & \textbf{range} & \textbf{domain} & \textbf{freedom} & \textbf{method} & \\
    \midrule \midrule
    \multirow{3}{*}{Cylinder} & $Re_D=200$ & $[32 D \times 16 D \times 4D]$ & 46,033,631 & m-AIA & \multirow{3}{*}{\begin{tabular}{c} drag reduction \\ wake stabilization \end{tabular}} \\
    & $Re_D=1,000$ & $[32 D \times 16 D \times 4D]$ & 82,611,117 & m-AIA & \\
    & $Re_D=3,900$ & $[32 D \times 16 D \times 4D]$ & 82,611,117 & m-AIA & \\\hline
    \multirow{2}{*}{Cylinder (MARL)} & $Re_D=1,000$ & $[32 D \times 16 D \times 4D]$ & 82,611,117 & m-AIA & \multirow{2}{*}{\begin{tabular}{c} drag reduction \\ wake stabilization \end{tabular}} \\
    & $Re_D=3,900$ & $[32 D \times 16 D \times 4D]$ & 82,611,117 & m-AIA & \\\hline
    \multirow{4}{*}{Pinball} & $Re_D=30$ & $[104 D \times 80 D \times 4D]$ & 16,465,722 & m-AIA & \multirow{4}{*}{drag reduction} \\
    & $Re_D=75$ & $[104 D \times 80 D \times 4D]$ & 16,465,722 & m-AIA & \\
    & $Re_D=100$ & $[104 D \times 80 D \times 4D]$ & 16,465,722 & m-AIA & \\
    & $Re_D=150$ & $[104 D \times 80 D \times 4D]$ & 16,465,722 & m-AIA & \\\hline
    \multirow{2}{*}{Cavity} & $Re_H=4,200$ & $[2.5 H \times 1.25 H \times 2.25 H]$ & 34,990,848 & m-AIA & \multirow{2}{*}{\begin{tabular}{c} shear \\ layer stabilization \end{tabular}} \\
    & $Re_H=7,500$ & $[2.5 H \times 1.25 H \times 2.25 H]$ & 34,990,848 & m-AIA & \\\hline
    \multirow{3}{*}{Square Cylinder} & $Re_H=200$ & $[32 H \times 16 H \times 4H]$ & 51,627,840 & m-AIA & \multirow{3}{*}{\begin{tabular}{c} drag reduction \\ wake stabilization \end{tabular}} \\
    & $Re_H=1,000$ & $[32 H \times 16 H \times 4H]$ & 97,531,712 & m-AIA & \\
    & $Re_H=3,900$ & $[32 H \times 16 H \times 4H]$ & 97,531,712 & m-AIA & \\\hline
    \multirow{7}{*}{NACA 0012} & $Re_c=100 \mid \alpha=20^\circ$ & $[32 c \times 16 c \times 4c]$ & 46,696,118 & m-AIA & \multirow{7}{*}{\begin{tabular}{c} gust mitigation \\ wake stabilization \\ drag reduction \end{tabular}} \\
    & $Re_c=100 \mid \alpha=40^\circ$ & $[32 c \times 16 c \times 4c]$ & 46,696,118 & m-AIA & \\
    & $Re_c=1,000 \mid \alpha=20^\circ$ & $[32 c \times 16 c \times 4c]$ & 69,698,591 & m-AIA & \\
    & $Re_c=1,000 \mid \alpha=40^\circ$ & $[32 c \times 16 c \times 4c]$ & 69,698,591 & m-AIA & \\
    & $Re_c=10,000 \mid \alpha=12^\circ$ & $[32 c \times 16 c \times 0.5c]$ & 93,070,696 & m-AIA & \\
    & $Re_c=50,000 \mid \alpha=12^\circ$ & $[32 c \times 16 c \times 0.5c]$ & 93,070,696 & m-AIA & \\
    & $Re_c=200,000 \mid \alpha=0^\circ$ & $[6 c \times 4 c \times 0.1c]$ & 219,456,000 & nek5000 & \\\hline
    \multirow{2}{*}{Cube} & $Re_H=300$ & $[32 H \times 16 H \times 16H]$ & 32,338,392 & m-AIA & \multirow{2}{*}{\begin{tabular}{c} drag reduction \\ wake stabilization \end{tabular}} \\
    & $Re_H=3,700$ & $[32 H \times 16 H \times 16H]$ & 115,965,412 & m-AIA & \\\hline
    \multirow{2}{*}{Sphere} & $Re_D=300$ & $[32 D \times 16 D \times 16D]$ & 40,798,880 & m-AIA & \multirow{2}{*}{\begin{tabular}{c} drag reduction \\ wake stabilization \end{tabular}} \\
    & $Re_D=3,700$ & $[32 D \times 16 D \times 16D]$ & 89,649,414 & m-AIA & \\\hline
    \multirow{1}{*}{Stenotic Pipe} & $Re_D=100$ & $[16D \times D \times D]$ & 1,000,000 & m-AIA & thermal regulation\\\hline
    \multirow{4}{*}{\shortstack[l]{Turbulent \\ Boundary Layer}} & $Re_{\tau}=200$ & $[300\theta \times 50\theta \times 40\theta]$ & 757,105,126 & m-AIA & \multirow{4}{*}{\begin{tabular}{c} drag reduction \\ net power saving \end{tabular}} \\
    & $Re_{\tau}=1,400$ & $[300\theta \times 50\theta \times 11\theta]$ & 5,667,772,251 & m-AIA & \\
    & $Re_{\tau}=1,550$ & $[300\theta \times 50\theta \times 11\theta]$ & 5,667,772,251 & m-AIA & \\
    & $Re_{\tau}=2,200$ & $[300\theta \times 50\theta \times 11\theta]$ & 5,667,772,251 & m-AIA & \\\hline
    Turbulent Channel & $Re_{\tau}=180$ & $2\pi h \times h \times \pi h $ & 368,640 & nek5000 & drag reduction \\\hline
    \multirow{2}{*}{DRA2303} & $Re_c=400,000 \mid \alpha=0^\circ \mid Ma=0.2$ & $[44.5c \times 50c \times 0.1c]$ & 1,627,468,000 & m-AIA & \multirow{2}{*}{\begin{tabular}{c} drag reduction \\ net power saving \end{tabular}} \\
    & $Re_c=400,000 \mid \alpha=0^\circ \mid Ma=0.7$ & $[44.5c \times 50c \times 0.1c]$ & 1,627,468,000 & m-AIA & \\\hline
    \multirow{2}{*}{\shortstack[l]{Turbulent Channel \\ (diff)}} & \multirow{2}{*}{$Re_{\tau} = 180$} & \multirow{2}{*}{$[2\pi \times 2 \times \pi]$} & \multirow{2}{*}{373,248} & \multirow{2}{*}{JAX} & \multirow{2}{*}{reduce wall shear stress} \\
    & & & & & \\\hline
    \multirow{2}{*}{\shortstack[l]{Single Divergent Nozzle \\ (diff)}} & \multirow{2}{*}{$Re = \infty \mid p_0 / p_\infty=4.6$} & \multirow{2}{*}{$[250D \times 150D \times 150D]$} & \multirow{2}{*}{225,000,000} & \multirow{2}{*}{JAX-Fluids} & \multirow{2}{*}{shock vector control} \\
    & & & & & \\ \bottomrule
\end{tabular}
\end{adjustbox}
\caption{Overview of HydroGym's 3D flow control environments available for reinforcement learning benchmarking. The table catalogs 3D flow configurations spanning bluff bodies (cylinder, square cylinder, cube, sphere, pinball), cavity flows, airfoils (NACA 0012, DRA2303), wall-bounded flows (stenotic pipe, turbulent boundary layer, turbulent channel flow), and differentiable physics environments (turbulent channel flow and single divergent nozzle), across a wide range of Reynolds numbers ($Re$ - based on the diameter $D$, cavity height $H$, cube/square height $H$, the chord length $c$, or the friction velocity $u_\tau$), angles of attack ($\alpha$), and Mach numbers ($Ma$). Additional environments include MARL scenarios. Differentiable environments (bottom section) support gradient-based policy optimization.}
\label{ExtDataTable2}
\end{table*}
\FloatBarrier
    \putbib[refs]

\onecolumn
\clearpage

\thispagestyle{empty}  
\vspace*{\fill}
\begin{center}
{\Huge \textbf{Appendices}}
\end{center}
\vspace*{\fill}
\clearpage

\flushbottom
\fontsize{11}{13.2}\selectfont
\renewcommand{\appendixname}{SI}
\begin{appendices}
    \renewcommand{\thesection}{\arabic{section}}
    \renewcommand{\thefigure}{SI~\arabic{figure}}
    \setcounter{figure}{0}
    \renewcommand{\thetable}{SI~\arabic{table}}
    \setcounter{table}{0}
        
        \section{Extended Introduction} 
The control of fluid flows is a critical challenge across virtually every major industry, from aerospace and automotive to energy production and manufacturing. Improved flow control has the potential to transform trillion-dollar sectors by dramatically increasing energy efficiency, transportation systems, and industrial processes. However, fluid flow control remains notoriously challenging due to the nonlinear, multiscale nature of turbulent flows, resulting in control formulations that are high-dimensional, non-convex, and computationally intractable using traditional approaches~\cite{Brunton2015amr,Brunton2020arfm}. The convergence of recent breakthroughs in reinforcement learning (RL) with advances in differentiable physics simulation \cite{drake, mittal2023orbit} and specialized hardware acceleration has created an unprecedented opportunity to revolutionize fluid dynamics control through artificial intelligence (AI).

This revolution encompasses both the practical implementation of flow control and the fundamental scientific understanding that underpins it. AI-driven differentiable physics environments represent a transformative scientific tool for advancing our fundamental understanding of fluid dynamics. The key advantage of differentiable simulations lies in their ability to compute exact gradients of objective functions (such as drag, lift, or mixing efficiency) with respect to control inputs, geometric parameters, or initial conditions through automatic differentiation. This enables gradient-based optimization and sensitivity analysis that would require prohibitively many simulations with traditional adjoint methods or finite-difference approximations, particularly for high-dimensional control spaces and long time horizons.

Moreover, the combination of differentiable physics with RL creates a unique scientific tool: agents can autonomously discover non-intuitive control strategies and flow manipulation techniques that human intuition might overlook, effectively serving as computational experiments that probe the limits of what is physically achievable. These discovered strategies often reveal unexpected connections between control actuation and flow response, inspiring new hypotheses about underlying physical mechanisms. The gradient information also enables systematic inverse design—working backward from desired flow features to determine the forcing or geometry required to achieve them—which is invaluable for developing physics-based reduced-order models and improving turbulence closure schemes. By providing efficient pathways to explore the control-response landscape and extract interpretable patterns from successful strategies, these platforms establish a new methodology for scientific discovery that complements traditional CFD while offering unique capabilities for optimization, inverse problems, and automated discovery of control principles.

To understand the potential of this technological convergence, consider the vast economic significance of fluid flow control, which extends across industries representing over \$10 trillion in global market value. The industrial flow control sector alone represents a \$225 billion market\footnote{McKinsey \& Company, "Flow control: Sector at a crossroads?" \url{https://www.mckinsey.com/industries/industrials-and-electronics/our-insights/flow-control-sector-at-a-crossroads}}, with direct applications spanning aerospace systems valued at \$3.2 billion and projected to reach \$30 billion by 2032\footnote{Global Market Insights, "Aerospace \& Defense Fluid Conveyance Systems Market Size, 2032." \url{https://www.gminsights.com/industry-analysis/aerospace-defense-fluid-conveyance-systems-market}}, energy systems where industrial pumping accounts for nearly 20\% of global electrical demand, and manufacturing processes across sectors responsible for 38\% of industrial CO$_2$ emissions\footnote{U.S. Department of Energy, "Transformative Pathways for US Industry," January 2025. \url{https://www.energy.gov/sites/default/files/2025-01/transformative-pathways-for-us-industry.pdf}}. 
Breakthrough successes in wind energy control have reduced costs from 55+ cents/kWh in 1980 to under 3 cents/kWh today\footnote{American Clean Power Association, "Wind Power Facts and Statistics." \url{https://cleanpower.org/facts/wind-power/}}, creating thousands of jobs and generating \$10 billion in annual investments.

This economic imperative coincides with a remarkable period of breakthrough achievements in reinforcement learning, particularly in complex scientific applications that mirror the challenges of fluid dynamics control. Reinforcement learning has demonstrated transformative capabilities across increasingly complex scientific domains, recently resulting in the 2024 Nobel Prizes in Physics for foundational neural network research that enables modern RL systems\footnote{The Nobel Prize, "Press release: The Nobel Prize in Physics 2024." \url{https://www.nobelprize.org/prizes/physics/2024/press-release/}}, and in Chemistry for AlphaFold's revolutionary protein structure prediction~\citep{jumper2021highly}. AlphaFold solved the 50-year protein folding challenge by achieving experimental-level accuracy within minutes rather than months, predicting structures for over 200 million proteins and fundamentally transforming biological research\footnote{Google DeepMind, "AlphaFold." \url{https://deepmind.google/science/alphafold/}}. Concurrent breakthroughs in plasma control demonstrate RL's capability to master complex physical systems: DeepMind's deployment of deep RL controllers on the TCV tokamak achieved real-time control of 100+ million°C plasmas at 10 kHz frequencies, enabling previously impossible configurations including sustained ``droplets'' and high-elongation shapes critical for fusion energy~\citep{degrave2022magnetic}. These systems overcame fundamental challenges including 19+ coupled magnetic coils, millisecond timescales with instability growth rates exceeding 1.4 kHz, and safety-critical constraints where failures could damage multi-billion dollar facilities.

These applications share common success factors that provide crucial insights for advancing fluid flow control. The success of RL across diverse domains has been catalyzed by standardized benchmark environments that democratize research access and accelerate algorithmic innovation. OpenAI Gym has attracted over 2 million users across 190 countries\footnote{OpenAI, "OpenAI Gym Beta." \url{https://openai.com/index/openai-gym-beta/}}, while specialized environments, such as MuJoCo for robotics and Atari for discrete control established evaluation frameworks that transformed entire research communities~\citep{brockman2016openai}. These platforms enabled fair algorithmic comparison, reproducible research, and systematic progress tracking that proved essential for breakthrough developments including AlphaGo, robotics deployments, and large language model alignment through reinforcement learning from human feedback.

However, despite these remarkable successes in analogous complex domains, fluid dynamics has remained largely resistant to similar breakthroughs due to a constellation of fundamental technical barriers. The primary obstacle is computational intractability: training effective RL agents typically requires thousands of environment interactions, but each evaluation demands expensive computational fluid dynamics (CFD) simulations that can require hours on high-performance computing clusters. This computational bottleneck is magnified by the inherent sample inefficiency of model-free RL methods, making it practically impossible for complex turbulent flow problems. Moreover, traditional CFD approaches face severe limitations in real-time control applications, with typical simulations requiring wall-clock times orders of magnitude longer than the physical processes they model.

Beyond computational constraints, fundamental challenges in generalization and adaptability have further impeded progress. Current RL approaches trained at specific Reynolds numbers fail to transfer to different flow regimes, and agents optimized for particular geometries cannot adapt to modified configurations~\citep{vinuesa2022enhancing}. The transition from laminar to turbulent conditions introduces chaotic, irregular dynamics that destabilize training algorithms, while the multiscale nature of turbulence---spanning from energy-containing large scales to dissipative Kolmogorov microscales---necessitates extensive data collection across multiple scales, which amplifies computational costs. Traditional control approaches face complementary limitations: PID controllers with fixed parameters cannot adapt to highly nonlinear turbulent flows, while optimal control theory encounters severe time horizon limitations due to instability in adjoint equations for chaotic systems~\citep{kochkov2021machine}.

Compounding these technical challenges, the fluid dynamics community has lacked the comprehensive benchmark platforms that have enabled other domains. While computer vision benefits from standardized datasets, such as ImageNet, that enable systematic algorithmic comparison, fluid dynamics researchers work with limited, problem-specific configurations that prevent fair evaluation and reproducible research. Most existing fluid dynamics RL efforts have concentrated on computationally inexpensive 2D configurations at moderate Reynolds numbers, with the extension to highly turbulent flows, complex 3D geometries, and multiphysics applications representing the current frontier of research.

Remarkably, the technological landscape has undergone a dramatic transformation that now makes it possible to overcome these historical barriers. GPU-accelerated differentiable fluid simulators now achieve orders-of-magnitude speedups over traditional CFD methods, with modern platforms enabling parallel environment training at scales previously requiring supercomputers~\citep{kochkov2021machine}. The ML4CFD competition at NeurIPS 2024, featuring over 240 teams, demonstrated 300-600x speedups over traditional solvers while maintaining accuracy\footnote{"NeurIPS 2024 ML4CFD Competition: Results and Retrospective Analysis."  \url{https://arxiv.org/abs/2506.08516} }, indicating the maturity of these approaches for practical applications. Differentiable reinforcement learning methods specifically designed for continuous control problems address sample efficiency challenges, with approaches using analytic gradients from differentiable simulation to dramatically reduce sample complexity in multiphysics environments~\cite{xing2024stabilizing}. Physics-informed neural networks provide principled ways to incorporate fluid dynamics knowledge into RL systems, achieving superior performance in strongly nonlinear problems with sparse data~\citep{Puri2024}.

The commercial sector has validated this technological readiness through concrete deployments and performance improvements. Major CFD vendors including Ansys and Siemens have launched GPU-accelerated solvers showing 20-33x performance improvements, while enterprise platforms provide production-grade tools for physics-informed machine learning. This convergence of mature differentiable simulation technology, breakthrough RL algorithms, and accessible hardware represents the first time that large-scale fluid dynamics RL training has become computationally feasible.

The urgency of capitalizing on this technological convergence is underscored by the massive economic stakes and immediate industrial needs for improved fluid control solutions. Advanced aerodynamic control in automotive and aerospace applications could generate billions in fuel savings across global fleets. For instance, removing side mirrors alone in vehicles (accounting for around 6\% of form drag~\cite{Ruettgers2019}) improves fuel economy by 1.5-2 miles per gallon, while comparable drag reductions in aircraft yield proportionally larger savings due to higher fuel consumption rate~\cite{lagemann2024extending, lagemann2024impact}. Coordinated wind turbine controls can increase farm output by 4-5\%\footnote{U.S. Department of Energy, "Next-Generation Wind Technology." \url{https://www.energy.gov/eere/wind/next-generation-wind-technology}}, representing substantial returns across the \$10 billion annual wind investment market. Industrial applications present even larger opportunities: pumping systems account for nearly 20\% of global electrical energy demand, with potential energy savings up to 75\% through advanced control. Manufacturing industries face direct competitive pressures where fluid control improvements translate to immediate economic advantages, from billion-dollar steel production operations to chemical processing facilities requiring millisecond response times for safety-critical applications.

Responding to this critical convergence of technological capability and economic opportunity, we introduce HydroGym, a benchmark flow control platform designed to bridge the gap between cutting-edge RL method development and fluid dynamics applications. HydroGym provides a flexible, solver-independent interface that seamlessly integrates with industry-standard RL frameworks while supporting both differentiable and non-differentiable fluid dynamics environments. Our platform abstracts computational fluid dynamics complexity from control algorithm development, enabling researchers from both communities to advance the field through standardized benchmarks and reproducible evaluation protocols.

The specific contributions of this work directly address each of the critical barriers that have prevented breakthrough progress in fluid dynamics RL:

\begin{itemize}
    \item \textbf{Comprehensive benchmark platform}: Introduction of the first scalable RL platform specifically designed for fluid dynamics control, featuring a solver-independent architecture that supports diverse CFD backends and computational configurations.
    \item \textbf{Diverse environment suite}: 61+ non-differentiable flow control environments with increasing complexity, from canonical flows past cylinders to advanced turbulent control scenarios on 3D aircraft wings, plus an initial set of differentiable flow control environments enabling gradient-based RL methods.
    \item \textbf{Systematic algorithmic evaluation}: Comprehensive performance assessment using industry-standard algorithms including Proximal Policy Optimization (PPO), Deep Deterministic Policy Gradient (DDPG), and Twin Delayed DDPG (TD3), establishing baseline performance metrics for future research.
    \item \textbf{Democratized research access}: Platform enables RL researchers and fluids experts to collaborate via standardized interfaces, lowering the barrier to entry and accelerating interdisciplinary progress.
    \item \textbf{Extensible foundation}: Readily extensible framework supports integration of new CFD solvers, environments, and RL algorithms, to evolve with advancing technology and research needs.
\end{itemize}

Through this comprehensive approach, HydroGym addresses the specific technical barriers that have prevented breakthrough progress -- computational intractability through state-of-the-art GPU-powered CFD solvers and efficient simulation interfaces, sample inefficiency through differentiable environments, generalization challenges through diverse benchmarks, and accessibility barriers through democratized platforms -- thereby enabling the transition from traditional engineering approaches to AI-driven fluid control systems. The platform arrives at a critical juncture where enabling technologies have matured and economic pressures intensify, positioned to catalyze transformative progress comparable to breakthroughs in protein folding, plasma control, and other complex scientific domains. Just as AlphaFold demonstrated that the convergence of deep learning, domain expertise, and computational resources could solve seemingly intractable scientific challenges, HydroGym establishes the foundation for similar breakthroughs in fluid dynamics control. Success in this domain promises a fundamental transformation of how we approach complex flow systems across trillion-dollar industries, establishing artificial intelligence as an essential tool for addressing the most challenging problems in fluid mechanics and beyond.

\section{Extended literature review}
\label{appendix:literature_review}
The intersection of reinforcement learning and flow control represents a paradigm shift in fluid system optimization. Traditional methods rely on predetermined strategies based on domain knowledge, while reinforcement learning enables autonomous discovery of optimal control policies through the direct interaction with flow environments. This capability is particularly valuable in turbulent flows, where traditional linear control theory fails due to high-dimensional, chaotic dynamics.
Since the first successful demonstration of deep reinforcement learning for cylinder wake control~\cite{rabault2019artificial}, the field has experienced rapid growth. The progression from proof-of-concept studies to practical implementations reveals the maturation of core methodologies and sophisticated frameworks capable of handling realistic engineering problems. This evolution has been driven by advances in deep learning architectures, increased computational resources, and growing recognition of traditional control limitations for complex fluid systems.

\paragraph{Canonical Flow Problems and Foundational Methods.}
The journey toward practical reinforcement learning in flow control began with carefully chosen problems that balanced complexity with computational feasibility. The application of reinforcement learning to circular cylinder wake control was a first step, establishing foundational principles for the field~\cite{rabault2019artificial}. The canonical von Kármán vortex street suppression provided a computationally tractable testbed, allowing researchers to develop and validate core methodologies before tackling more challenging applications.

Building on these initial successes, numerous studies consistently achieved $8-15\%$ drag reduction with minimal actuation energy (0.5-1\% of momentum deficit), demonstrating that neural agents could autonomously discover effective control strategies without prior knowledge of flow physics~\cite{chatzimanolakis2024learning, tokarev2020deep, xu2020active, rabault2019accelerating, wang2024dynamic}. These results were encouraging because they showed that reinforcement learning could match or exceed the performance of carefully tuned traditional controllers without prior understanding of the underlying flow physics.
Crucially, experimental validation eliminated concerns about simulation-to-reality transfer that had plagued earlier computational studies. Laboratory studies using rotating cylinders achieved similar performance improvements~\cite{fan2020reinforcement}, with drag reductions of 25-30\% at Reynolds numbers up to 140,000. This success proved that reinforcement learning can handle noise, measurement uncertainties, and modeling errors inherent in real fluid systems. 

Attention has since turned to more complex geometries that better represent practical engineering challenges. Square cylinder flows present additional complexity due to fixed separation points and pronounced pressure fluctuations~\cite{chen2023deep, yan2023stabilizing, yousif2023optimizing, yan2024deep, xia2024active}. Reinforcement learning policies often learn to employ multi-frequency patterns differing significantly from natural shedding frequencies, highlighting the discovery potential of learning-based methods. 

The field has rapidly matured towards industrially relevant flows. Studies at Reynolds numbers exceeding 200,000 have achieved substantial drag reduction through multi-frequency control strategies~\cite{jia2024robust}. Further, multi-agent reinforcement learning has shown superior performance compared to single-actuator approaches~\cite{suarez2025active, suarez2025flow}, setting the stage for more complex applications requiring distributed control.

\paragraph{Shape Optimization and Advanced Applications.}
A natural next step in RL for fluids was to explore design optimization where the geometry itself becomes the control parameter. Reinforcement learning for shape optimization represents a major step toward autonomous design generation~\cite{viquerat2021direct,Ruettgers2021}. This transition from controlling existing geometries to generating new ones opened new possibilities for fluid engineering.
Modern frameworks parameterize complex shapes via Bézier curves or B-splines with hundreds of design variables~\cite{lou2023aerodynamic, bhola2023multi}; optimization in these spaces was previously intractable.  The key advantage over conventional optimization lies in reinforcement learning's ability to navigate high-dimensional design spaces without getting trapped in local optima, while simultaneously handling multiple competing objectives.

This multi-objective capability has proven particularly valuable for airfoil design, where competing requirements must be balanced across flight conditions. Reinforcement learning frameworks generate continuous Pareto fronts representing optimal trade-offs between lift and drag across ranges of Reynolds numbers and angles of attack~\cite{dussauge2023reinforcement}. Performance comparisons consistently demonstrate superiority over traditional optimization methods, with superior resistance to local minima and better generalization to untested operating conditions.

The success in two-dimensional applications naturally led to more ambitious and industrially relevant three-dimensional flows. Automotive and marine applications have leveraged reinforcement learning for complex three-dimensional geometries~\cite{patel2024enhancing,tran2024aerodynamics}, where traditional optimization approaches become computationally prohibitive. Marine applications have achieved significant reductions in acoustic signatures while maintaining hydrodynamic performance~\cite{yeo2024deep}, while hull optimization frameworks generate designs meeting multiple constraints simultaneously~\cite{oh2024reinforcement}, demonstrating the technology's readiness for real-world engineering problems.

\paragraph{Separation Control and Turbulent Flow Applications.}
 Flow separation control represents one of the most demanding applications, requiring precise manipulation of boundary layer dynamics in highly unsteady environments. Reinforcement learning has revealed sophisticated strategies surpassing traditional periodic actuation~\cite{font2025deep, xiang2024experimental}, often discovering control mechanisms that human experts had not previously considered.
Plasma actuators enable precise boundary layer manipulation~\cite{shimomura2020closed}, with learned policies discovering multi-frequency control strategies that simultaneously modify large-scale structures while fine-tuning smaller-scale phenomena. This multi-scale control represents a significant advance over traditional approaches that typically target single frequency ranges or specific flow scales.

The complexity of separation control is perhaps best illustrated in dynamic stall applications, where reinforcement learning has achieved complete stall prevention rather than mere mitigation~\cite{garcia2025deep}. These applications require real-time adaptation to rapidly changing flow conditions, demonstrating the technology's potential for autonomous flight control. Multi-actuator systems coordinated by multi-agent frameworks enable spatially distributed manipulation with precision unattainable through centralized approaches.

The application of reinforcement learning to fully turbulent channel flow control has emerged as a canonical problem for high Reynolds number evaluation~\cite{guastoni2023deep,lee2023turbulence, zhou2025reinforcement, sonoda2023reinforcement}, representing conditions typical of many engineering applications. Successful agents achieve drag reduction levels of 20-40\% across Reynolds numbers spanning two orders of magnitude, with performance scaling favorably with system size.
The RL control mechanisms often differ fundamentally from traditional opposition control, employing sophisticated strategies that redistribute turbulent kinetic energy across different components and locations. Multi-agent reinforcement learning has addressed scalability for spatially distributed control, enabling parallelization across multiple locations while avoiding exponential complexity growth, pointing toward practical implementations in large-scale systems.

\paragraph{Bio-Inspired and Bio-medical Applications.}
Researchers have also explored how RL can provide insights into biological systems, while simultaneously investigating bio-inspired approaches for artificial applications. These studies yield insights into natural propulsion while revealing opportunities for artificial systems to potentially exceed biological performance.
Self-propelled swimming studies reproduce complex behaviors including rheotaxis and Kármán gating~\cite{novati2017synchronisation, verma2018efficient, qin2023reinforcement, xiong2025chemotactic}, demonstrating that reinforcement learning can discover sophisticated control strategies evolved by natural swimmers. Multi-agent applications to collective swimming reveal coordination emerging from individual optimization without explicit communication~\cite{zhu2022learning,walchli2025inverse}, suggesting new approaches for autonomous swarms.

Reinforcement learning for bio-medical fluids has mainly been used for respiration and hemodynamics.
In the respiratory system, patient-specific respiratory flow simulations have been coupled with reinforcement learning to propose surgical interventions in the nasal cavity that optimize nasal resistance and improve the airway’s capacity to heat incoming air~\cite{Ruettgers2024,Ruettgers2025a}.
In the cardiovascular system, reinforcement learning agents have been trained to regulate nerve stimulation to maintain target heart rate and mean arterial pressure~\cite{Parisa2024}, to control the motor speed of cardiovascular pumps in order to restore pulsatility under varying heart failure conditions~\cite{Shi2025}, or to design optimal stent placements in patient-specific aneurysm geometries~\cite{Hachem2023}.
Together, these studies highlight the potential of reinforcement learning and physics-based simulations to enable personalized medicine and support clinical decision-making.

\paragraph{Computational Enhancement and Algorithmic Advances.}
Parallel to advances in direct flow control applications, reinforcement learning has begun transforming the computational tools used to study fluid mechanics itself. RL for CFD solver optimization represents a paradigm shift toward intelligent computational fluid dynamics, where the simulation process itself becomes adaptive and self-optimizing.
Adaptive mesh refinement applications treat grid optimization as sequential decision problems~\cite{yang2023reinforcement, foucart2023deep, zhu2025unstructured}, with learned strategies demonstrating superior performance compared to conventional error-based approaches. High-order method integration enables simultaneous optimization of mesh topology and polynomial order distributions~\cite{huergo2024reinforcement}, potentially revolutionizing how CFD simulations are constructed and executed.

Concurrently, reinforcement learning has emerged as a transformative approach for turbulence modeling, addressing long-standing challenges in Reynolds-Averaged Navier-Stokes (RANS) closure and Large-Eddy subgrid scale modeling. By treating model parameter optimization as an adaptive learning problem where coefficients adjust based on local flow conditions~\cite{duraisamy2019turbulence, kurz2023deep, novati2021automating, zhang2025application}, these approaches overcome the fundamental limitation of traditional turbulence models that rely on universal constants derived from canonical flows.

Multi-agent reinforcement learning frameworks have been successfully applied to wall-model development for large eddy simulations, enabling distributed optimization of near-wall turbulence representations while maintaining computational efficiency~\cite{bae2022scientific,zhou2024wall}. Physics-informed reinforcement learning approaches have integrated governing differential equations directly into the learning framework, ensuring that learned turbulence models satisfy fundamental conservation laws while improving predictive accuracy across diverse flow configurations~\cite{hu2024efficient, rodwell2023physics}.
Data-driven model development leveraging high-fidelity simulation databases has demonstrated superior performance compared to traditional fixed-parameter approaches, with learned models providing better uncertainty quantification and adaptability to new flow regimes. Model-consistent training strategies that incorporate CFD solvers directly into the learning process have addressed the distribution shift problems that previously limited machine learning applications to turbulence modeling, enabling effective deployment in practical engineering simulations.

These advances represent a paradigm shift toward intelligent turbulence modeling where algorithms continuously adapt their closure strategies based on evolving flow physics rather than relying on universal constants derived from canonical flows. This adaptability is particularly valuable for complex engineering flows that deviate significantly from the canonical configurations used to calibrate traditional models.

Another promising direction targets model-based reinforcement learning approaches in flow control learning surrogate models of fluid dynamics to enable more sample-efficient policy optimization~\cite{ye2025model, weiner2025model, zolman2024sindy,lagemann2026learning}. These lightweight, physics-informed models provide significant computational advantages over high-fidelity CFD simulations during policy training while maintaining sufficient accuracy for effective control strategy development, enabling practical deployment of reinforcement learning in scenarios where extensive interaction with expensive fluid solvers would otherwise be prohibitive.

The convergence of reinforcement learning with fluid mechanics has evolved from isolated demonstrations to a comprehensive framework capable of addressing fundamental challenges across the discipline, from turbulence closure modeling to autonomous system design. This progression demonstrates not just the maturation of individual techniques, but the emergence of a new paradigm for approaching fluid mechanics problems that emphasizes adaptation, learning, and autonomous optimization. Looking forward, the integration of physics-informed learning algorithms with high-performance computing promises to unlock previously intractable optimization problems in fluid systems, potentially revolutionizing fields ranging from renewable energy harvesting to biomedical device design. As these methodologies mature beyond laboratory settings toward real-world implementation, they represent a paradigmatic shift toward intelligent, adaptive fluid systems that autonomously optimize their performance in response to changing environmental conditions, fundamentally changing how we design and operate fluid mechanical systems.

\section{HydroGym - Numerical Backends and Implementation} 
\label{appendix:Implementation}

HydroGym provides multiple computational backends to accommodate diverse research
requirements and computational resources in flow control applications. The framework's
modular architecture supports six distinct solver implementations: a high-performance
lattice Boltzmann method (LBM) solver built on the m-AIA framework for large-scale
simulations (Section~\ref{appendix:Implementation:MAIA}), a compressible
finite-volume (FV) solver also within m-AIA targeting wall-bounded turbulent flows at
higher Mach numbers (Section~\ref{appendix:Implementation:MAIA:FV}), a
spectral-element solver based on Nek5000 for incompressible high-Reynolds-number
configurations requiring spectral accuracy
(Section~\ref{appendix:Implementation:nek5000}), a differentiable incompressible
JAX-based implementation optimized for gradient-enhanced machine learning workflows
(Section~\ref{appendix:Implementation:JAX}), a differentiable compressible solver
based on JAX-Fluids for high-order flow control in compressible and multiphase regimes
(Section~\ref{appendix:Implementation:JAXFluids}), and a Firedrake finite element
backend designed for maximum code transparency and rapid prototyping
(Section~\ref{appendix:Implementation:Firedrake}). Each implementation targets
specific use cases while maintaining a unified interface for reinforcement learning
integration, enabling researchers to select the most appropriate computational
approach based on their performance requirements, hardware constraints, and research
objectives.

\subsection{m-AIA - Lattice-Boltzmann solver implementation}
\label{appendix:Implementation:MAIA}

The direct numerical simulations are conducted using a lattice Boltzmann method (LBM) embedded within the m-AIA solver framework, which has been continuously developed at the Institute of Aerodynamics of RWTH Aachen University for over two decades~\cite{maia}. This method solves a discretized form of the Boltzmann equation, which describes the evolution of the particle distribution function $f(\mathbf{x}, \mathbf{v}, t)$ in time $t$ and space $x=(x,y,z)$:
\begin{equation}
    \frac{\partial f}{\partial t} + \mathbf{v} \cdot \nabla f = \Omega(f).
\end{equation}
Here, $\Omega(f)$ is the collision operator, which models the redistribution of particle populations due to intermolecular collisions. Through a first-order Chapman-Enskog expansion of this equation, the Navier-Stokes equations can be recovered, establishing the method's connection to macroscopic fluid dynamics.

The discretization of the continuous Boltzmann equation results in the lattice Boltzmann equation (LBE), which evolves discrete particle populations $f_i$ on a computational grid. The LBE update rule follows a two-step process consisting of collision and streaming phases:
\begin{equation}
    f_i(\mathbf{x} + \mathbf{c}_i \Delta t, t + \Delta t) = f_i(\mathbf{x}, t) + \Omega_i(f).
\end{equation}
For the present test cases, a three-dimensional lattice with 27 discrete velocity vectors ($\mathbf{c}_i$) is employed, which is known as the D3Q27 model. From these discrete particle distributions, macroscopic fluid variables such as density $\rho$ and velocity $\mathbf{u}$ are computed from the moments:
\begin{equation}
    \rho = \sum_i f_i \quad \text{and} \quad \rho\mathbf{u} = \sum_i \mathbf{c}_i f_i.
    \label{eq:macros}
\end{equation}

Overall, m-AIA's LBM solver offers two different collision operators. The first approach is the widely-used Bhatnagar-Gross-Krook (BGK) model~\cite{bhatnagar1954model}, which relaxes the particle distributions towards a local Maxwellian equilibrium $f_i^{\text{eq}}$ with a single relaxation frequency $\omega_{\text{BGK}}$:
\begin{equation}
    \Omega_i = \omega_{\text{BGK}}(f_i^{\text{eq}} - f_i).
\end{equation}
The relaxation frequency is related to the effective kinematic viscosity $\nu_{\text{eff}}$ of the fluid by $\omega_{\text{BGK}} = (\Delta t c_s^2) / (v_{\text{eff}} + 0.5 \Delta t c_s^2)$, where $c_s$ is the lattice speed of sound. For isothermal and low-Mach number flows, the equilibrium distribution is given by:
\begin{equation}
    f_i^{\text{eq}} = w_i \rho \left[ 1 + \frac{\mathbf{c}_i \cdot \mathbf{u}}{c_s^2} + \frac{(\mathbf{c}_i \cdot \mathbf{u})^2}{2c_s^4} - \frac{\mathbf{u} \cdot \mathbf{u}}{2c_s^2} \right],
\end{equation}
with $w_i$ being the lattice-dependent weighting factors.

For enhanced stability in simulations of higher Reynolds number flows, an alternative cumulant-based collision operator is employed~\cite{geier2015cumulant}. In this approach, the collision is performed in the space of statistical cumulants $c_\alpha$, which are relaxed towards their equilibrium values with a set of distinct relaxation rates $\omega_\alpha$. The post-collision state in cumulant space is expressed as $c_\alpha^* = c_\alpha + \omega_\alpha(c_\alpha^{\text{eq}} - c_\alpha)$. Following the methodology of Geier et al.~\cite{geier2015cumulant}, all relaxation rates are set to unity except for one, which is set equal to $\omega_{\text{BGK}}$ to control the fluid viscosity.

For thermal simulations, a second set of particle distributions $g_i$ is solved.
To compute the thermal equilibrium distributions, three approaches have been validated and employed in respiratory flow simulations with m-AIA: (1) treating the temperature ($T$) solely as a passive scalar~\cite{Lintermann2013,Ruettgers2021}, (2) including internal energy effects~\cite{Ruettgers2022}, and (3) considering total energy~\cite{Ruettgers2024,Ruettgers2025a}. 
Similar to Eq.~\ref{eq:macros}, $T$ is computed from the moments of $g_i$.     

Furthermore, local grid refinement is implemented based on the cell-centered interface layout proposed by Eitel-Amor et al.~\cite{EitelAmor2013} to efficiently resolve flow features across different length scales. This refinement strategy involves adjusting the relaxation frequency on different grid levels to maintain a constant kinematic viscosity across the entire domain. In order to keep the Mach number constant across the grid transition, the time step is scaled proportionally to the local grid spacing $\Delta t \sim \Delta$. The viscosity is kept constant by setting the local relaxation frequency according to:
\begin{equation}
    v_{\text{eff}} = \Delta t c_s^2 \left( \frac{1}{\omega_{\text{BGK}}} - \frac{1}{2} \right).
\end{equation}
While the macroscopic quantities and thus the equilibrium state are interpolated in space and time, the non-equilibrium part is additionally rescaled ensuring consistency of the viscous stress tensor, see~\cite{dupuis2003theory}.

\paragraph{Boundary Conditions and actuation control.}
Solid boundaries within the fluid domain are handled using an interpolated bounce-back scheme~\cite{Bouzidi2001}. This method provides a second-order accurate representation of the no-slip condition for arbitrarily shaped geometries. For the far-field boundaries, inlet and outlet conditions are imposed by setting the particle distributions to an equilibrium state based on prescribed velocity, density, or temperature values, while using second-order extrapolation for any unknown variables from the domain's interior.

Flow control is implemented by changing the inflow temperature in the thermal benchmark cases, and by imposing time-varying Dirichlet boundary conditions on the surface of the immersed body in the remaining cases. These conditions modify the standard no-slip wall condition by specifying a non-zero velocity vector, $\mathbf{u}_w$, which is then enforced through the interpolated bounce-back scheme. Two distinct types of actuation are employed: synthetic jet actuation and surface rotation. Both methods utilize a temporal ramping mechanism to ensure smooth transitions between control actions, which is critical for numerical stability.\\

\textit{Temporal Smoothing of Control Actions.}
To prevent numerical instabilities arising from discontinuous changes in boundary conditions, the control action is never applied instantaneously. Instead, the actuation magnitude $A(t)$ (representing either jet velocity or angular velocity) is smoothly interpolated from its value at the previous control step, $A^{\text{old}}$, to the new target value, $A^{\text{new}}$, over a predefined number of sub-iterations $N_r$.

Hence, the instantaneous actuation magnitude is calculated as:
\begin{equation}
    A(t) = A^{\text{old}} + (A^{\text{new}} - A^{\text{old}}) \cdot R(\tau),
    \label{eq:ramp_general}
\end{equation}
where $\tau = (t - t_{\text{update}}) / (N_r \Delta t)$ is the normalized time within the ramping interval, such that $\tau \in [0, 1]$. The function $R(\tau)$ is a normalized ramp function that maps from 0 to 1. Two different ramp functions are available to tune the actuation response:
\begin{itemize}
    \item Hyperbolic Tangent Ramp: This function provides a smooth, S-shaped transition that is gentle at the beginning and end of the ramp period. It is defined as:
\begin{equation}
    R_{\text{tanh}}(\tau) = \frac{1}{2} \left( \tanh\left(b(\tau - 0.5)\right) + 1 \right).
    \label{eq:ramp_tanh}
\end{equation}
The parameter $b$ controls the steepness of the ramp.

\item Exponential Ramp: This function provides a ramp that starts slowly and accelerates, allowing for a more aggressive actuation towards the end of the interval. It is defined as:
\begin{equation}
    R_{\text{exp}}(\tau) = \frac{e^{k\tau} - 1}{e^k - 1}.
    \label{eq:ramp_exp}
\end{equation}
The parameter $k$ controls the curvature of the exponential ramp.
\end{itemize}

\textit{Synthetic Jet Actuation.}
This boundary condition models synthetic jets by prescribing a velocity profile over specified regions of the wall. Each jet $j$ is defined by its center position $\mathbf{x}_j$, blowing angle $\alpha_j$, and width $W_j$. The velocity profile is parabolic, ensuring it is maximal at the centerline and smoothly decays to zero at the jet's edges. The velocity vector $\mathbf{u}_w(\mathbf{x}, t)$ for a point $\mathbf{x}$ within the jet's active region is determined by the product of a spatial profile function $\mathbf{S}_j(\mathbf{x})$ and the time-varying jet velocity magnitude $A_j(t)$:
\begin{equation}
    \mathbf{u}_w(\mathbf{x}, t) = A_j(t) \cdot \mathbf{S}_j(\mathbf{x}).
\end{equation}
The spatial profile vector $\mathbf{S}_j(\mathbf{x})$ combines the jet's shape and direction:
\begin{equation}
    \mathbf{S}_j(\mathbf{x}) = \left( 1 - \left( \frac{2d_\perp}{W_j} \right)^2 \right) \mathbf{t}_j,
\end{equation}
where $\mathbf{t}_j = (\cos\alpha_j, \sin\alpha_j)$ is the unit vector in the direction of the jet, and $d_\perp$ is the perpendicular distance from the point $\mathbf{x}$ to the jet's centerline. This profile is applied for locations where $d_\perp \le W_j/2$. The jet magnitude $A_j(t)$ is supplied by the control agent and interpolated using one of the temporal smoothing functions described by Eq.~\ref{eq:ramp_general}.\\

\textit{Surface Rotation Actuation.}
This boundary condition simulates the rotation of solid surfaces, such as a cylinder. The actuation is defined by an angular velocity $\omega(t)$ around a specified center of rotation $\mathbf{x}_c$ and axis of rotation $\hat{\mathbf{a}}$. The resulting tangential velocity at any point $\mathbf{x}$ on the boundary is given by the cross product:
\begin{equation}
    \mathbf{u}_w(\mathbf{x}, t) = \omega(t) \hat{\mathbf{a}} \times (\mathbf{x} - \mathbf{x}_c).
\end{equation}
For a 2D simulation in the $x-y$ plane, the rotation axis is fixed to $\hat{\mathbf{a}} = (0, 0, 1)$, and the control action is the scalar angular velocity $\omega_z(t)$, which serves as the actuation magnitude $A(t)$. The velocity vector at the wall simplifies to:
\begin{equation}
    \mathbf{u}_w(x, y, t) = \omega_z(t) (- (y - y_c), (x - x_c)).
\end{equation}
The angular velocity $\omega_z(t)$ provided by the control agent is temporally smoothed between control updates.

\paragraph{MPI communications and GPU acceleration.}
HydroGym's C\texttt{++} code implementation benefits from hybrid parallelization based on MPI and shared memory models, i.e., OpenMP and higher-level parallelism features introduced by C\texttt{++}20 parallel algorithms. This allows for a hardware-agnostic implementation on both CPU and GPU-based architectures, relying on GPU backends provided in, e.g., NVIDIA HPC SDK or AMD ROCm HIPSTDPAR. The hierarchical unstructured Cartesian grids are generated using a massively parallel grid generator. Previous work has achieved favorable strong and weak scaling on modern HPC systems~\cite{lagemann2024hydrogym}. To facilitate reproducibility and ease of deployment, we provide (pre-built) containerized environments and pre-configured setup files that enable users to readily deploy the HydroGym framework with minimal configuration overhead.

\subsection{m-AIA -- Finite-volume solver implementation}
\label{appendix:Implementation:MAIA:FV}

Beyond the lattice Boltzmann method, m-AIA also includes a compressible finite-volume (FV) solver targeting direct numerical simulation and large-eddy simulation of wall-bounded turbulent flows. The Navier-Stokes equations are discretized on structured, body-conforming curvilinear grids with a cell-centered finite-volume scheme. This grid topology is advantageous for configurations such as flat plates and airfoils, where strong velocity gradients are concentrated near solid surfaces and can be efficiently captured by anisotropic mesh refinement in the wall-normal direction.

\paragraph{Governing equations.}
The FV solver advances the compressible Navier-Stokes equations cast in an arbitrary Lagrangian-Eulerian (ALE) formulation, accommodating domains whose boundaries may deform over time. In integral form, these conservation laws for mass, momentum, and energy read
\begin{equation}
    \frac{d}{dt} \int_{V(t)} \mathbf{Q} \, dV + \oint_{A(t)} \left( \mathbf{H}_{\mathrm{inv}} + \mathbf{H}_{\mathrm{vis}} \right) \cdot \mathbf{n} \, dA = 0,
\end{equation}
with the state vector $\mathbf{Q} = (\rho, \rho \mathbf{u}, \rho E)^\top$ collecting the conserved quantities. Here, $\rho$ denotes the density, $\mathbf{u} = (u, v, w)^\top$ the velocity, and $E$ the specific total energy, related to the specific internal energy $e$ by $\rho E = \rho e + \frac{\rho}{2} |\mathbf{u}|^2$. The tensors $\mathbf{H}_{\mathrm{inv}}$ and $\mathbf{H}_{\mathrm{vis}}$ represent inviscid and viscous fluxes through the surface $A(t)$ whose outward unit normal is $\mathbf{n}$. Their combined contribution reads
\begin{equation}
    \mathbf{H} = \mathbf{H}_{\mathrm{inv}} - \mathbf{H}_{\mathrm{vis}} = \begin{pmatrix} \rho(\mathbf{u} - \mathbf{u}_A) \\ \rho\mathbf{u}(\mathbf{u} - \mathbf{u}_A) + p\mathbf{I} \\ \rho E(\mathbf{u} - \mathbf{u}_A) + p\mathbf{u}_A \end{pmatrix} + \begin{pmatrix} 0 \\ \boldsymbol{\tau} \\ \boldsymbol{\tau}\mathbf{u} + \mathbf{q} \end{pmatrix},
\end{equation}
with $\mathbf{u}_A$ the velocity of the moving control-volume surface, $\boldsymbol{\tau}$ the viscous stress tensor, $\mathbf{q}$ the heat flux vector, and $p$ the pressure. Assuming a Newtonian fluid with vanishing bulk viscosity, the stress tensor takes the form $\boldsymbol{\tau} = 2\mu \mathbf{S} - \frac{2}{3}\mu(\nabla \cdot \mathbf{u})\mathbf{I}$, where $\mathbf{S} = \frac{1}{2}[\nabla\mathbf{u} + (\nabla\mathbf{u})^\top]$ is the rate-of-strain tensor and $\mu$ is obtained from Sutherland's law. Heat conduction obeys Fourier's law at a fixed Prandtl number $Pr_0 = 0.72$, and the thermodynamic closure consists of the ideal gas relation $T = \gamma p / \rho$ together with $e = p / [(\gamma - 1)\rho]$, using $\gamma = 1.4$ for air.

\paragraph{Curvilinear coordinate transformation.}
To resolve the steep wall-normal gradients characteristic of boundary layer flows, the equations are formulated on structured body-fitted grids. A mapping $(x,y,z) \to (\xi,\eta,\zeta)$ with unit spacing in computational space aligns coordinate lines with curved solid boundaries. Transforming the differential Navier-Stokes equations into this coordinate system and making use of the metric-identity relations~\cite{thomas1979geometric} recovers a fully conservative form,
\begin{equation}
    \frac{\partial \hat{\mathbf{Q}}}{\partial t} + \frac{\partial \hat{\mathbf{E}}}{\partial \xi} + \frac{\partial \hat{\mathbf{F}}}{\partial \eta} + \frac{\partial \hat{\mathbf{G}}}{\partial \zeta} = 0,
\end{equation}
with the scaled state vector $\hat{\mathbf{Q}} = J\mathbf{Q}$, where $J$ is the reciprocal of the Jacobian determinant. The transformed fluxes combine contributions from all physical directions weighted by the metric coefficients, e.g.,
\begin{equation}
    \hat{\mathbf{E}} = J(\xi_t \mathbf{Q} + \xi_x \mathbf{E} + \xi_y \mathbf{F} + \xi_z \mathbf{G}),
\end{equation}
and $\hat{\mathbf{F}}$, $\hat{\mathbf{G}}$ follow analogously.

When the grid moves in time, as is the case for surface-actuation problems, the geometric conservation law (GCL) must be enforced to avoid non-physical mass, momentum, or energy production in the discrete equations~\cite{thomas1979geometric}. This is achieved by evaluating the temporal metrics $(J\xi_t)$, $(J\eta_t)$, $(J\zeta_t)$ from the volumes swept by individual cell faces between successive time levels.

\paragraph{Inviscid flux discretization.}
The inviscid fluxes at cell interfaces are evaluated with the advection upstream splitting method (AUSM)~\cite{liou1993new}. AUSM separates each interface flux into two contributions: a convective component, which is upwind-biased according to the local flow direction, and a pressure component, which is centrally averaged. Specifically, the discretized inviscid flux in the $\xi$-direction reads
\begin{equation}
    \hat{\mathbf{E}}_{\mathrm{inv}} = \frac{1}{2}\left[M_{L/R}\left(\hat{\mathbf{E}}^c_L + \hat{\mathbf{E}}^c_R\right) + |M_{L/R}|\left(\hat{\mathbf{E}}^c_L - \hat{\mathbf{E}}^c_R\right)\right] + \hat{\mathbf{E}}^p_{L/R},
\end{equation}
where the interface Mach number $M_{L/R} = \frac{1}{2}(M_L + M_R)$ determines whether the left or right convective flux is used, and the pressure flux is computed as $\hat{\mathbf{E}}^p_{L/R} = \frac{1}{2}(p_L + p_R) J (\xi_x, \xi_y, \xi_z, \xi_t)^\top$. Second-order accuracy is achieved through a monotonic upstream-centered scheme for conservation laws (MUSCL)~\cite{vanleer1979towards} extrapolation of the primitive variables to the cell interfaces. The left and right states at an interface $i+\frac{1}{2}$ are reconstructed from neighboring cell-center values using a distance-weighted interpolation that accounts for non-uniform grid spacing.

\paragraph{Viscous flux discretization.}
Because the viscous terms contain nested first-order derivatives, their discretization produces second-order operators that propagate information omnidirectionally. A central scheme is therefore appropriate, and a modified cell-vertex (MCV) approach is adopted. Each interface flux is obtained by averaging the fluxes evaluated at the four vertices of that face:
\begin{equation}
    \left.\frac{\partial \hat{\mathbf{E}}_{\mathrm{vis}}}{\partial \xi}\right|_{i,j,k} = \frac{1}{4}\sum_{m,n \in \{-\frac{1}{2},+\frac{1}{2}\}} \hat{\mathbf{E}}_{\mathrm{vis},i+\frac{1}{2},j+m,k+n} - \frac{1}{4}\sum_{m,n \in \{-\frac{1}{2},+\frac{1}{2}\}} \hat{\mathbf{E}}_{\mathrm{vis},i-\frac{1}{2},j+m,k+n},
\end{equation}
where the velocity and temperature gradients at each vertex are computed with standard central differences. Compared with a naive evaluation at cell centers, this vertex-based averaging lowers the number of metric-term evaluations while retaining second-order accuracy.

\paragraph{Temporal integration.}
Discretizing the spatial operators converts the transformed Navier-Stokes equations into a system of ordinary differential equations,
\begin{equation}
    \frac{\partial \hat{\mathbf{Q}}}{\partial t} = \mathrm{RHS}\left(t; \hat{\mathbf{Q}}\right)
\end{equation}
which is marched forward in time with a five-stage, low-storage Runge-Kutta method whose stage coefficients are $\alpha_k = (1/4, 1/6, 3/8, 1/2, 1)$, providing second-order accuracy in time. When the grid deforms, all geometric quantities---metric terms, Jacobians, and interface volume fluxes---are updated at every Runge-Kutta stage so that temporal accuracy is preserved.

\paragraph{Implicit large-eddy simulation.}
When the mesh is too coarse to resolve all turbulent scales, the solver relies on a monotonically integrated large-eddy simulation (MILES) strategy~\cite{stolz1999approximate,schlatter2004transitional}. Rather than adding an explicit subgrid-scale model, MILES exploits the fact that the upwind-biased AUSM scheme already introduces controlled numerical diffusion that acts preferentially on the smallest resolved scales, effectively mimicking the energy drain of the unresolved turbulent cascade. Extensive validation across a range of turbulent configurations has confirmed that this implicit modelling yields predictions of comparable quality to those obtained with explicit subgrid closures~\cite{meinke2002numerical}.

\paragraph{MPI parallelization and GPU acceleration.}
For parallel execution, the structured grid is partitioned into blocks of roughly equal size via a balanced cut-tree algorithm that simultaneously minimizes the surface area shared between neighboring blocks, thereby reducing communication overhead. Ghost-cell layers whose depth matches the stencil width surround each block, so that all numerical operators can be applied uniformly across the interior without special near-boundary treatment. Ghost-cell values are synchronized across blocks through MPI after every Runge-Kutta stage. As described for the LBM solver, the FV code also supports hybrid parallelism via OpenMP and C\texttt{++}20 parallel algorithms, allowing it to run on GPU-accelerated architectures through the NVIDIA HPC SDK or AMD ROCm HIPSTDPAR backends.

\paragraph{Boundary conditions and actuation control.}
At solid walls, adiabatic no-slip conditions enforce the fluid velocity to match the wall velocity and prescribe vanishing wall-normal gradients of pressure and temperature. Far-field in- and outflow boundaries employ a characteristic condition based on the local contravariant Mach number~\cite{whitfield1984}, supplemented by sponge layers on the pressure field to absorb acoustic reflections. Where required, periodic conditions are applied in the spanwise direction, and synthetic turbulence generation following the reformulated method of Roidl et~al.~\cite{roidl2013} provides fully turbulent inflow data from precursor RANS solutions.
 
For flow control, the FV solver supports the same synthetic jet actuation described in Section~\ref{appendix:Implementation:MAIA}, in which a parabolic velocity profile is imposed on designated wall segments and the jet magnitude is smoothly ramped between successive control updates. In addition, a streamwise traveling transversal surface wave boundary condition is available, following the formulation introduced by Albers et~al.~\cite{albers2024,lagemann2024impact,mateling2022analysis,mateling2023spanwise,mateling2020detection} for compressible turbulent boundary layer drag reduction. In the original fixed-parameter formulation, the wall is deflected in the wall-normal direction according to
\begin{equation}
    y\big|_{\mathrm{wall}}(x,t) = g(x)\, A \cos\!\left(\frac{2\pi}{\lambda}\, x - \frac{2\pi}{T}\, t\right),
    \label{eq:travelingwave}
\end{equation}
where $A$ is the wave amplitude, $\lambda$ the streamwise wavelength, $T$ the period, and $g(x)$ is a piecewise cosine envelope that smoothly ramps the wave on and off over prescribed streamwise intervals. The corresponding wall velocity $v|_{\mathrm{wall}} = \mathrm{d}y/\mathrm{d}t|_{\mathrm{wall}}$ is imposed consistently, and the body-fitted grid deforms to track the moving surface while satisfying the geometric conservation law.
 
For integration with the HydroGym reinforcement learning framework, this boundary condition is generalized so that the wave amplitude~$A$, wavelength~$\lambda$, and phase speed~$c = \lambda / T$ become time-varying control parameters set by the RL agent at each control step. Because abrupt changes in any of these quantities would introduce discontinuities in the grid velocity field and compromise numerical stability, a cross-fade transition strategy is employed. When the agent issues new target values $(A^{\mathrm{new}}, \lambda^{\mathrm{new}}, c^{\mathrm{new}})$, the actuation is not switched instantaneously. Instead, a blending factor
\begin{equation}
    \alpha(\tau) = \tfrac{1}{2}\bigl(1 - \cos(\pi\tau)\bigr), \qquad \tau = n_{\mathrm{step}} / N_{\mathrm{trans}} \in [0,1],
    \label{eq:crossfade_alpha}
\end{equation}
is advanced over $N_{\mathrm{trans}}$ sub-iterations, where $n_{\mathrm{step}}$ counts the elapsed transition steps. The instantaneous wave parameters are then obtained by interpolation between their previous and target values, e.g., $A(t) = A^{\mathrm{old}} + \alpha\,(A^{\mathrm{new}} - A^{\mathrm{old}})$, and analogously for the phase speed. Rather than interpolating the wavelength directly, the wavenumbers $k^{\mathrm{old}} = 2\pi/\lambda^{\mathrm{old}}$ and $k^{\mathrm{new}} = 2\pi/\lambda^{\mathrm{new}}$ are blended, which avoids singularities for large wavelengths and yields a smoother spatial transition. The wall displacement during the cross-fade is computed as a weighted superposition of the old and new waveforms,
\begin{equation}
    y\big|_{\mathrm{wall}}(x,t) = g(x)\, A(t) \bigl[(1-\alpha)\cos(k^{\mathrm{old}}\, \xi) + \alpha\cos(k^{\mathrm{new}}\, \xi)\bigr],
    \label{eq:crossfade_displacement}
\end{equation}
where $\xi(x,t) = x - x_0 - d(t)$ is the position relative to the wave origin shifted by the accumulated phase distance $d(t) = \int_0^t c(t')\,\mathrm{d}t'$, which is integrated numerically to ensure phase continuity across parameter changes. The grid velocity and acceleration fields required by the ALE formulation are obtained by consistent analytical time differentiation of Eq.~\eqref{eq:crossfade_displacement}, accounting for the time dependence of $\alpha$, $A$, and $d$. This cross-fade mechanism guarantees $C^1$-continuous grid motion in time and permits the RL agent to explore the full three-dimensional parameter space $(A, \lambda, c)$ without risking solver instabilities from sudden actuation changes.

\subsection{Nek5000 implementation}
\label{appendix:Implementation:nek5000}

\paragraph{Mathematical Formulation and Discretization}

Nek5000~\citep{nek5000} is an incompressible Navier--Stokes solver that employs the spectral-element method (SEM) to achieve high numerical accuracy while maintaining computational efficiency~\citep{demoura_semadvantage_2024}.
The computational domain is partitioned into hexahedral elements, within which the velocity and pressure fields are approximated by high-order Lagrange interpolants following the $\mathbb{P}_N\mathbb{P}_{N-2}$ formulation~\citep{maday_spectral_1989}.
For a polynomial order of $N-1$, the velocity field is represented on $N^3$ grid points per element distributed according to the Gauss--Lobatto--Legendre (GLL) quadrature rule.
The pressure field is defined on a staggered grid comprising ${(N-2)}^3$ points per element using Gauss--Legendre (GL) quadrature.

Temporal integration is performed using a third-order explicit extrapolation scheme (EXT3) for the nonlinear convective terms and a third-order implicit backward differentiation scheme (BDF3) for the viscous contributions.
To mitigate aliasing errors, overintegration is applied by oversampling the evaluation of nonlinear terms with a factor of $3/2$ relative to the adopted polynomial order $N$ in each spatial direction.

HydroGym currently provides two Nek5000-based environments: (1) a three-dimensional turbulent channel flow configuration solved using direct numerical simulation (DNS), and (2) a three-dimensional NACA~0012 wing section simulated using high-resolution large-eddy simulation (LES).
Detailed descriptions of the turbulent channel flow and the NACA~0012 wing configurations are presented in Secs.~\ref{appendix:Environments:TCF} and~\ref{appendix:Environments:NACA0012}, respectively.

\paragraph{Boundary Conditions and actuation control}
A time-dependent wall-normal velocity component $v_n$ is prescribed at the wall as the control actuation through a Dirichlet boundary condition.
The spatial mean of the actuation signal is removed to enforce the zero-net-mass-flux (ZNMF) constraint, thereby ensuring global mass conservation.
This boundary treatment models localized jet actuation that induces blowing or suction at the solid surface.

\paragraph{MPI communications and parallelization}
All Nek5000 environments are executed on CPU architectures, with parallelization achieved through a hybrid distributed-memory and shared-memory strategy based on OpenMPI.
By taking advantage of the structured meshes and the tensor-product formulation inherent to the SEM, the solver exhibits strong scalability and robust parallel performance, as demonstrated in prior studies.
The coupling with reinforcement learning agents is implemented through dynamic MPI-based communication, enabling stable and efficient interaction with large-scale simulations, such as three-dimensional wing configurations at high Reynolds numbers described in Sec.~\ref{appendix:Environments:NACA0012}.
To enhance reproducibility and facilitate deployment, we provide pre-built containerized environments together with pre-configured setup files, allowing users to deploy the HydroGym framework with minimal configuration overhead.

\subsection{Incompressible JAX implementation}
\label{appendix:Implementation:JAX}
Physics simulators built with modern automatic differentiation libraries make it possible to compute gradients of output quantities with respect to desired input parameters, even with complex physical processes in the middle. These gradients can then be integrated into gradient-based control schemes, one of them being gradient-enhanced reinforcement learning~\cite{son2023gippo}. The HydroGym framework explores this capability by offering flow solvers written in JAX ---a differentiable programming language with GPU acceleration--- to model flows described by the incompressible Navier–Stokes equations:
\begin{equation}
  \frac{\partial \mathbf{u}}{\partial t} + (\mathbf{u} \cdot \nabla) \mathbf{u}  = -\nabla p + \nu \nabla^2 \mathbf{u} + \mathbf{f}, \quad \nabla \cdot \mathbf{u} = 0,
\label{eq:navier_stokes}
\end{equation}

where $\mathbf{u}$ denotes the velocity, $p$ is the pressure, $\nu$ is the kinematic viscosity and $\mathbf{f}$ is an external forcing term. HydroGym currently provides two differentiable environments which model Equation \ref{eq:navier_stokes}: a three-dimensional turbulent channel flow and a two-dimensional forced turbulence problem known as the Kolmogorov flow~\cite{mokbel2025controlling}. A more detailed implementation description of both environments is given in Secs. \ref{appendix:Environments:Kolmogorov} \& \ref{appendix:Environments:Channel}.

\paragraph{Software Architecture.}
Each environment employs a modular architecture that separates physical problem definitions from the reinforcement learning interface, following the design principles of HydroGym's Firedrake environments (Sec. \ref{appendix:Implementation:Firedrake}) to ensure code clarity, extensibility, and ease of use. The differentiable environments differ by implementing a Gymnax interface~\cite{gymnax2022github} and inheriting reinforcement learning frameworks from PureJaxRL~\cite{purejaxrl2023github}, enabling a fully synchronous training pipeline compatible with JAX's automatic differentiation. This architecture allows users to compute gradients of any flow output with respect to tunable input parameters at any point during environment execution or RL training via JAX's \texttt{grad} function, facilitating gradient-based optimization and control strategies.

\paragraph{GPU acceleration and parallelization.}
The JAX environments leverage GPU acceleration and parallelization to enable rapid RL training. The end-to-end framework supports parallel execution across multiple random seeds using JAX's automatic vectorization functions such as \texttt{vmap} for batch processing. This enables training hundreds of agents simultaneously in a fraction of the time required for single-agent training in PyTorch~\cite{purejaxrl2023github, huang2022cleanrl}. Additionally, just-in-time compilation via \texttt{jit} automatically optimizes code execution on GPU hardware.

\subsection{JAX-Fluids implementation}
\label{appendix:Implementation:JAXFluids}

JAX-Fluids is a fully-differentiable, high-order computational fluid dynamics (CFD) solver designed for compressible single- and two-phase flows \cite{bezgin2023jax,bezgin2025jax}. Written entirely in Python using the JAX library, it leverages automatic differentiation (AD) and hardware acceleration to seamlessly bridge traditional CFD with machine learning workflows. Within the HydroGym framework, JAX-Fluids provides a powerful backend for optimizing active flow control strategies in compressible regimes, enabling exact gradient computation through complex, highly-resolved flow physics.

\paragraph{Mathematical Formulation and Discretization.}
Unlike the incompressible solvers described previously, JAX-Fluids solves the compressible Navier-Stokes equations. The fluid state is described by the vector of conservative variables $\mathbf{U} = [\rho, \rho\mathbf{u}, E]^T$, representing density, momentum, and total energy per unit volume, respectively. The governing equations are formulated as:
\begin{equation}
    \frac{\partial \mathbf{U}}{\partial t} + \nabla \cdot \mathbf{F}^c(\mathbf{U}) = \nabla \cdot \mathbf{F}^d(\mathbf{U}) + \mathbf{S}(\mathbf{U}),
\end{equation}
where $\mathbf{F}^c$ and $\mathbf{F}^d$ denote the convective and diffusive fluxes, and $\mathbf{S}$ represents source terms such as external forcings or body forces. The system is closed using an appropriate equation of state, typically the ideal or stiffened gas law.

The spatial domain is discretized using a finite-volume method on structured Cartesian grids, which support arbitrary one-dimensional mesh stretching. Convective fluxes are evaluated using a high-order Godunov-type approach, pairing advanced shock-capturing spatial reconstruction schemes (e.g., WENO5-Z, TENO6-A) with an approximate HLLC Riemann solver. Diffusive fluxes are approximated via high-order central finite differences. Time integration of the semi-discrete system is performed using explicit total-variation-diminishing Runge-Kutta (TVD-RK) schemes. Furthermore, JAX-Fluids natively supports multiphase flows, offering both a sharp-interface level-set method and a five-equation diffuse-interface model. Positivity-preserving limiters are used to ensure robust integration in the presence of strong shocks or large density ratios.

The level-set implementation in JAX-Fluids additionally functions as a conservative immersed boundary method (IBM), enabling flow simulations around complex geometries. The IBM is particularly well suited for active flow control problems, as control actions (e.g., blowing or suction) can be naturally incorporated through interface exchange terms.

\paragraph{Software Architecture and Machine Learning Integration.}
JAX-Fluids departs from traditional loop-based programming in favor of the array programming paradigm, utilizing JAX's NumPy-like API. By adhering strictly to functional programming constraints—such as pure functions and static control flow operations—the entire simulation pipeline is end-to-end differentiable. 

This architecture allows researchers to compute exact gradients of arbitrary scalar quantities of interest (e.g., time-averaged drag, aerodynamic efficiency, or an RL reward function) with respect to control inputs, boundary conditions, or neural network weights. Gradients are backpropagated through the complete temporal simulation trajectory using JAX's \texttt{grad} or \texttt{value\_and\_grad} transformations. To mitigate memory bottlenecks associated with unrolling long rollout trajectories during the backward pass, JAX-Fluids employs gradient checkpointing (\texttt{jax.checkpoint}). This makes it highly straightforward to couple the solver with deep learning libraries like Flax and Optax, facilitating the joint optimization of parameterized control policies directly within the CFD environment.

\paragraph{HPC and Parallelization.}
To accommodate the vast computational resources required for three-dimensional compressible turbulence and complex flow control scenarios, JAX-Fluids incorporates a hardware-agnostic parallelization strategy tailored for modern HPC clusters. The computational domain is partitioned using a homogeneous domain decomposition strategy, distributing grid blocks across multiple XLA devices (GPUs or TPUs) via the \texttt{jax.pmap} transformation. 

Crucially, inter-block communications for face, edge, and vertex halo updates are executed exclusively through JAX's collective permutation primitives (\texttt{jax.lax.ppermute}). This purely JAX-native communication ensures that the automatic differentiation graph remains intact across distributed, multi-node compute clusters. This enables highly efficient weak scaling on supercomputers—such as NVIDIA A100 GPU clusters and Google TPU pods—while simultaneously supporting the parallel evaluation of AD gradients for large-scale differentiable flow control tasks.

\subsection{Firedrake implementation}
\label{appendix:Implementation:Firedrake}
The HydroGym framework provides an alternative implementation using Firedrake~\cite{rathgeber2016firedrake}, a finite element framework built on PETSc that offers automatic code generation and optimization for variational problems. This implementation targets researchers requiring full access to solver internals and complete Python-based extensibility, prioritizing code transparency and flexibility over computational performance. The Firedrake backend implements two-dimensional flow control problems using mixed finite element methods for the incompressible Navier-Stokes equations.

\paragraph{Mathematical Formulation and Discretization.}
The solver employs Taylor-Hood elements with second-order continuous Galerkin (CG) elements for velocity and first-order CG elements for pressure, ensuring inf-sup stability. The discretized system takes the form:
\begin{equation}
    \mathbf{M}\frac{\partial \mathbf{q}}{\partial t} + \mathbf{R}(\mathbf{q}) = \mathbf{0},
\end{equation}
where $\mathbf{q} = (\mathbf{u}, p)$ represents the mixed velocity-pressure state, $\mathbf{M}$ is the mass matrix, and $\mathbf{R}(\mathbf{q})$ is the nonlinear residual of the incompressible Navier-Stokes equations:
\begin{equation}
    \mathbf{R}(\mathbf{q}) = \int_\Omega \left[ -(\mathbf{u} \cdot \nabla)\mathbf{u} \cdot \mathbf{v} - \boldsymbol{\sigma}(\mathbf{u}, p) : \boldsymbol{\epsilon}(\mathbf{v}) + (\nabla \cdot \mathbf{u}) s \right] d\Omega,
\end{equation}
where $(\mathbf{v}, s)$ are velocity and pressure test functions, $\boldsymbol{\sigma}(\mathbf{u}, p) = 2\nu\boldsymbol{\epsilon}(\mathbf{u}) - p\mathbf{I}$ is the Newtonian stress tensor, and $\boldsymbol{\epsilon}(\mathbf{u}) = \frac{1}{2}(\nabla\mathbf{u} + (\nabla\mathbf{u})^T)$ is the strain rate tensor. Time integration is performed using implicit schemes with automatic selection of solver parameters based on Reynolds number and grid resolution.

\paragraph{Software Architecture.}
The implementation follows a three-tier modular architecture that separates physical problem definition from numerical methods and reinforcement learning interfaces. This design enables extensibility while maintaining code clarity and facilitating rapid prototyping of new flow control scenarios.\\

The \texttt{PDEBase} class serves as an abstract template defining the core functionality required for PDE-based control problems. It encapsulates the physical configuration including domain discretization, boundary conditions, and objective function evaluation, while remaining solver-agnostic. Key abstract methods include mesh handling (\texttt{load\_mesh}, \texttt{initialize\_state}), state management (\texttt{set\_state}, \texttt{copy\_state}, \texttt{reset}), and problem-specific definitions (\texttt{get\_observations}, \texttt{evaluate\_objective}). The class provides a standardized interface through which control agents interact with the underlying fluid physics, handling the conversion between control inputs and boundary condition modifications. Additionally, it manages actuator dynamics and maintains checkpoint functionality for simulation restart capabilities.\\

The \texttt{TransientSolver} class implements the time-stepping algorithms for evolving the PDE state. It provides a \texttt{solve} method that integrates the flow equations over specified time intervals, supporting optional callback functions for data collection and visualization. The core \texttt{step} method advances the simulation by one time increment, accepting control inputs and updating the PDE state accordingly. This separation allows for different time integration schemes to be implemented without modifying the physical problem definition. The solver supports both open-loop time integration and closed-loop operation with feedback controllers of the form $u(t, y)$.\\

The \texttt{FlowEnv} class bridges the PDE framework with reinforcement learning by implementing the Farama Foundation Gymnasium interface. It wraps \texttt{PDEBase} and \texttt{TransientSolver} instances, translating between RL concepts (observations, actions, rewards) and CFD operations (state queries, boundary condition updates, objective evaluation). The environment handles episode management through reset functionality, computes rewards as the negative of the time-integrated objective function, and manages termination conditions. This design allows any combination of PDE problem and solver to be automatically instantiated as a Gym-compatible environment through configuration dictionaries.

\paragraph{Available Environments.}
The framework includes three benchmark flow configurations: flow around a circular cylinder with rotary control, flow around a cylinder with blowing/suction actuation, cavity flow with leading-edge control, and the three-cylinder pinball configuration with independent rotary actuation for each cylinder.

\paragraph{Computational Infrastructure}
The implementation supports MPI parallelization through Firedrake's distributed mesh capabilities and leverages PETSc's parallel linear algebra routines. While computational efficiency is secondary to code accessibility, the framework maintains reasonable performance for research applications through automatic optimization of generated finite element code and efficient sparse linear system solvers. The modular architecture enables researchers to extend the framework with new flow configurations by implementing problem-specific methods while inheriting the complete finite element infrastructure and RL integration capabilities. Similar to the m-AIA environments, we also provide pre-built container environments.

\section{Benchmarked RL Agents} 
\label{appendix:rl_agents}

The HydroGym platform conducts comprehensive performance evaluation using four distinct reinforcement learning algorithms, representing both established model-free approaches and emerging gradient-enhanced methodologies. For the non-differentiable environments, we employ TorchRL~\cite{bou2023torchrl} as a stable baseline implementation for Proximal Policy Optimization (PPO), Deep Deterministic Policy Gradient (DDPG), and Twin Delayed Deep Deterministic Policy Gradient (TD3). Additionally, we utilize in-house implementations of PPO and Gradient-Enhanced Proximal Policy Optimization (GPPO) specifically designed for JAX-based differentiable environments~\cite{bradbury2018jax}. This dual implementation strategy ensures optimal performance across both traditional CFD solvers and modern differentiable simulation frameworks while maintaining algorithmic consistency and reproducibility.

\subsection{Proximal Policy Optimization (PPO)}

PPO implements a trust region policy optimization approach that constrains policy updates through a clipped surrogate objective function to prevent destructive parameter changes during training~\cite{schulman2017proximal}. The core innovation lies in its clipped probability ratio mechanism, which limits the extent of policy updates based on the ratio between new and old policy probabilities. The algorithm optimizes the following clipped objective:

\begin{equation}
L^{CLIP}(\theta) = \mathbb{E}_t \left[ \min(r_t(\theta) A_t, \text{clip}(r_t(\theta), 1-\epsilon, 1+\epsilon) A_t) \right]
\end{equation}

\noindent where $r_t(\theta) = \frac{\pi_\theta(a_t|s_t)}{\pi_{\theta_{old}}(a_t|s_t)}$ represents the probability ratio, $A_t$ denotes the advantage estimate, and $\epsilon$ controls the clipping range (typically 0.2).

A Generalized Advantage Estimation (GAE) with $\lambda = 0.95$ for variance reduction in advantage calculations~\cite{schulman2015high} is employed, combined with a value function loss term and entropy regularization to maintain exploration. Network architectures utilize fully connected layers with ReLU activations, featuring separate actor and critic networks. The algorithm demonstrates consistent convergence across most HydroGym flow configurations, successfully learning control policies for drag reduction tasks, cavity flow stabilization, and multi-body coordination in the fluidic pinball. Training hyperparameters are adapted per environment, with learning rates ranging from $5 \times 10^{-5}$ to $1 \times 10^{-3}$, batch sizes of 16-48 experiences, and episode lengths matching the physical timescales of each flow configuration.

\subsection{Deep Deterministic Policy Gradient (DDPG)}

DDPG extends the deterministic policy gradient theorem to high-dimensional continuous action spaces through an actor-critic architecture combined with experience replay and target networks~\cite{lillicrap2015continuous}. The algorithm involves four neural networks: an actor network $\mu(s|\theta^\mu)$ that deterministically maps states to actions, a critic network $Q(s,a|\theta^Q)$ that estimates action-value functions, and corresponding networks $\mu'$ and $Q'$ to provide stable learning targets~\cite{mnih2015human}. The critic is trained using temporal difference learning with the Bellman equation:
\begin{equation}
L(\theta^Q) = \mathbb{E}_{s,a,r,s' \sim \mathcal{D}} \left[ (Q(s,a|\theta^Q) - y)^2 \right]
\end{equation}
where $y = r + \gamma Q'(s', \mu'(s'|\theta^{\mu'})|\theta^{Q'})$ represents the target value computed using target networks. The actor network is updated using the deterministic policy gradient~\cite{silver2014deterministic}:
\begin{equation}
\nabla_{\theta^\mu} J \approx \mathbb{E}_{s \sim \mathcal{D}} \left[ \nabla_{\theta^\mu} \mu(s|\theta^\mu) \nabla_a Q(s,a|\theta^Q)|_{a=\mu(s)} \right]
\end{equation}
Exploration during training is achieved through additive Ornstein-Uhlenbeck noise $\mathcal{N}$ with parameters $\mu_{OU} = 0.15$, $\sigma = 0.2$, providing temporally correlated exploration in the action space~\cite{uhlenbeck1930theory}. Target networks are updated using soft updates with $\tau = 0.005$ to ensure stable learning dynamics. We employ replay buffers of size $10^4$ to $10^5$ experiences, with batch sampling of 256 transitions per update step. Network architectures feature 256-512 hidden units across two hidden layers, utilizing batch normalization for improved training stability~\cite{ioffe2015batch}.

\subsection{Twin Delayed Deep Deterministic Policy Gradient (TD3)}

TD3 addresses the overestimation bias inherent in DDPG through three key modifications: twin critic networks, delayed policy updates, and target policy smoothing~\cite{fujimoto2018addressing}. The twin critic architecture maintains two independent Q-functions $Q_{\theta_1}$ and $Q_{\theta_2}$, taking the minimum estimate for target value computation to reduce overestimation~\cite{hasselt2010double}:

\begin{equation}
y = r + \gamma \min_{i=1,2} Q_{\theta_i'}(s', \tilde{a})
\end{equation}

where $\tilde{a} = \mu'(s') + \text{clip}(\epsilon, -c, c)$ represents the smoothed target action with clipped noise $\epsilon \sim \mathcal{N}(0, \sigma)$. Policy updates occur less frequently than critic updates (typically every second iteration) to allow critics to converge before policy modification. This delayed update mechanism prevents policy updates from exploiting temporary critic errors. Target policy smoothing adds regularized noise to target actions during critic training, smoothing the value function over similar actions and reducing the impact of function approximation errors~\cite{fujimoto2018addressing}.

The complete TD3 update procedure alternates between critic training using both networks and periodic actor updates using the first critic network only. Hyperparameter selection follows $\sigma = 0.2$ for target noise, clipping bounds $c = 0.5$, and policy delay factor of 2. 

\subsection{Gradient-Enhanced Proximal Policy Optimization (GPPO)}

GPPO represents a hybrid approach that integrates analytic gradient information from differentiable fluid simulations into the standard PPO framework~\cite{kochkov2021machine}. This method exploits the differentiability of JAX-based flow solvers to compute exact policy gradients with respect to the flow dynamics, providing more informative gradient signals than traditional finite-difference estimates or likelihood ratio methods~\cite{williams1992simple}.

The algorithm combines analytic gradients $\nabla_\theta \mathbb{E}[R(\tau)]$ computed directly through the differentiable simulator with PPO's clipped surrogate objective. The enhanced gradient computation leverages automatic differentiation through the entire simulation trajectory~\cite{griewank2008evaluating}:

\begin{equation}
\nabla_\theta J = \nabla_\theta \mathbb{E}_{\tau \sim \pi_\theta} \left[ \sum_{t=0}^T R(s_t, a_t) \right]
\end{equation}

where $\theta$ represents the tunable policy parameters, $J$ is the objective, $R$ is the reward, and the expectation gradient is computed analytically through the deterministic fluid dynamics, eliminating variance associated with policy gradient estimation. To provide a more clear example, consider the 3D channel flow, where the objective is to minimize the wall-shear stress, $\tau_w$, and the reward signal was a function of $\tau_w$. At a point in time $t$, the environment backward pass can compute:  

\begin{equation}
\frac{\partial \tau_w}{\partial \theta} = \frac{\partial \tau_w}{\partial s_t} \frac{\partial s_t}{\partial \theta} + \frac{\partial \tau_w}{\partial a_t} \frac{\partial a_t}{\partial \theta} 
\end{equation}

which can be included in the optimization by adding $\tau_w$ to the loss term. An alternative approach to leveraging gradient information is to incorporate it directly into the reward function. Taking into account this specific objective, the wall shear stress gradient with respect to a previous action can be computed by a backward pass of the environmental step, giving $\nabla_{a_t} \tau_{w}$. This value can then augment the reward function as follows: 

\begin{equation}
R(s_t, a_t) = - [ \tau_{w} + (\nabla_{a_t} \tau_{w} \cdot a_t )],  
\end{equation}

which can guide the policy towards actions that minimize future wall shear stress. These  methods were applied to both the 3D channel flow and the 2D Kolmogorov flow, and demonstrated both sample efficiency and improved performance in a variety of flow scenarios.

The implementation of GPPO maintains PPO's trust region constraints while incorporating these enhanced gradients through a weighted combination scheme. The gradient weighting balances analytic precision with exploration requirements, using adaptive coefficients based on gradient magnitude and policy entropy. Network architectures are specifically designed for JAX compatibility, utilizing Flax layers with consistent initialization schemes across differentiable and non-differentiable components~\cite{heek2020flax}.

\section{Environments} 
\label{appendix:envs}
The HydroGym platform provides a comprehensive suite of fluid dynamics environments specifically designed for reinforcement learning benchmarking and active flow control research. These environments span a diverse range of canonical flow configurations that capture fundamental fluid mechanics phenomena while maintaining computational tractability for iterative RL training processes. The framework encompasses 61 distinct flow scenarios across 14 different geometric configurations, ranging from two-dimensional cylinder flows to complex three-dimensional turbulent channel cases. The environments are systematically organized to cover multiple Reynolds number regimes within each configuration, enabling the investigation of flow control strategies across different physical regimes—from laminar vortex shedding at lower Reynolds numbers to transitional and turbulent dynamics at higher values.

The benchmark suite includes both classical bluff body flows (cylinders, squares, cubes, spheres) that exhibit well-documented vortex shedding phenomena as well as more specialized configurations such as cavity flows with shear layer instabilities, NACA 0012 airfoils under gust interactions, shock-vector control for divergent nozzle flows, extreme-event-driven Kolmogorov flows and biomedically-relevant stenotic pipe flows. Additionally, the framework incorporates multi-agent reinforcement learning (MARL) scenarios for spatially-distributed control strategies and differentiable physics environments that enable gradient-based optimization approaches.

Each environment follows a standardized framework that includes configurable reward functions targeting specific control objectives (typically drag reduction and flow stabilization), flexible actuation mechanisms (jet-based control, surface rotation and displacement, or temperature modulation), and adaptable sensor configurations through point probe measurements of flow quantities. The temporal coupling between CFD solvers and reinforcement learning agents is carefully calibrated to match characteristic flow timescales while maintaining computational efficiency for training.

The systematic organization of these environments provides researchers with a progressive pathway from fundamental flow control problems to increasingly complex multi-physics scenarios, supporting both algorithm development and practical flow control applications. The following subsections provide detailed descriptions of each environment category, including the underlying flow physics, numerical implementation specifics, validation against literature benchmarks, and the specific RL environment setup including state spaces, action spaces, and reward formulations.

\subsection{Circular cylinder flow}
\label{appendix:Environments:Cylinder}

\paragraph{Characteristic physics.}
The flow around circular cylinders in the Reynolds number range 200-3,900 based on the cylinder diameter and the free-stream velocity exhibits rich dynamical behavior characterized by distinct regime transitions and complex three-dimensional instabilities that have profound implications for both fundamental fluid mechanics research and practical flow control applications. This transitional range encompasses the evolution from two-dimensional laminar vortex shedding to fully three-dimensional turbulent wakes, presenting unique challenges for modeling, prediction, and control.

The flow physics within this range is governed by two critical transitions that fundamentally alter wake dynamics. The first occurs at Reynolds numbers around 190-200, where the initially two-dimensional von Kármán vortex street undergoes its first three-dimensional instability, known as Mode A. This elliptic instability manifests as large-scale spanwise undulations with characteristic wavelengths of approximately 3-4 cylinder diameters, originating in the primary vortex cores and representing the onset of three-dimensional wake behavior~\cite{williamson1996vortex, williamson1996three}. As the Reynolds number continues to increase beyond $Re = 230-260$, a second transition emerges through Mode B instability. Unlike Mode A, this hyperbolic instability develops in the braid shear layers and exhibits much smaller spanwise wavelengths of approximately 0.8-1.0 cylinder diameters~\cite{aleksyuk2024shortwavelength, bhattacharya2015investigation}. Together, these successive transitions mark the beginning of a complex regime where two-dimensional approximations become progressively inadequate for accurate flow prediction.
These three-dimensional instabilities create fundamental differences between 2D and 3D flow configurations that extend well beyond simple geometric considerations. While two-dimensional simulations remain computationally efficient and provide reasonable approximations for basic flow characteristics below $Re = 189$, they fundamentally cannot capture the governing physics once three-dimensional instabilities develop~\cite{ren2023three}. That is, the spanwise variations introduced by Mode A and Mode B instabilities significantly affect drag coefficients, vortex shedding frequencies, and heat transfer characteristics, with these effects becoming increasingly dominant as the Reynolds number increases. In the subcritical regime ($Re > 1,000$), the wake transitions to fully turbulent characteristics while maintaining coherent large-scale vortex shedding patterns. This creates a complex interplay between organized and chaotic motion that further challenges traditional modeling approaches~\cite{parnaudeau2008experimental, kravchenko2000numerical}.

Overall, the rich dynamical behavior has established cylinder wakes as particularly valuable canonical test cases for developing and validating flow control strategies. The combination of the relatively simple geometry with complex physics provides an ideal benchmark for both traditional control methods and emerging machine learning approaches. Recent developments in deep reinforcement learning have demonstrated remarkable success, achieving drag reductions of $8-25\%$ using various actuator configurations including synthetic jets, rotational oscillation, and plasma actuators~\cite{rabault2019artificial, tokarev2020deep}. The Reynolds number range 200-3,900 proves especially well-suited for these applications because the flow exhibits sufficient complexity to challenge control algorithms while remaining computationally tractable for iterative training processes.

However, the practical implementation of flow control in this Reynolds number range faces several specific challenges stemming from the complex transitional flow physics. Three-dimensional instabilities introduce spanwise variations that complicate sensor placement and actuator design, while the coexistence of laminar and turbulent regions creates difficulties in developing robust control strategies. Energy efficiency considerations become particularly critical, as the power required for active control can exceed the benefits gained through drag reduction, especially at lower Reynolds numbers within this range~\cite{han2022deep}. Furthermore, the sensitivity of the wake behavior to boundary conditions and geometric details makes experimental validation of numerical control studies challenging, often requiring careful attention to computational domain size and boundary condition implementation.
To address these challenges, multi-agent deep reinforcement learning frameworks have been designed to target the three-dimensional control problems, though most successful demonstrations remain limited to lower Reynolds numbers within the transitional range~\cite{suarez2025flow}.

To summarize, the established nature of cylinder wake physics makes this configuration particularly valuable as emerging computational methods and experimental techniques continue to provide new insights into flow control possibilities. The comprehensive understanding of fundamental mechanisms enables researchers to confidently assess whether novel approaches capture the essential physics or merely exploit computational artifacts~\cite{brunton2015annual,Brunton2022book, lagemann2023invariance}. Therefore, the well-documented cylinder wake serves as a valuable benchmark scenario for validating new innovations before applying them to more complex geometries and flow physics.

\begin{figure}[t!]
    \centering
    \includegraphics[width=\linewidth]{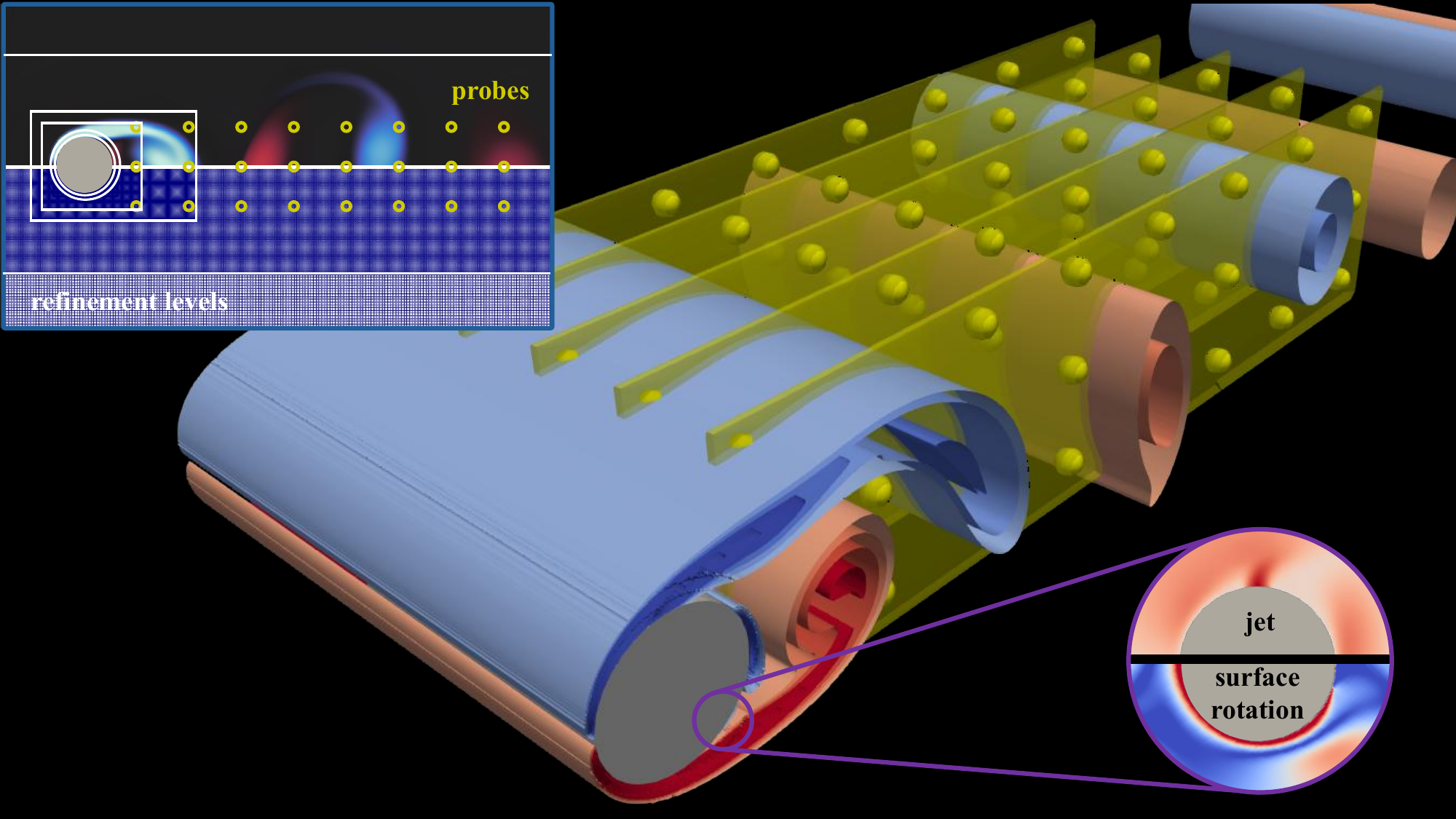}
    \caption{Computational domain setup, sensor probe distribution and actuation strategy for the circular cylinder flows: Three-dimensional view showing the probe distribution around the cylinder setup with vortical structures visualized through Q-criterion isosurfaces. The yellow dots in the zoom-in highlight sensor probe locations in a single spanwise plane. Active flow control is either performed using jet actuators positioned at the top and bottom of the cylinder or via surface rotation.}
    \label{appendix:fig:cylinder_actuation}
\end{figure}

\paragraph{Numerical setup and validation.}
The cylinder benchmark scenario comprises 2D and 3D low Mach number flows ($Ma<0.2$) over circular cylinders at different Reynolds number, i.e. $Re \in [100, ~200, ~1,000, ~3,900]$.
For the 2D cases, the computational domain is defined by $(x/D,y/D) \in [0,40] \times [-10,10]$ with the cylinder being centered at $(x/D,y/D) = (10,0)$ and $x$ representing the streamwise coordinate direction.
The grid spacing is determined by the grid refinement level $n$ via $\Delta_n=\Delta_0/2^n$ with $\Delta_0^{(2D)}=40D$. Three refinement patches decrease the grid spacing up to a level of $n=14$, see Fig. \ref{appendix:fig:cylinder_actuation}, corresponding to more than 320 cells per cylinder diameter.

In 3D the domain is $(x/D,y/D,z/D) \in [0,32] \times [0,16] \times [-2,2]$, being periodic in the $z$-direction while the center of the cylinder is located at $(x/D,y/D)=(8,8)$ and its axis is aligned in $z$-direction. The grid spacing is defined by $\Delta_0^{(3D)}=32D$ and refinement levels of $n \in [9,14]$ resolving the cylinder by 500 grid points.

At the inlet, a Dirichlet boundary condition is implemented with a prescribed constant velocity profile. The velocity vector is read from the properties file and applied uniformly across the inlet cross-section. The density at the inlet boundary cells is extrapolated from the interior domain using second-order accurate extrapolation to maintain consistency with the low Mach number assumption. The equilibrium distribution functions are calculated based on the prescribed velocity and extrapolated density values.

At the outlet, non-reflecting boundary conditions are applied minimizing spurious wave reflections. The velocity components are extrapolated from the interior domain using von Neumann conditions (zero normal gradient). The density is computed using a non-reflecting formulation:

\begin{equation}
\rho = \frac{
    \rho_{\text{old}}
    + \rho_{\text{ref}} / c_s (|\vec{v}|-|\vec{v}_{\text{old}}|)
    + \alpha \Delta t \rho_{\text{ref}}
}{
    1+\alpha \Delta t }
\end{equation}

where $\rho_{\text{old}}$ and $\vec{v}_{\text{old}}$ are values from the previous time step, $c_s$ is the speed of sound, $\rho_{\text{ref}}$ is a reference density term that can be specified as either density fluctuations or absolute density depending on the simulation requirements, and $\alpha=1.0$ is the relaxation parameter.

Periodic boundary conditions are applied to the lateral boundaries (side walls) of the computational domain. This treatment ensures mass conservation and eliminates the need for additional boundary condition specifications at these surfaces, effectively creating an infinite domain in the lateral directions.

Based on a comprehensive validation study, our numerical simulations demonstrate good agreement with established literature values across multiple Reynolds numbers and flow configurations. For 2D cylinder flows, our results closely match reference data, with Strouhal numbers within $1-3\%$ of published values and drag coefficients typically within $5\%$ of established benchmarks. The 3D simulations show similarly strong validation, with our Strouhal numbers and drag coefficients falling well within the scatter of literature values across all tested Reynolds numbers. The close correspondence between our computed flow parameters and the extensive reference database confirms the accuracy and reliability of our numerical approach for cylinder flow simulations.

\begin{table}[t!]
\centering

\begin{subtable}[t]{0.32\textwidth}
\centering
\caption{$Re = 200$}
\label{tab:2D_cylinder_Re200}
\adjustbox{width=\textwidth,center}{%
\begin{tabular}{@{}lccc@{}}
\toprule
Reference & St & $\overline{C_D}$ & $\overline{C_L}$ \\
\midrule
\textbf{Present study} & \textbf{0.196} & \textbf{1.341} & $\bm{\pm 0.659}$ \\
Meneghini et al.\cite{meneghini2001numerical} & 0.196 & 1.30 & -- \\
Ding et al.\cite{ding2007numerical} & 0.196 & 1.348 & $\pm 0.659$ \\
Braza et al.\cite{braza1986numerical} & 0.200 & 1.40 & $\pm 0.75$ \\
Posdziech \& Grundmann\cite{posdziech2007systematic} & 0.194 & 1.311 & $\pm 0.66$ \\
Harichandan \& Roy\cite{harichandan2010numerical} & 0.192 & 1.32 & $\pm 0.602$ \\
\bottomrule
\end{tabular}
}
\end{subtable}
\hfill
\begin{subtable}[t]{0.32\textwidth}
\centering
\caption{$Re = 1,000$}
\label{tab:2D_cylinder_Re1000}
\adjustbox{width=\textwidth,center}{%
\begin{tabular}{@{}lccc@{}}
\toprule
Reference & St & $\overline{C_D}$ & $\overline{C_L}$ \\
\midrule
\textbf{Present study} & \textbf{0.234} & \textbf{1.533} & $\bm{\pm 1.449}$ \\
Colagrossi et al.\cite{colagrossi2013particles} & 0.231 & 1.507 & $\pm 1.457$ \\
Durante et al.\cite{durante2017numerical} & 0.235 & 1.491 & $\pm 1.46$ \\
Scott \& Durst\cite{scott2024chaotic} & 0.2276 & 1.54 & $\pm 1.225$ \\
Chatzimanolakis et al.\cite{chatzimanolakis2024learning} & -- & 1.61 & -- \\
von Wahl et al.\cite{von2025benchmark} & 0.23 & 1.53 & -- \\
\bottomrule
\end{tabular}
}
\end{subtable}
\hfill
\begin{subtable}[t]{0.32\textwidth}
\centering
\caption{$Re = 3,900$}
\label{tab:2D_cylinder_Re3900}
\adjustbox{width=\textwidth,center}{%
\begin{tabular}{@{}lccc@{}}
\toprule
Reference & St & $\overline{C_D}$ & $\overline{C_L}$ \\
\midrule
\textbf{Present study} & \textbf{0.21} & \textbf{1.68} & $\bm{\pm 1.61}$ \\
Chatzimanolakis et al.\cite{chatzimanolakis2024learning} & -- & 1.75 & -- \\
Durante et al.\cite{durante2017numerical} & 0.19 & 1.98 & $\pm 1.73$ \\
\multicolumn{4}{@{}l@{}}{\footnotesize (Re = $1 \times 10^4$)} \\
\bottomrule
\end{tabular}
}
\end{subtable}

\vspace{0.5em}

\caption{Validation of 2D cylinder flow results at different Reynolds numbers. St: Strouhal number; $\overline{C_D}$: Time-averaged drag coefficient; $\overline{C_L}$: Time-averaged lift coefficient; $\pm$: Oscillation amplitude; --: Data not reported or not applicable}
\label{tab:cylinder_validation_all}
\end{table}

\begin{table}[t!]
\centering
\begin{subtable}[t]{0.32\textwidth}
\centering
\caption{$Re = 200$}
\label{tab:3D_cylinder_Re200}
\adjustbox{width=\textwidth,center}{%
\begin{tabular}{@{}lccc@{}}
\toprule
Reference & St & $\overline{C_D}$ & $C_{L'}$ \\
\midrule
\textbf{Present study} & \textbf{0.192} & \textbf{1.318} & \textbf{0.418} \\
Qu et al.\cite{qu2013quantitative} & 0.183 & 1.25 & 0.35 \\
Posdziech \& Grundmann\cite{posdziech2001numerical} & 0.182 & 1.24 & -- \\
Carmo \& Meneghini\cite{carmo2006numerical} & 0.184 & 1.28 & -- \\
Labbé \& Wilson\cite{labbe2007numerical} & 0.195 & 1.318 & -- \\
Behara \& Mittal\cite{behara2009parallel} & 0.195 & 1.38 & -- \\
Rajani et al.\cite{rajani2009numerical} & 0.1936 & 1.338 & 0.4276 \\
\bottomrule
\end{tabular}
}
\end{subtable}
\hfill
\begin{subtable}[t]{0.32\textwidth}
\centering
\caption{$Re = 1,000$}
\label{tab:3D_cylinder_Re1000}
\adjustbox{width=\textwidth,center}{%
\begin{tabular}{@{}lccc@{}}
\toprule
Reference & St & $\overline{C_D}$ & $C_{L'}$ \\
\midrule
\textbf{Present study} & \textbf{0.213} & \textbf{1.102} & \textbf{0.374} \\
Tong et al.\cite{tong2015numerical} & 0.215 & 1.08 & 0.2 \\
Lei et al.\cite{lei2001spanwise} & -- & 1.112 & 0.325 \\
Labbé \& Wilson\cite{labbe2007numerical} & 0.215 & 1.200 & -- \\
Zhao et al.\cite{zhao2009direct} & 0.21 & 1.17 & 0.335 \\
\bottomrule
\end{tabular}
}
\end{subtable}
\hfill
\begin{subtable}[t]{0.32\textwidth}
\centering
\caption{$Re = 3,900$}
\label{tab:3D_cylinder_Re3900}
\adjustbox{width=\textwidth,center}{%
\begin{tabular}{@{}lccc@{}}
\toprule
Reference & $L_z$ & St & $\overline{C_D}$ \\
\midrule
\textbf{Present study} & \textbf{4.0} & \textbf{0.217} & \textbf{1.074} \\
Suárez et al.\cite{suarez2025active} & $\pi$ & 0.220 & 1.08 \\
Lehmkuhl et al.\cite{lehmkuhl2013low} & $\pi$ & 0.215 & 1.015 \\
Parnaudeau et al.\cite{parnaudeau2008experimental} & 23 & 0.208 &  -- \\
Tremblay et al.\cite{tremblay2002flow} & $\pi$ & 0.22 & 1.03 \\
Kravchenko \& Moin\cite{kravchenko2000numerical} & $\pi$ & 0.21 &  1.04 \\
\bottomrule
\end{tabular}
}
\end{subtable}

\vspace{0.5em}

\caption{Comparison of 3D cylinder flow results at different Reynolds numbers. St: Strouhal number; $\overline{C_D}$: Time-averaged drag coefficient; $C_{L'}$: RMS lift coefficient; $L_z$: Span length; $L_r/D$: Recirculation length to diameter ratio; --: Data not reported or not applicable}
\label{tab:3D_cylinder_comparison_all}

\end{table}

\paragraph{Environment setup.}

The default reward objective $r$ for the cylinder flow environment targets drag minimization and the reduction of lift force oscillations and reads $r = -|C_D| - \omega |C_L|$, where $C_D, C_L$ denote the normalized drag and lift coefficients, and $\omega$ is a Reynolds number-dependent scaling factor with values of 0.1, 0.2, 1.0, and 1.0 for $Re = 100, ~200, ~1,000, ~3,900$, respectively. Since the unstable symmetric state is also the minimum-drag configuration, this objective corresponds to stabilizing this unstable fixed point. This reward formulation serves as the default configuration, though additional reward shaping terms can be easily implemented by users according to their specific research objectives.

Two different actuation strategies are implemented to provide the RL agent with control authority over the flow dynamics. First, a zero-net mass-flux blowing/suction strategy is provided, comprised of two equal and opposite jets - one at the top and one at the bottom of the cylinder. The RL agent controls the mass flux of the upper jet, which automatically specifies the lower jet through the zero-net-mass constraint. In the default configuration, the jets are placed at $90^\circ$ and $270^\circ$ positions and cover $10^\circ$ of the cylinder circumference each, though the HydroGym framework is designed to allow for flexible user-specified adjustments of the jet number, placement angles, and coverage areas. Second, a rotating surface actuator is implemented for which the RL agent directly controls the angular velocity of the cylinder surface. Both action spaces are normalized to the range $[-1, 1]$ to ensure consistent RL algorithm performance across different actuation modes.

The environment supports both two-dimensional and three-dimensional cylinder flow configurations across multiple Reynolds numbers ($Re = 100, ~200, ~1,000, ~3,900$), enabling the investigation of flow control across different flow regimes from laminar vortex shedding to fully turbulent conditions. For three-dimensional cases at $Re = 1,000$ and $3,900$, the framework additionally supports multi-agent reinforcement learning through spanwise domain decomposition, where the cylinder is divided into 4 and 10 subcylinder segments respectively, each individually controlled by a dedicated agent. This multi-agent approach enables the investigation of cooperative control strategies and spanwise flow coordination. The framework provides flexibility in multi-agent coordination, allowing agents to share gradient information and experience buffer data if desired, though independent agent operation is also supported based on user requirements.

The state space representation consists of user-configurable point probes distributed throughout the computational domain, providing measurements of velocity components, pressure, density, and body forces at specified locations. By default, the state observations consist of the raw solver output, but different normalization strategies are implemented to normalize the state space in the range $[-1, 1]$ for consistent RL training performance. This flexible sensor configuration allows the investigation of minimal sensing strategies and optimal sensor placement for effective flow control.

Each training episode spans 200 control actions, corresponding to approximately 10 vortex shedding periods, providing sufficient time for the agent to observe and respond to flow dynamics. The temporal coupling between the CFD solver and RL agent employs Reynolds number-specific action frequencies: 200, 350, and 600 CFD timesteps per control action for 2D cases at $Re = 200, ~1,000, ~3,900$ respectively, and 400, 500, and 1,200 timesteps for corresponding 3D cases. These ratios are designed to match the characteristic timescales of vortex shedding dynamics while maintaining computational efficiency and control authority.

The environment framework provides a comprehensive testbed for investigating reinforcement learning-based active flow control strategies, supporting both fundamental research into control mechanisms and practical development of sensing and actuation approaches for cylinder wake management. All described configurations represent default settings that can be easily modified by users to accommodate specific research requirements, including custom reward functions, alternative actuation strategies, modified sensor placements, and adjusted temporal coupling parameters.


\subsection{Fluidic pinball flow}
\label{appendix:Environments:Pinball}

\paragraph{Characteristic physics.}
The fluidic pinball serves as an exemplary test case for investigating complex wake dynamics and flow control strategies in transitional flow regimes. This benchmark configuration consists of three equal-diameter circular cylinders positioned at the vertices of an equilateral triangular arrangement, with one vertex directed toward the incoming flow~\cite{noack2017pinball}. Although geometrically simple, this arrangement demonstrates a diverse spectrum of nonlinear flow behaviors, including steady-state solutions, time-periodic oscillations, quasi-periodic motion, and fully chaotic dynamics across different operating parameters.

The fluidic pinball exhibits distinct flow regime transitions as the Reynolds number varies from 30 to 150, with critical thresholds that differ from classical single-cylinder wake behavior. Initial investigations by~\cite{deng2020loworder} identified specific transition points unique to this multi-cylinder configuration. For Reynolds numbers below Re < 18, the flow field maintains global stability with symmetric recirculation zones forming downstream of the cylinder cluster. The transition from this steady state occurs via a supercritical Hopf bifurcation at approximately $Re = 18$~\cite{deng2020loworder}. This early onset of instability results from the enhanced wake interactions between the three cylinders.
The Hopf bifurcation introduces time-periodic vortex shedding characterized by complex conjugate eigenvalues crossing the stability boundary in the linearized flow equations. The resulting flow exhibits von Kármán-type vortex shedding with dimensionless frequency (Strouhal number) of approximately 0.2. Importantly, this initial instability preserves the mirror-plane symmetry about the configuration centerline, maintaining symmetric shedding patterns until $Re \approx 68$.

At $Re \approx 68$, the fluidic pinball undergoes a second critical transition through a supercritical pitchfork bifurcation~\cite{pastur2019pinball}. This transition fundamentally alters the wake dynamics by breaking the mirror symmetry preserved during the initial periodic regime. The physical mechanism involves the destabilization of the jet flow between the two downstream cylinders, which becomes susceptible to preferential deflection in either the upward or downward direction. This symmetry-breaking creates two mirror-symmetric stable solutions - one with upward jet deflection and another with downward deflection. The system exhibits bistable behavior, where initial conditions determine which asymmetric state the flow adopts~\cite{pastur2019pinball}. This represents a qualitative change from the previous symmetric periodic shedding to asymmetric limit cycle dynamics.

Beyond $Re = 104$, the fluidic pinball system develops additional complexity through a secondary Neimark-Sacker bifurcation~\cite{deng2021route}. This mathematical bifurcation introduces quasi-periodic motion by creating torus-like attractors in the system's phase space. The primary vortex shedding frequency becomes modulated by secondary low-frequency oscillations, establishing multi-frequency dynamics characteristic of weakly chaotic flows. The transition to fully chaotic motion occurs above $Re \approx 130$, following what was identified by~\cite{deng2021route} as a Newhouse-Ruelle-Takens scenario. In this $Re$ number regime, the quasi-periodic torus structure becomes unstable and breaks down, leading to aperiodic motion with sensitive dependence on initial conditions - the hallmark of deterministic chaos.

Moreover, three-dimensional effects can introduce additional complexity beyond the two-dimensional analysis. Following the foundational work of~\cite{williamson1996vortex} on cylinder wake transitions, the fluidic pinball exhibits spanwise instabilities originating from hyperbolic regions within the braid shear layers. These instabilities manifest as short-wavelength structures with characteristic spanwise dimensions comparable to the cylinder diameter. The three-dimensional modes create streamwise vortical structures that interact with the primary cross-stream vortex shedding. Multiple instability modes can coexist and compete, leading to complex scenarios including cooperative destabilization and vortex reconnection phenomena that cannot be captured by two-dimensional computational models.

To summarize, the fluidic pinball's value as a benchmark derives from the combination of geometric simplicity and dynamical richness, making it suitable for testing advanced control strategies while remaining computationally tractable. The independent rotation capability of each cylinder provides multiple actuation inputs, enabling implementation of various flow control mechanisms established in the literature. Classical flow control approaches applicable to this configuration include boundary layer manipulation through surface blowing (analogous to base bleeding techniques), high-amplitude periodic forcing for vortex synchronization, and circulation control via cylinder rotation utilizing Magnus force effects. Additionally, modern approaches such as opposition control and model-based feedback strategies can be evaluated. 

Hence, the fluidic pinball presents several fundamental challenges for flow control design. The succession of bifurcations means that control strategies must adapt to qualitatively different flow physics as operating conditions change. Systems designed for the symmetric periodic regime may prove ineffective or counterproductive when applied to the asymmetric bistable regime. The multi-frequency dynamics create coupling between different temporal scales, where actuation at one frequency can influence the entire spectral content of the flow response. This cross-frequency coupling necessitates control architectures that account for broadband nonlinear interactions rather than targeting individual frequency components. Furthermore, the presence of multiple coexisting attractors (in the bistable regime) and hysteresis effects between flow states require control strategies capable of managing state transitions and maintaining desired operating points despite external disturbances.

Extension to three-dimensional control introduces requirements for spanwise-distributed actuation to address the various three-dimensional instability modes. The transition between different three-dimensional regimes creates scenarios where two-dimensional control approaches become inadequate, necessitating fundamentally different control frameworks. Control strategies optimized for one three-dimensional mode may lose effectiveness or become destabilizing when the flow transitions to alternative modal states. This highlights the importance of robust adaptive control approaches that maintain performance across regime boundaries.

\paragraph{Numerical setup and validation.}
The fluidic pinball configuration consists of three circular cylinders arranged in an equilateral triangle formation, creating a fundamentally different flow topology compared to the single cylinder case. Relative to the pinball center at 
\begin{wraptable}{r}{0.5\textwidth}
\centering
\begin{adjustbox}{width=0.5\textwidth,center}
\begin{tabular}{@{}lcccccc@{}}
\toprule
$Re$ & Reference & Dim & $f_0$ & $\overline{C_D}$ & $\overline{C_{L,2}}$ & $\overline{C_{L,3}}$ \\
\midrule
\multirow{3}{*}{30} & \textbf{Present study} & \textbf{2D} & \textbf{0.064} & \textbf{2.651} & \textbf{-0.361} & \textbf{0.361} \\
& \textbf{Present study} & \textbf{3D} & \textbf{0.061} & \textbf{2.597} & \textbf{-0.358} & \textbf{0.359} \\
& Wang et al.\cite{wang2023cluster} & 2D & 0.065 & 2.656 & -0.365 & 0.365 \\
\hline \hline
\multirow{3}{*}{100} & \textbf{Present study} & \textbf{2D} & \textbf{0.088} & \textbf{2.904} & \textbf{-0.079} & \textbf{0.110} \\
& \textbf{Present study} & \textbf{3D} & \textbf{0.087} & \textbf{2.795} & \textbf{-0.0107} & \textbf{0.119} \\
& Wang et al.\cite{wang2023cluster} & 2D & 0.090 & 2.874 & -0.0745 & 0.108 \\
\hline \hline
\multirow{3}{*}{150} & \textbf{Present study} & \textbf{2D} & \textbf{0.120} & \textbf{2.922} & \textbf{-0.0441} & \textbf{0.0330} \\
& \textbf{Present study} & \textbf{3D} & \textbf{0.113} & \textbf{2.949} & \textbf{-0.0467} & \textbf{0.0297} \\
& Wang et al.\cite{wang2023cluster} & 2D & 0.118 & 2.960 & -0.043 & 0.0319 \\
\bottomrule
\end{tabular}
    
\end{adjustbox}
\caption{Validation of the fluidic pinball flow results at different Reynolds numbers. Dim: Dimensionality (2D/3D); $f_0$: Natural frequency; $\overline{C_D}$: Time-averaged drag coefficient; $\overline{C_{L,i}}$: Time-averaged lift coefficient for cylinder i}
\label{tab:fluidic_pinball_validation}
\end{wraptable}
$(x/D,y/D) = (10.0,0.0)$ for the 2D simulations and $(8.0,0.0)$ for the 3D counterpart, the cylinders are positioned with centers at $(x/D,y/D) = (-1.299,0.0)$, $(0.0,0.75)$, and $(0.0,-0.75)$, where the upstream cylinder interacts with two downstream cylinders to generate complex wake dynamics and vortex interactions. The computational setup employs the same domain dimensions and boundary condition treatments as described for the single cylinder benchmark. However, the fluidic pinball operates at lower Reynolds numbers $Re \in [30,100,150]$, allowing for a uniform refinement strategy with $n=12$ across all cases. This refinement level provides adequate resolution to capture the intricate flow features arising from the multi-cylinder interaction, including wake merging, vortex pairing, and the characteristic frequency modulation effects. The triangular arrangement leads to inherently asymmetric flow patterns, where the downstream cylinders experience different flow conditions due to their positioning relative to the upstream cylinder's wake. This asymmetry manifests in the lift coefficient distributions and requires careful validation against established benchmarks to ensure accurate representation of the flow physics. Validation results are summarized in Table~\ref{tab:fluidic_pinball_validation}, demonstrating good agreement with the reference data of Wang et al.~\cite{wang2023cluster}. The simulations accurately predict the natural frequencies across all Reynolds numbers. The time-averaged drag coefficients show good agreement (within 2\% deviation), while the asymmetric lift coefficients on the downstream cylinders are well-captured, confirming a proper resolution of the wake interaction dynamics. The 3D results maintain similar accuracy levels, validating the three-dimensional extension of the fluidic pinball configuration. 

\begin{figure}[t!]
    \centering
    \includegraphics[width=\linewidth]{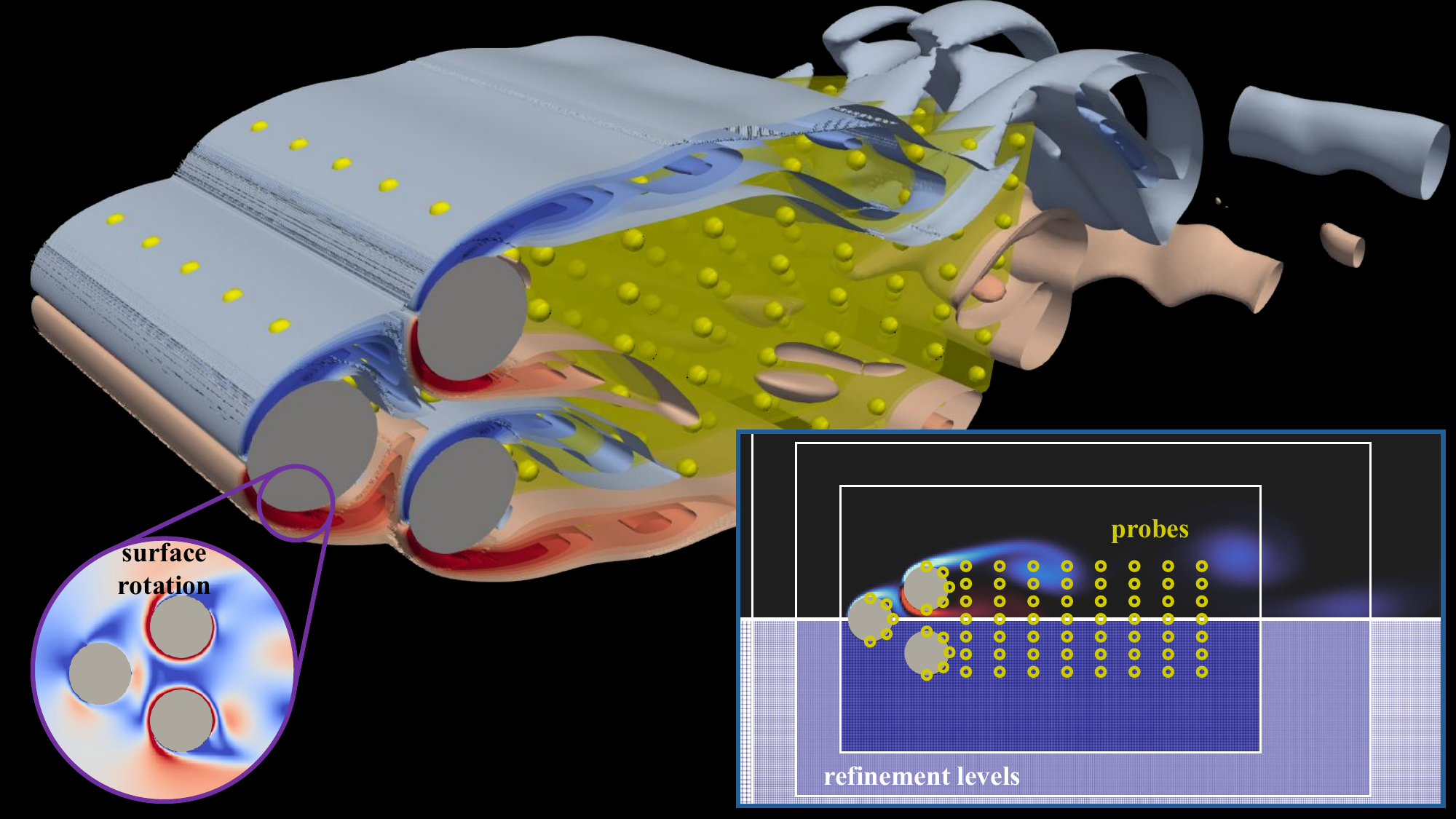}
    \caption{Computational domain setup, sensor probe distribution and actuation strategy for the fluidic pinball environment: Three-dimensional view showing the probe distribution around the fluidic pinball with vortical structures visualized through Q-criterion isosurfaces. The yellow dots in the zoom-in highlight sensor probe locations in a single spanwise plane. Active flow control is performed by individual surface rotation of each cylinder.}
    \label{appendix:fig:pinball}
\end{figure}

\paragraph{Environment setup.}

The fluidic pinball environment features a multi-cylinder flow control setup comprising three circular cylinders positioned in an equilateral triangular arrangement. The primary reward function $r$ for the fluidic pinball targets total drag reduction across all three cylinders and is defined as $r = -|\sum_{i=1}^{3} C_{D,i}| - \omega |\sum_{i=1}^{3} C_{L,i}|$, where $C_{D,i}$ represents the normalized drag coefficient of cylinder $i$ and $\omega = 1.0$ serves as the scaling parameter consistently applied across all Reynolds numbers to avoid control policies that introduce a significant lift generation. Users can readily adapt this reward structure to explore different objectives including thrust control or independent cylinder optimization.

The control mechanism employs surface rotation actuation, where the RL agent manages the angular velocity of each cylinder individually (see Fig. \ref{appendix:fig:pinball}). This creates a three-dimensional continuous action space with components bounded within $[-1, 1]$, facilitating exploration of cooperative multi-cylinder control approaches. Although jet-based actuation is not part of the standard setup, the framework design accommodates user integration of blowing/suction mechanisms for extended research applications or comparative analysis.

The framework encompasses both two-dimensional and three-dimensional fluidic pinball flow scenarios across various Reynolds numbers ($Re = 30, 75, 100, 150$), covering flow regimes from steady symmetric configurations through periodic vortex shedding to bistable asymmetric behavior and quasi-periodic dynamics. This Reynolds number spectrum encompasses the distinctive bifurcation characteristics of the fluidic pinball, including the initial Hopf bifurcation at $Re \approx 18$, the symmetry-breaking pitchfork bifurcation at $Re \approx 68$, and progressive transitions toward chaotic flow patterns. The setup utilizes a single-agent configuration, where one agent manages all three cylinder rotations concurrently, promoting research into unified control methodologies for multi-body flow applications.

The observation space utilizes the same adaptable probe-based methodology as established in other environments, featuring user-definable point measurements of velocity components, pressure, density, and body forces positioned throughout the computational domain. Training episodes encompass 200 control actions, equivalent to roughly 10 vortex shedding cycles, offering an adequate temporal scope for agents to perceive and react to the intricate multi-frequency behavior inherent in the fluidic pinball. The CFD-RL coupling utilizes Reynolds number-dependent action intervals tailored to match evolving flow characteristics: 230, 190, 170, and 100 CFD timesteps per control action for 2D scenarios at $Re = 30, 75, 100, 150$ respectively, with corresponding 3D cases using 250, 235, 225, and 200 timesteps. These intervals adapt to the changing temporal scales and flow complexity encountered as the system progresses through distinct dynamical states, maintaining suitable control responsiveness while preserving computational tractability. All outlined specifications serve as baseline configurations that users can readily adjust to meet particular research needs, encompassing modified reward structures, alternative control mechanisms, customized sensor layouts, and tailored temporal coupling settings.


\subsection{Open cavity flow}
\label{appendix:Environments:Cavity}

\paragraph{Characteristic physics.}
The flow around open cavities in the Reynolds number range 4,000-7,500 based on the cavity height and the freestream velocity exhibits complex transitional behavior characterized by distinct regime evolutions and sophisticated three-dimensional instabilities that have profound implications for both fundamental aeroacoustics research and practical flow control applications. This transitional range encompasses the evolution from predominantly steady two-dimensional recirculation patterns to complex three-dimensional unsteady dynamics with strong acoustic coupling, presenting unique challenges for modeling, prediction, and control that distinguish it from both lower and higher Reynolds number regimes.

Within this Reynolds number range, the flow physics is governed by several critical transitions that fundamentally alter the cavity dynamics and acoustic characteristics. At Reynolds numbers around 4,000-5,000, the initially stable two-dimensional recirculation undergoes its first significant three-dimensional instability through centrifugal mechanisms arising from curved streamlines in the recirculation region~\cite{sipp2007global, faure2007visualization}. This transition manifests as weakly unsteady flow behavior with emerging spanwise variations that mark the onset of complex temporal dynamics. As the Reynolds number continues to increase, a second critical transition occurs around Re = 7,000-8,000, where strong temporal instabilities emerge through Hopf bifurcation characteristics, introducing multiple competing temporal modes and mode-switching phenomena where different frequencies compete for energy~\cite{barbagallo2009closed}. These successive transitions create a regime where simplified two-dimensional approximations become progressively inadequate for an accurate prediction of both aerodynamic and acoustic behavior.

The emergence of these transitional instabilities creates fundamental differences between 2D and 3D cavity flow configurations that extend far beyond geometric considerations. While two-dimensional simulations remain computationally efficient and provide valuable insights into fundamental feedback mechanisms, they systematically overpredict cavity tone amplitudes by up to 15 dB compared to three-dimensional cases and miss critical mixing processes that govern realistic flow physics~\cite{devicente2015linear}. The three-dimensional effects introduce end-wall influences that cause flow spillage over cavity sides, development of Taylor-Görtler-like vortices, and breakdown of coherent shear layer structures due to spanwise instabilities. In compressible flows within this Reynolds number range, the acoustic feedback mechanisms become particularly complex, with the classical Rossiter formula requiring depth-sensitized modifications to account for three-dimensional cavity effects and modified convection velocities~\cite{barone2018cavity, rossiter1964wind}.

This rich transitional behavior makes cavity flows particularly valuable as canonical test cases for developing and validating advanced flow control strategies, especially those employing machine learning approaches. The geometric simplicity combined with complex physics provides an ideal benchmark that bridges fundamental research and practical aerodynamic applications. Various frameworks have provided theoretical foundations for understanding the underlying bifurcation mechanisms and sensitivity characteristics that govern cavity flow transitions~\cite{sipp2007global}. However, the practical implementation of flow control in this Reynolds number range faces several specific challenges that stem from the complex transitional flow physics and multi-frequency acoustic coupling. The strong nonlinear coupling between Rossiter modes creates frequency cross-talk and triadic interactions that complicate traditional linear control approaches, while mode-switching regimes introduce time-varying dynamics that challenge adaptive control systems~\cite{rowley2006dynamics}. Multiple time scales ranging from fast Kelvin-Helmholtz instabilities to slow mode-switching phenomena create complex control requirements, with real-time implementation demanding computational capabilities supporting control frequencies up to 10 kHz. Energy efficiency considerations become particularly critical in practical applications, as the spatially disjoint measurement and control locations inherent in cavity flows introduce time-delay effects that can destabilize feedback control systems if not properly addressed~\cite{cattafesta2008active}. As a result, the rich transitional physics and computational accessibility make cavity flows in this Reynolds number range an enduring benchmark for flow control innovation, particularly as machine learning approaches mature and computational resources continue to expand.

\paragraph{Numerical setup and validation.}
The open cavity benchmark comprises 2D and 3D low Mach number flows ($Ma < 0.2$) over square cavities at Reynolds numbers $Re_D \in [4,200, ~7,500]$, based on the cavity depth $D$ and free-stream velocity $U_\infty$, following the configuration of Barbagallo et al.~\cite{barbagallo2009closed}.

\begin{wrapfigure}[20]{r}{0.55\textwidth}
   \centering
    \vspace{-1.2em}
    \includegraphics[clip, trim=0.0cm 0.0cm 0.0cm 0.0cm, width=0.55\textwidth]{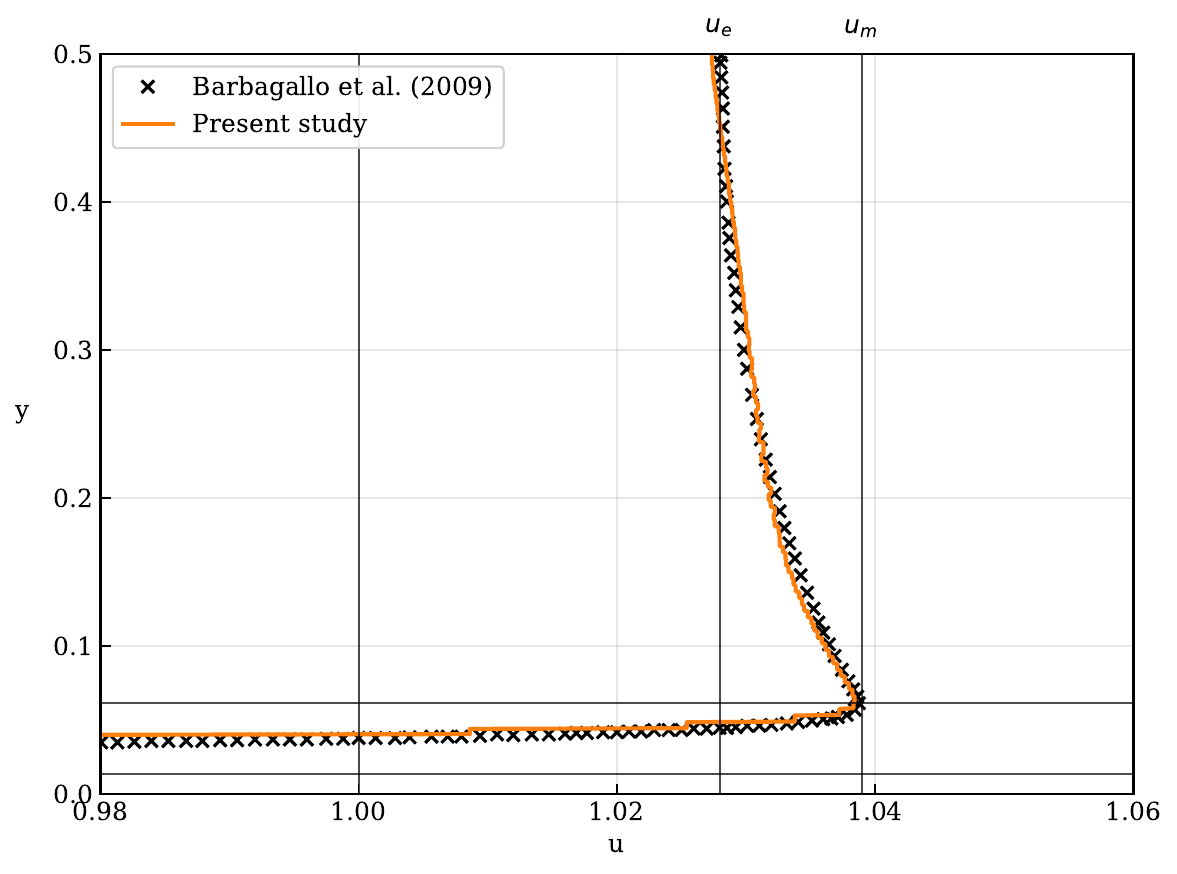}
    \caption{Time-averaged leading edge velocity profiles for 2D cavity flow at $Re = 7,500$. Present numerical results (solid line) compared with Barbagallo et al. (2009)~\cite{barbagallo2009closed} reference data (crosses). Velocities normalized by free-stream velocity; vertical coordinate normalized by cavity length.}
    \label{appendix:2D_cavity_validation}
\end{wrapfigure}
\begin{figure}[t!]
    \centering
    \includegraphics[width=\linewidth]{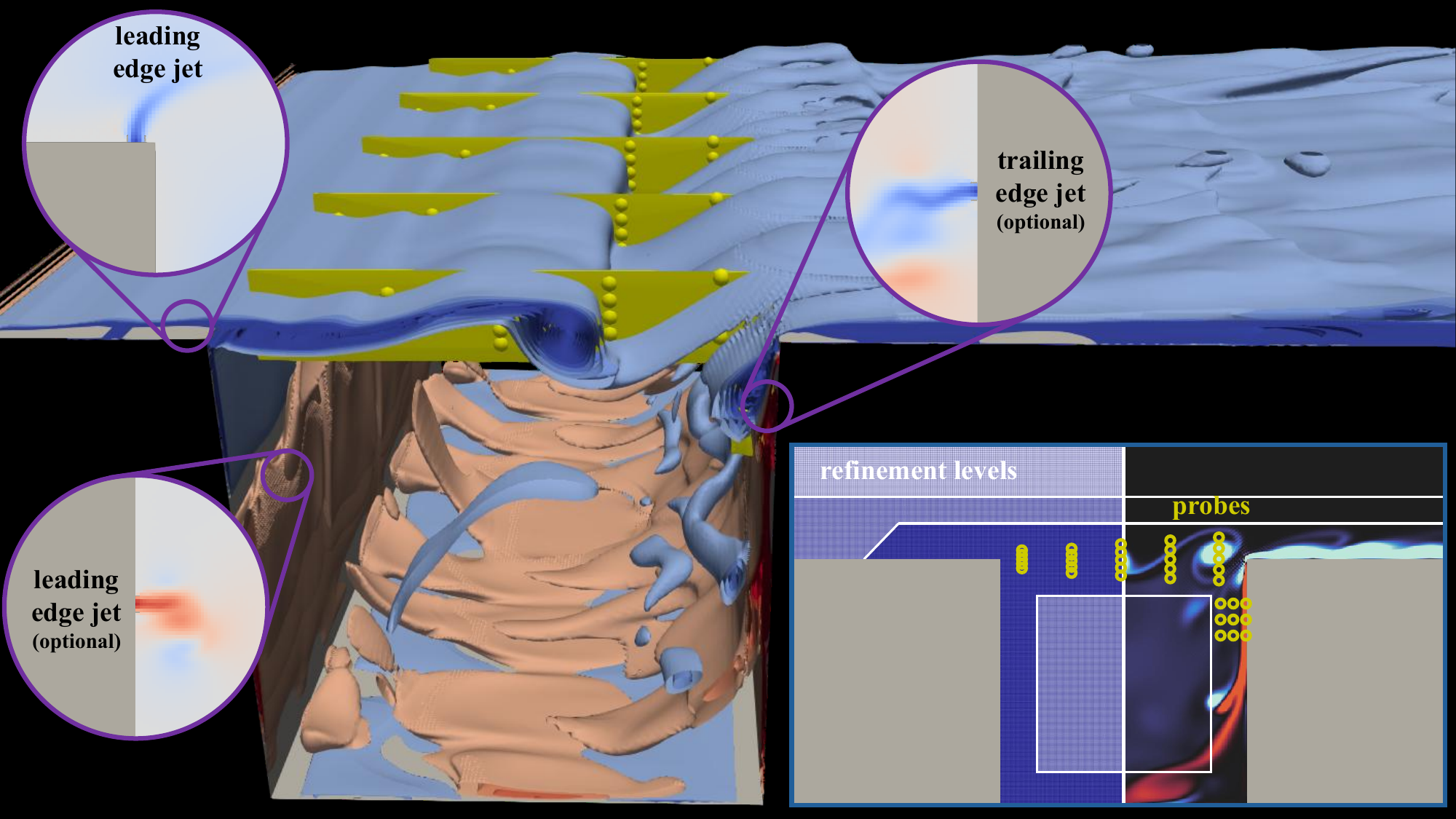}
    \caption{Computational domain setup, sensor probe distribution and actuation strategy for the open cavity flow: Three-dimensional view showing the probe distribution in the open cavity environment with vortical structures visualized through Q-criterion isosurfaces. The yellow dots highlight sensor probe locations in a single spanwise plane. Active flow control is performed using up to three independent jet actuators positioned at the leading edge  and inside the cavity.}
    \label{appendix:fig:cavity}
\end{figure}

For the 2D case, the computational domain is defined by $(x/D, y/D) \in [0.0, 2.5] \times [0, 1.5]$ with the square cavity located at $(x/D, y/D) \in [0.27, 1.27] \times [0, 1]$. The grid spacing is determined by the grid refinement level $n$ via $\Delta_n = \Delta_0/2^n$ with $\Delta_0^{(2D)} = 2.5D$. Multiple refinement patches are employed with maximum refinement levels of $n = 12$ near the cavity edges and shear layer regions, corresponding to approximately 227 cells per cavity depth (see Fig. \ref{appendix:fig:cavity}). Special attention is given to resolving the boundary layer upstream of the cavity, targeting a displacement thickness of $\delta_1 \approx 0.012D$. This numerical setup demonstrates good agreement with established literature~\cite{barbagallo2009closed} as shown in Fig. \ref{appendix:2D_cavity_validation}. In 3D, the domain extends to $(x/D, y/D, z/D) \in [0.0, 2.5] \times [0, 1.5] \times [0, 2.25]$ with periodic boundary conditions applied in the spanwise $z$-direction. A similar grid compared to the 2D test case is used to maintain adequate resolution of three-dimensional instabilities while keeping computational costs manageable.
At the inlet, a Dirichlet boundary condition implements a uniform velocity profile $(u, v, w) = (U_\infty, 0, 0)$. The top boundary implements symmetry conditions $(\partial_y u = 0, v = 0, \partial_y w = 0)$. No-slip conditions $(u = v = w = 0)$ are applied on the cavity walls and the bottom surface from $x/D = 0.1777$ onwards, while symmetry conditions are used for $x/D <0.1777$ to generate appropriate boundary layer development.

\paragraph{Environment setup.}
The open cavity environment features a flow control setup designed to stabilize the inherent shear layer instability of cavity flows. This configuration provides a consistent framework across both 2D and 3D flow scenarios, enabling comprehensive investigation of the cavity flow dynamics and active control strategies.

The primary objective centers on minimizing flow fluctuations through targeted stabilization of the shear layer that develops across the cavity opening. The reward function $r$ quantifies this stabilization goal as:

\begin{equation}
r = -\sum_{i} \left(\frac{\text{obs}_i - \text{target}_{mean,i}}{\text{target}_{std,i}}\right)^2
\end{equation}

where $\text{obs}_i$ represents the observed quantity at measurement location $i$, $\text{target}_{mean,i}$ denotes the target reference value, and $\text{target}_{std,i}$ provides normalization scaling. This formulation encourages the control system to maintain observed flow quantities close to their target states, thereby reducing the amplitude of shear layer oscillations and associated cavity resonance phenomena. For simplicity, the target system state is approximated by a temporal mean over 1,000 instability cycles. 

The control mechanism employs jet-based actuation strategically positioned to influence the shear layer development and cavity flow dynamics. The standard configuration utilizes a single jet oriented in the $y$-direction and located at the cavity leading edge, providing direct interaction with the separating boundary layer. For enhanced control authority, an extended configuration incorporates additional jets positioned within the cavity at coordinates $(1.277, 0.75)$ and $(0.277, 0.5)$ that act in streamwise direction, enabling multi-point flow manipulation and improved stabilization capabilities.

The jet actuation creates a continuous action space where the RL agent controls the jet momentum coefficients or velocity ratios. This setup facilitates exploration of both passive flow stabilization through steady jet injection and active control strategies involving time-varying actuation patterns. The framework design readily accommodates user modifications to jet positioning, orientation, and actuation parameters to investigate alternative control approaches or cavity geometries.

The CFD-RL coupling employs 400 CFD timesteps per control action for 2D test cases and 375 timesteps for 3D counterparts, providing sufficient temporal resolution to capture the essential flow physics while maintaining computational efficiency. This interval balances the need to resolve the characteristic timescales of cavity flow instabilities with practical training considerations, allowing the control agent adequate time to observe the effects of its actions on the flow field evolution. Training episodes are designed to encompass multiple characteristic flow timescales, enabling agents to learn both short-term stabilization responses and longer-term flow management strategies. The environment supports the investigation of various Reynolds numbers and cavity aspect ratios, accommodating research into the Reynolds number dependence of control effectiveness and the geometric sensitivity of stabilization approaches.


\subsection{Square cylinder flow}
\label{appendix:Environments:Square_cylinder}

\paragraph{Characteristic physics.}
The flow around a square cylinder represents a canonical configuration for studying bluff body aerodynamics, providing fundamental insights relevant to both scholarly inquiry and a wide range of engineering applications, including the design of buildings, bridges, and offshore structures. The interaction of the fluid with the sharp corners of the cylinder forces flow separation, leading to the formation of an unsteady wake dominated by the periodic shedding of large-scale vortices \citep{williamson1996vortex}. This alternating vortex shedding generates fluctuating aerodynamic forces on the body, which can induce structural vibrations—a phenomenon termed vortex-induced vibration—posing a significant risk of material fatigue and creating acoustic noise.

The dynamics of the flow are highly dependent on the Reynolds number, and within the range of 200 to 3,900, the wake undergoes a series of complex transitions from a two-dimensional laminar state to a fully three-dimensional turbulent one. The initial transition from a purely two-dimensional to a three-dimensional wake occurs at a critical Reynolds number between approximately 150 and 200 \citep{sohankar1999simulation, bai2018dependence}. This transition is marked by the emergence of spanwise instabilities, akin to those observed in circular cylinder wakes. The first of these is the Mode A instability, which appears around Re $\approx$ 165 and is characterized by the formation of large-scale streamwise vortex loops with a spanwise wavelength of about four to five times the cylinder diameter \citep{jiang2018three, bai2018dependence}. As the Reynolds number increases further, a transition to the Mode B instability occurs around Re $\approx$ 210, which involves smaller-scale, rib-like streamwise vortices with a much shorter spanwise wavelength of approximately one cylinder diameter \citep{jiang2018three}. Beyond these initial instabilities, for Reynolds numbers approaching and exceeding 1,000, the separated shear layers themselves become unstable, leading to the roll-up of Kelvin-Helmholtz vortices which subsequently break down into finer turbulent structures \citep{Trias2015}. At higher Reynolds numbers the wake is fully turbulent and characterized by a broad spectrum of interacting eddies \citep{Trias2015, Srinivas2006}.

Significant distinctions exist between the flow around square and circular cylinders, primarily stemming from the nature of the flow separation. The sharp corners of the square cylinder fix the separation points, whereas for a circular cylinder, these points are mobile and their location depends on the Reynolds number \citep{zhu2021numerical}. This fixed separation generally leads to a wider, more pronounced wake and a higher overall drag coefficient for the square cylinder under identical inflow conditions \citep{Yuce2016}. While the wake instability modes are qualitatively similar, their onset and development occur at different critical Reynolds numbers. 
The differences between 2D and 3D configurations are also profound; two-dimensional simulations are incapable of capturing the physics of spanwise instabilities and the resulting energy cascade to smaller turbulent scales. Consequently, 2D models often yield inaccurate predictions of aerodynamic forces and wake dynamics for Reynolds numbers beyond the onset of three-dimensionality \citep{sohankar1999simulation}.

The complex, multi-scale nature of this flow presents considerable challenges for numerical modeling. Reynolds-Averaged Navier-Stokes (RANS) models often fail to accurately predict the highly unsteady and separated flow, particularly the subtle dynamics of vortex shedding and turbulent transition \citep{rodi1997comparison}. Scale-resolving approaches, such as Large Eddy Simulation (LES) and Direct Numerical Simulation (DNS), are better suited but are computationally intensive. Achieving accurate results with these methods requires not only extremely fine computational meshes but also large domain sizes and long integration times to ensure that the statistics are fully converged and free from boundary condition influences \citep{Trias2015, sohankar1998low}.

These well-defined yet complex flow characteristics make the square cylinder an important test case for the development and validation of flow control strategies, including those leveraging advanced RL techniques. Key challenges for applying RL in this context include the high computational cost of simulations required for training, which can limit the exploration of the solution space, and the persistent difficulty of transferring control policies learned in simulation to real-world experimental setups.

\paragraph{Numerical setup and validation.}
The square cylinder benchmark follows the same computational setup as the circular cylinder case described above, with the key difference being the bluff body geometry. The square cylinder has a side length $D$ and the Reynolds number is based on this characteristic length, i.e. $Re \in [200,~1,000,~3,900]$.

\begin{figure}[t!]
    \centering
    \includegraphics[width=\linewidth]{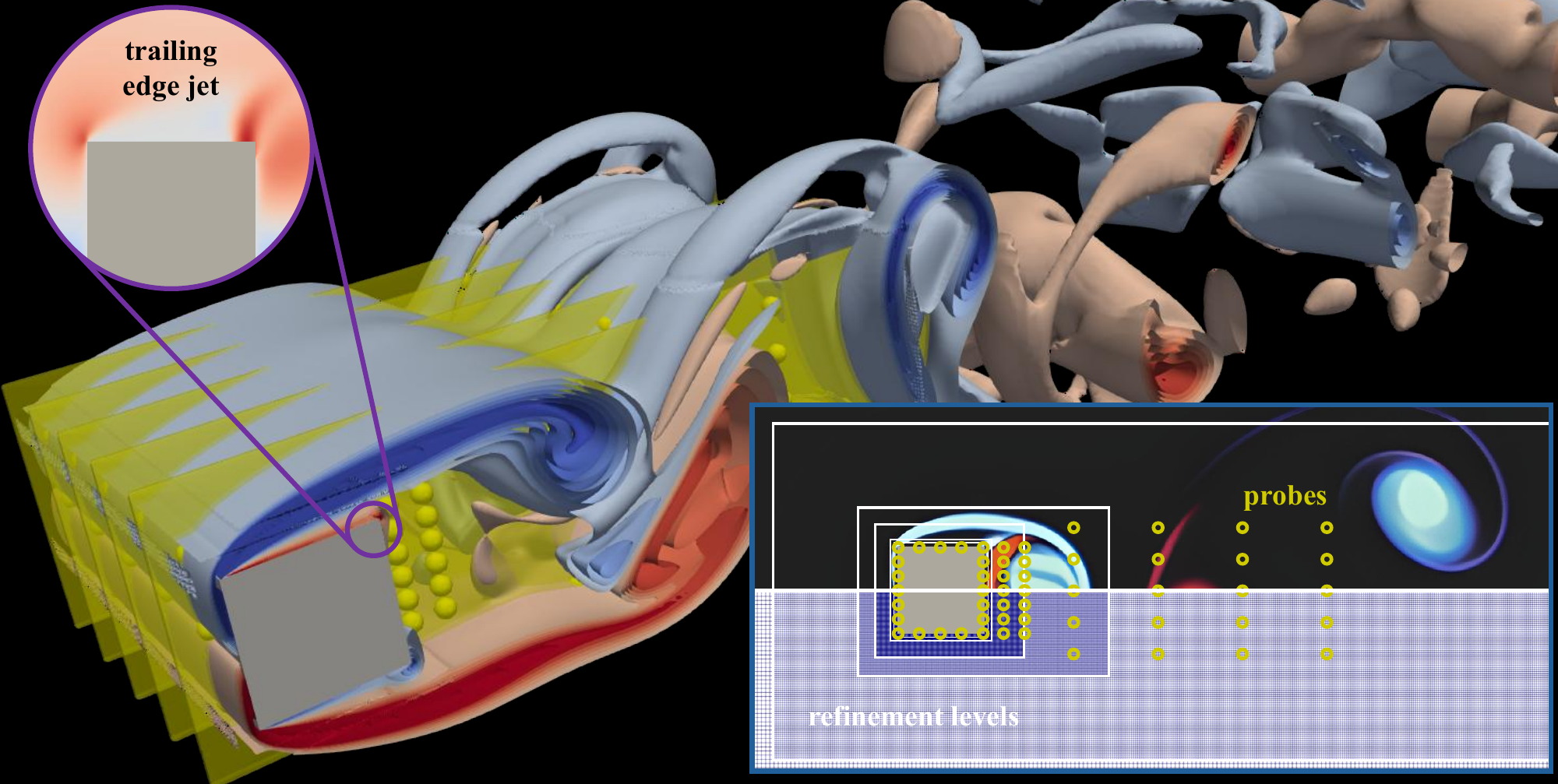}
    \caption{Computational domain setup, sensor probe distribution and actuation strategy: Three-dimensional view showing the probe distribution around the square cylinder with vortical structures visualized through Q-criterion isosurfaces. The yellow dots in zoom-in highlight sensor probe locations in a single spanwise plane. Active flow control is performed through jet actuators positioned at the leading and/or trailing edge.}
    \label{fig:square_cylinder_actuation}
\end{figure}

The computational domains, grid spacing definitions, and refinement strategies remain identical to the circular cylinder configuration. For 2D cases, the domain spans $(x/D,y/D) \in [0,40] \times [-10,10]$ with the square cylinder centered at $(x/D,y/D) = (10,0)$. The 3D domain is $(x/D,y/D,z/D) \in [0,32] \times [0,16] \times [-2,2]$ with the square cylinder centered at $(x/D,y/D)=(8,8)$ and periodic boundary conditions in the $z$-direction as shown in Fig.~\ref{fig:square_cylinder_actuation}.

All boundary condition implementations---inlet Dirichlet conditions with prescribed velocity profiles, non-reflecting outlet conditions, and periodic lateral boundaries---are applied identically to the circular cylinder setup. The grid refinement levels up to $n=14$ and the resolution criteria also follow the same specifications as detailed in the previous subsection resulting in 512 cells per side length $D$.

Our validation results for the square cylinder demonstrate comparable accuracy to the circular cylinder cases, with computed Strouhal numbers and drag coefficients showing good agreement with established literature values across available Reynolds numbers in literature (see Tab. \ref{tab:square_cylinder_comparison_all}).

\begin{table}[t!]
\centering
\begin{subtable}[t]{0.32\textwidth}
\centering
\caption{2D, $Re = 200$}
\label{tab:2D_square_cylinder_Re200}
\adjustbox{width=\textwidth,center}{%
\begin{tabular}{@{}lccc@{}}
\toprule
Reference & St & $\overline{C_D}$ & $C_{L'}$ \\
\midrule
\textbf{Present study} & \textbf{0.151} & \textbf{1.469} & \textbf{0.391} \\
Sohankar et al.\cite{sohankar1999simulation} & 0.170 & 1.46 & 0.32 \\
Sohankar et al.\cite{sohankar1998low} & 0.15 & 1.462 & 0.377 \\
Cheng et al.\cite{cheng2007numerical} & 0.15 & 1.45 & 0.372 \\
Okajima\cite{okajima1982strouhal} & 0.148 & 1.45 & -- \\
Jiang \& Cheng\cite{jiang2018hydrodynamic} & 0.152 & 1.443 & 0.412 \\
\bottomrule
\end{tabular}
}
\end{subtable}
\hfill
\begin{subtable}[t]{0.32\textwidth}
\centering
\caption{2D, $Re = 1,000$}
\label{tab:2D_square_cylinder_Re1000}
\adjustbox{width=\textwidth,center}{%
\begin{tabular}{@{}lccc@{}}
\toprule
Reference & St & $\overline{C_D}$ & $std(C_L)$ \\
\midrule
\textbf{Present study} & \textbf{0.121} & \textbf{2.302} & \textbf{1.405} \\
Yan et al.\cite{yan2023stabilizing} & 0.120 & 2.34 & 1.49 \\
Okajima\cite{okajima1982strouhal} & 0.124 & -- & -- \\
Norberg\cite{norberg1993flow} & 0.124 & 2.02--2.33 & -- \\
Cao \& Tamura\cite{cao2012shear} & 0.123 & 2.07 & -- \\
Bai \& Alam\cite{bai2018dependence} & 0.120 & 2.10--2.33 & -- \\
\bottomrule
\end{tabular}
}
\end{subtable}
\hfill
\begin{subtable}[t]{0.32\textwidth}
\centering
\caption{3D, $Re = 300$}
\label{tab:3D_square_cylinder_Re300}
\adjustbox{width=\textwidth,center}{%
\begin{tabular}{@{}lccc@{}}
\toprule
Reference & St & $\overline{C_D}$ & $C_{L'}$ \\
\midrule
\textbf{Present study} & \textbf{0.146} & \textbf{1.441} & \textbf{0.222} \\
Sohankar et al.\cite{sohankar1999simulation} & 0.153 & 1.47 & 0.2 \\
Yoon et al.\cite{yoon2012three} & 0.145 & 1.431 & 0.205 \\
Jiang \& Cheng\cite{jiang2018hydrodynamic} & 0.146 & 1.434 & 0.183 \\
\bottomrule
\end{tabular}
}
\end{subtable}
\vspace{0.5em}
\caption{Comparison of 2D and 3D square cylinder flow results at different Reynolds numbers. St: Strouhal number; $\overline{C_D}$: Time-averaged drag coefficient; $C_{L'}$: RMS lift coefficient; $std(C_L)$: Standard deviation of lift coefficient; --: Data not reported or not applicable}
\label{tab:square_cylinder_comparison_all}
\end{table}

\paragraph{Environment setup.}
The square cylinder environment follows the same fundamental framework as the circular cylinder configuration described above, with identical reward formulation ($r = -|C_D| - \omega |C_L|$), state space representation through configurable point probes, and episode structure spanning 200 control actions over approximately 10 vortex shedding periods.

The key distinction lies in the actuation strategy, where jet-based control is exclusively employed. Unlike the circular cylinder's zero-net mass-flux configuration, the square cylinder utilizes individual jets that can be positioned at four discrete locations: leading edge upper, leading edge lower, trailing edge upper, and trailing edge lower surfaces. The RL agent controls the mass flux of each active jet independently, with actions normalized to the range $[-1, 1]$ for consistent algorithm performance.

The temporal coupling between the CFD solver and RL agent employs Reynolds number-specific action frequencies: 225, 750, and 375 CFD timesteps per control action for 2D cases at $Re = 200, ~1,000, ~3,900$ respectively, and 700, 900, and 875 timesteps for corresponding 3D cases. This environment configuration enables the investigation of asymmetric control strategies and the comparative effectiveness of leading versus trailing edge actuation for square cylinder wake management, while maintaining the same comprehensive sensing and reward framework established for the circular cylinder testbed.


\subsection{Flow around a NACA 0012 airfoil}
\label{appendix:Environments:NACA0012}

\paragraph{Characteristic physics.}
The flow around a NACA 0012 airfoil in the Reynolds number range of 100 to 200,000 presents a rich variety of fluid dynamics phenomena, transitioning from steady, laminar flow to complex unsteady, transitional, and separated regimes. At the lower end of this spectrum, for Reynolds numbers on the order of $10^2$, the flow is characterized by thick laminar boundary layers that are susceptible to separation, even at high angles of attack~\cite{mateescu2010analysis}. As the angle of attack is increased, the wake is often dominated by unsteady vortex shedding.

With an increase in the Reynolds number into the range of $10^3$, the flow dynamics are frequently governed by the formation of a laminar separation bubble (LSB) on the suction side of the airfoil~\cite{eljack2021simulation, almutairi2017dynamics}. This phenomenon occurs when the laminar boundary layer separates due to an adverse pressure gradient, transitions to a turbulent state within the separated shear layer, and then reattaches to the airfoil surface. The formation, size, and behavior of the LSB are highly sensitive to both the Reynolds number and the angle of attack, which can lead to significant nonlinearities in the aerodynamic characteristics of the airfoil~\cite{eljack2021simulation}. Near-stall conditions can induce self-sustained, low-frequency oscillations in the flow field, which are linked to the quasi-periodic bursting and reforming of the laminar separation bubble~\cite{almutairi2017dynamics}. The various flow regimes can be categorized based on these dynamics, from steady attached flow at very low angles of attack, to the formation of a stable LSB, and ultimately to large-scale separation and stall, which is characterized by the shedding of large vortices~\cite{rodriguez2013direct}.

A primary challenge in the computational modeling of these flow regimes is the accurate prediction of the laminar-to-turbulent transition. This transition is a critical element as it determines whether the separated flow will reattach. Standard RANS models often struggle to accurately capture the complex physics within the separation bubble. Consequently, higher-fidelity methods such as LES or DNS are necessary to resolve the fine-scale turbulent structures and transition mechanisms, albeit at a much greater computational expense. The intricate interplay between the Kelvin-Helmholtz instability of the separated shear layer, the amplification of disturbances, and the eventual turbulent breakdown and reattachment complicates the fundamental understanding of the physics.

Significant differences are observed between 2D and 3D configurations. While 2D simulations can capture many of the fundamental aspects of LSB formation and vortex shedding, they do not account for spanwise instabilities. In a 3D flow, the separated shear layer is subject to three-dimensional instabilities that can lead to a more rapid and complex transition to turbulence~\cite{shan2005direct}. These three-dimensional effects can alter the size and shape of the separation bubble and, as a result, the overall aerodynamic forces~\cite{martinez2016comparison}. This generally leads to a decrease in aerodynamic performance when compared to idealized two-dimensional predictions.

The NACA 0012 airfoil within the considered Reynolds number range is a crucial benchmark case for the development and testing of flow control applications, including those that utilize RL. The simple, symmetric geometry of the airfoil produces fundamentally important and challenging flow phenomena, which are directly relevant to the performance of micro air vehicles (MAVs), drones, and small wind turbine blades. The high sensitivity of the flow to small perturbations makes it an excellent platform for evaluating active flow control strategies aimed at suppressing separation, reducing drag, and enhancing lift. To this end, RL is a particularly promising approach as it can discover complex, non-intuitive, time-dependent control strategies, such as using blowing and suction jets, without requiring a precise analytical model of the flow physics. Nonetheless, the inherent instabilities of the shear layer and the broadband nature of turbulence necessitate control systems with high bandwidth and precise actuation. The effectiveness of a control strategy can be highly dependent on the specific flow state, which poses a challenge for the development of robust controllers that can operate effectively across a range of angles of attack and Reynolds numbers. 

\paragraph{Characteristic physics in gust-airfoil interactions.}

Understanding the aerodynamic consequences of an airfoil encountering an atmospheric gust is a pivotal challenge in aeronautics, with direct implications for aircraft structural integrity, safety, and flight stability. The issue is especially pronounced for smaller aerial platforms, such as drones and MAVs, whose lower operating speeds make them highly susceptible to disturbances. When the flight speed of such a vehicle is on the same order as the gust velocity, a condition of near-resonance can occur, which dramatically magnifies the gust's aerodynamic impact \citep{zarovy2010experimental}. The increasing deployment of autonomous aircraft in unpredictable environments, such as urban settings or mountainous terrain, has intensified the need for robust technologies capable of mitigating these adverse gust-induced effects.

The nature of this aerodynamic interaction can be broadly classified based on the gust ratio, $G$, which compares the gust's velocity to that of the aircraft's freestream. For large commercial aircraft, their high cruise speeds ensure the gust ratio is typically low ($G < 1$). In these scenarios, provided the gust amplitude is not excessive, the aerodynamic response can often be approximated using linear theoretical models \citep{kussner1936zusammenfassender}. However, the validity of these linear models degrades as nonlinearities become more prominent. In sharp contrast, smaller vehicles frequently operate in conditions where the gust velocity matches or exceeds their flight speed ($G \geq 1$). This high-G regime, often termed 'extreme aerodynamics', introduces a set of physical phenomena far more complex than those in milder encounters, rendering linear theories inadequate \citep{fukami2024data}.

Physically, a high-G interaction is dominated by highly unsteady and often violent aerodynamic events. As the gust alters the oncoming flow, the airfoil experiences a rapid and substantial change in its effective angle of attack. This can trigger the formation of a large, coherent leading-edge vortex. Concurrently, the rapid adjustment in the airfoil's overall circulation leads to the shedding of vorticity from the trailing edge, forming a trailing-edge vortex. If the effective angle of attack surpasses a critical threshold, the flow can undergo dynamic stall, a process distinct from its steady-state counterpart. During dynamic stall, large-scale vortical structures detach from the airfoil's surface and convect downstream, causing dramatic and often unpredictable shifts in lift and moment characteristics \citep{granlund2014airfoil}. The resulting aerodynamic loads can momentarily spike to levels $200-300\%$ greater than the maximum steady-state loads. This behavior is also accompanied by significant hysteresis in the force and moment curves, meaning the aerodynamic response depends on the history of the motion, which presents a major complication for control system design.

To summarize, the development of effective flow control strategies for gust load alleviation is the ultimate engineering goal. Conventional control systems, which rely on the deflection of large control surfaces, often have actuation delays and bandwidth limitations that make them too slow to counteract the rapid onset of high-frequency gusts. This has spurred innovation in active flow control technologies. The inherent complexity and unpredictability of gust encounters make them an ideal application area for reinforcement learning as the agent learns an optimal control strategy through direct trial-and-error interaction with its environment, making it well-suited to mastering the nonlinear dynamics of gust-airfoil interactions. However, the transient nature of the gust event demands a control policy that can respond almost instantaneously. To systematically develop and validate these advanced data-driven models and control algorithms, the aerospace community urgently needs well-documented benchmark test cases, such as flow over the NACA 0012 airfoil.

\begin{figure}[b]
    \centering
    \begin{subfigure}{0.49\textwidth}
        \centering
        \includegraphics[width=\textwidth]{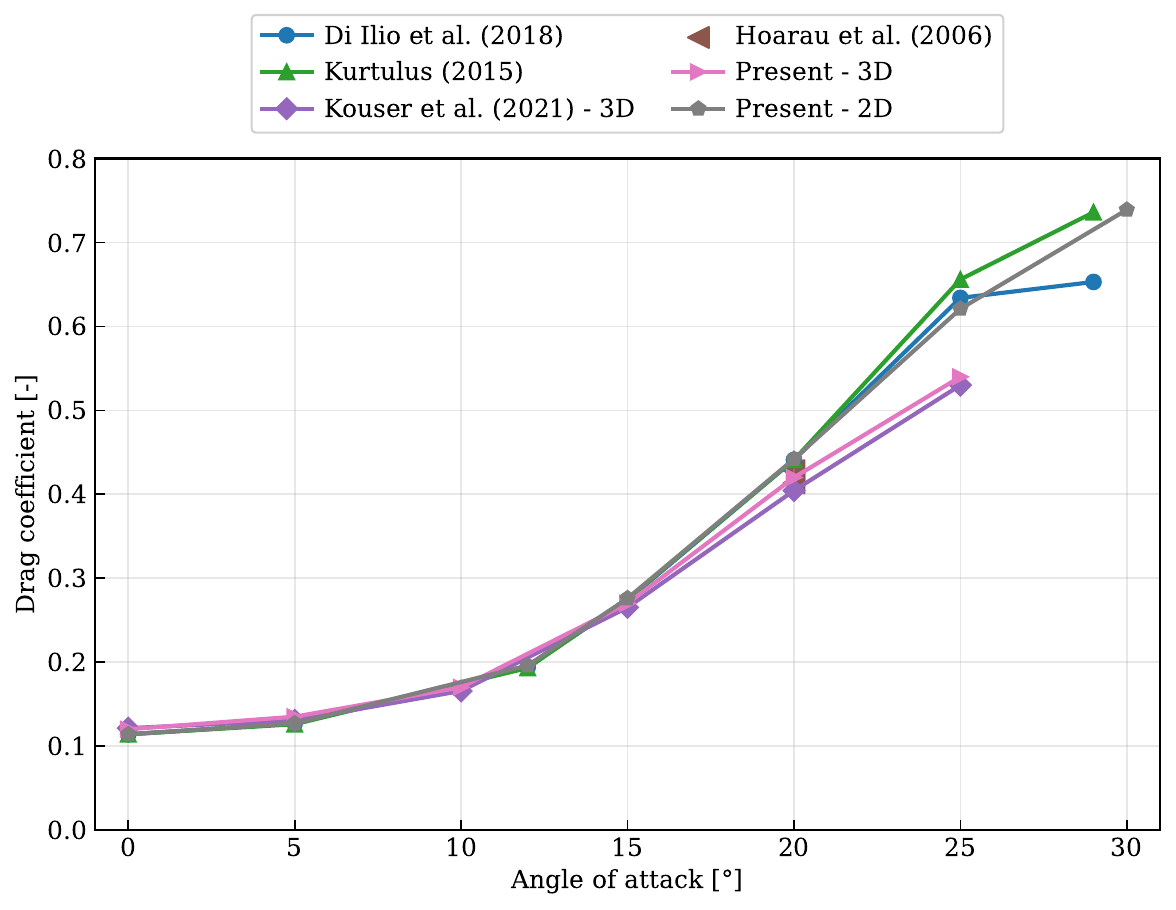}
        \caption{drag coefficient}
        \label{appendix:naca_validation_drag}
    \end{subfigure}
    \hfill
    \begin{subfigure}{0.49\textwidth}
        \centering
        \includegraphics[width=\textwidth]{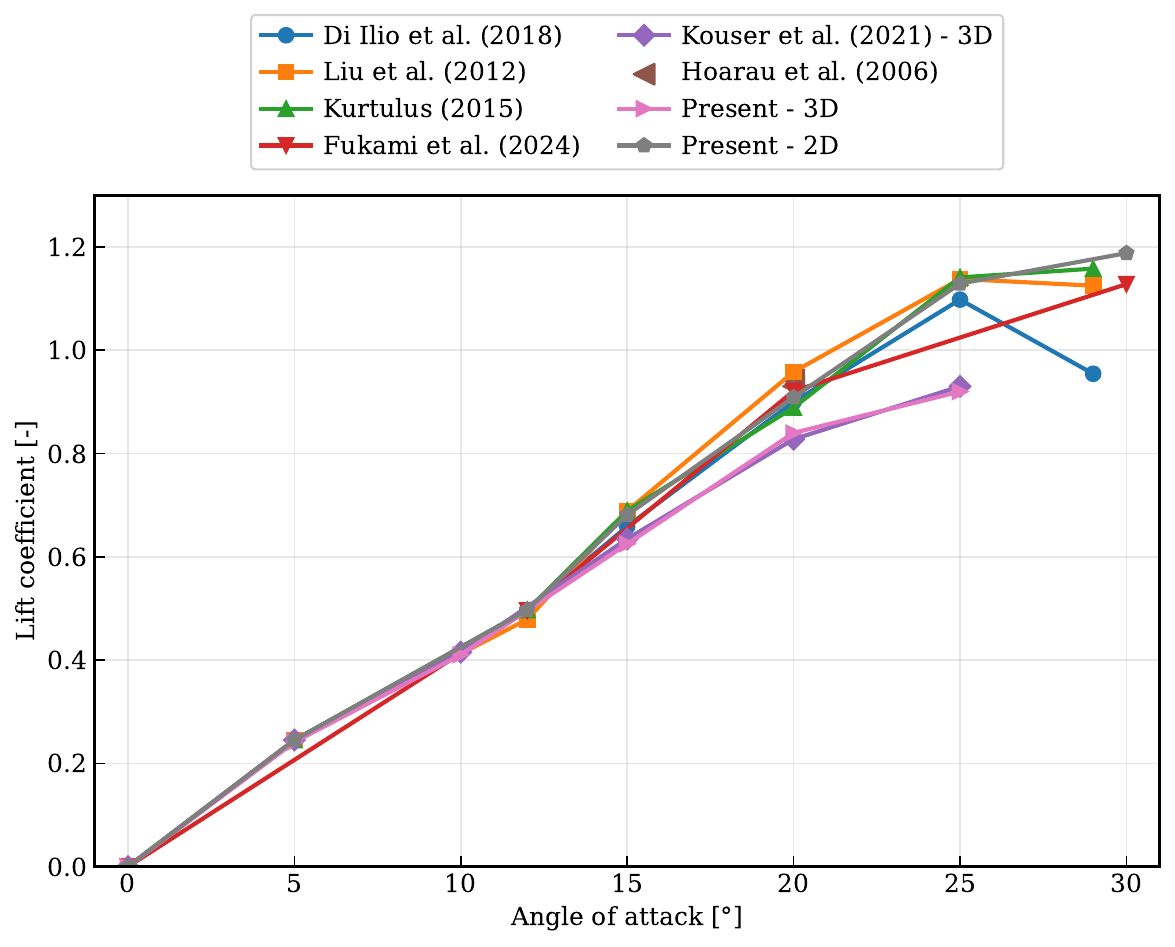}
        \caption{lift coefficient}
        \label{appendix:naca_validation_lift}
    \end{subfigure}
    \caption{Comparison of (a) drag coefficient and (b) lift coefficient as a function of the angle of attack for a NACA 0012 airfoil at $Re = 1,000$. Results from the present study (2D and 3D simulations) are compared with experimental and numerical data from the literature.}
    \label{appendix:naca_validation_drag_lift}
\end{figure}

\paragraph{Numerical setup and validation - m-AIA.}
The NACA 0012 airfoil simulations are conducted in both 2D and 3D configurations to capture the full spectrum of flow physics relevant to gust-airfoil interactions. 
For the 2D cases, the computational domain is defined by $(x/c, y/c) \in [0, 40] \times [-10, 10]$ with the NACA 0012 airfoil centered at $(x/c, y/c) = (10, 0)$. This 2D configuration allows for efficient exploration of parameter spaces and validation against classical airfoil theory, while serving as a baseline for understanding the fundamental flow mechanisms before extending to 3D effects. The reduced computational cost of 2D simulations enables systematic studies across wider ranges of angles of attack and Reynolds numbers.

For the 3D cases, the computational domain spans $[32 \times 16 \times 4]$ chord lengths $c=1.0$ for $Re\leq1,000$ and $[32 \times 16 \times 0.5]$ chord lengths for $Re\in [10,000, ~50,000]$, with the NACA 0012 airfoil positioned 8c from the inflow boundary. The domain extent is designed to minimize boundary reflections while maintaining computational efficiency. The airfoil surface is treated as a no-slip boundary condition implemented through an interpolated bounce-back scheme. Periodic boundary conditions are imposed in the spanwise direction to simulate a nominally two-dimensional configuration while capturing three-dimensional flow instabilities.

The grid resolution is chosen to ensure that the smallest turbulent scales are adequately captured, particularly in the separated flow regions. The computational mesh employs an adaptive refinement approach with 14 refinement levels, with the grid spacing determined by $\Delta_n = \Delta_0/2^n$ where $\Delta_0^{(3D)} = 32c$. For 2D cases, the base grid spacing is $\Delta_0^{(2D)} = 40c$ with refinement levels up to $n = 12$, providing approximately 200 grid points per chord length in the highest resolution regions around the airfoil. Similar to~\cite{gondrum2024porous}, multiple refinement patches are strategically positioned to capture the boundary layer, wake region, and areas of expected flow separation with sufficient resolution. The numerical setup is validated against benchmark data from the literature~\cite{di2018fluid, liu2012numerical, kurtulus2015unsteady, fukami2024data, kouser2021direct, hoarau2006first} showing good agreement in time-averaged lift and drag coefficients across various angles of attack $0^\circ < \alpha < 30^\circ$ for both 2D and 3D configurations (see Fig. \ref{appendix:naca_validation_drag_lift}).

\paragraph{Numerical setup and validation - Nek5000.}
{High-resolution} large-eddy simulation (LES) is employed to resolve the turbulent boundary layers (TBLs) developing over the NACA0012 wing sections at a chord-based Reynolds number of ${Re}_c = 200{,}000$ and an angle of attack of $0^{\circ}$. The present computations adopt a high-resolution LES strategy, with a grid density approaching that typically associated with direct numerical simulation (DNS)~\cite{vinuesa_turbulent_2018}.

The computational domain is constructed using a C-type mesh, with streamwise and vertical extents of $L_x = 6c$ and $L_y = 4c$, respectively. The spanwise width of the NACA0012 configuration is set to $L_z = 0.1c$, resulting in a discretization comprising 220{,}000 spectral elements~\cite{tanarro_effect_2020}. The leading and trailing edges of the airfoil are positioned at distances of $2c$ and $3c$ from the inflow boundary, respectively. This configuration has been validated in previous {investigations~\cite{vinuesa_turbulent_2018}}, demonstrating that the selected spanwise extent is sufficient to capture both the energetically relevant turbulent scales and the largest coherent structures characterizing the wing boundary layer.

The near-wall spatial resolution, expressed in viscous units, satisfies $\Delta x^+_t < 18.0$, $\Delta y^+_n < (0.64, 11.0)$, and $\Delta z^+ < 9.0$ in the wall-tangential, wall-normal, and spanwise directions, respectively. The viscous length scale is defined as $l^* = \nu / u_{\tau}$, where the friction velocity is given by $u_{\tau} = \sqrt{\tau_{w} / \rho}$. The mean wall-shear stress is expressed as $\tau_{w} = \rho \nu \left( \mathrm{d} U_t / \mathrm{d} y_n \right)|_{y_n = 0}$. To achieve the prescribed resolution, a polynomial order of $N = 7$ is adopted, resulting in a total number of grid points of approximately $N{\rm grid} \approx 1.1 \times 10^{8}$. In the wake region, the mesh resolution satisfies $\Delta x / \eta < 9$, where $\eta = (\nu^3 / \epsilon)^{1/4}$ denotes the Kolmogorov length scale and $\epsilon$ represents the local isotropic dissipation rate.

The smallest unresolved turbulent scales are represented through a subgrid-scale (SGS) model based on a time-independent relaxation-term filter~\cite{negi_unsteady_2018}. This filtering operation is implemented implicitly via a volume-force formulation that dissipates unresolved fluctuations while preserving mass conservation. The LES filter is a high-pass filter that restricts the LES dynamics to a subset of spectral modes within each element. The implementation of this approach in Nek5000 has been extensively validated in previous studies~\cite{vinuesa_turbulent_2018}.

Dirichlet boundary conditions are imposed at the inflow, upper, and lower boundaries using a far-field velocity distribution obtained from an auxiliary Reynolds-averaged Navier–Stokes (RANS) computation. The RANS solution is performed with the $k$–$\omega$ shear-stress transport (SST) model~\cite{menter_kmSST_1994} in a circular domain of radius $200c$. At the outflow boundary, the formulation proposed by~\cite{dong_robustoutlet_2014} is employed to prevent the unphysical influx of kinetic energy. Boundary-layer transition is enforced at $x/c = 0.1$ on both the suction and pressure sides through a localized wall-normal body force~\cite{schlatter_tripping_2012}, which generates intense, time-dependent streaks that subsequently break down and promote transition to turbulence~\cite{vinuesa_wingTBL_2017}.

One flow-over time ($T_{\rm FO}$) is defined as the duration required for a fluid particle convecting at the free-stream velocity $U_{\infty}$ to travel one chord length $c$. The procedure adopted to generate the uncontrolled reference simulations is described {in Ref.~\cite{wang_opposition_2025}}. Controlled simulations are initialized from fully developed turbulent fields obtained from the corresponding uncontrolled cases. Statistical convergence of the flow quantities is achieved after approximately 20 $T_{\rm FO}$ for the NACA0012 configuration.

\begin{figure}[t!]
    \centering
    \includegraphics[width=\linewidth]{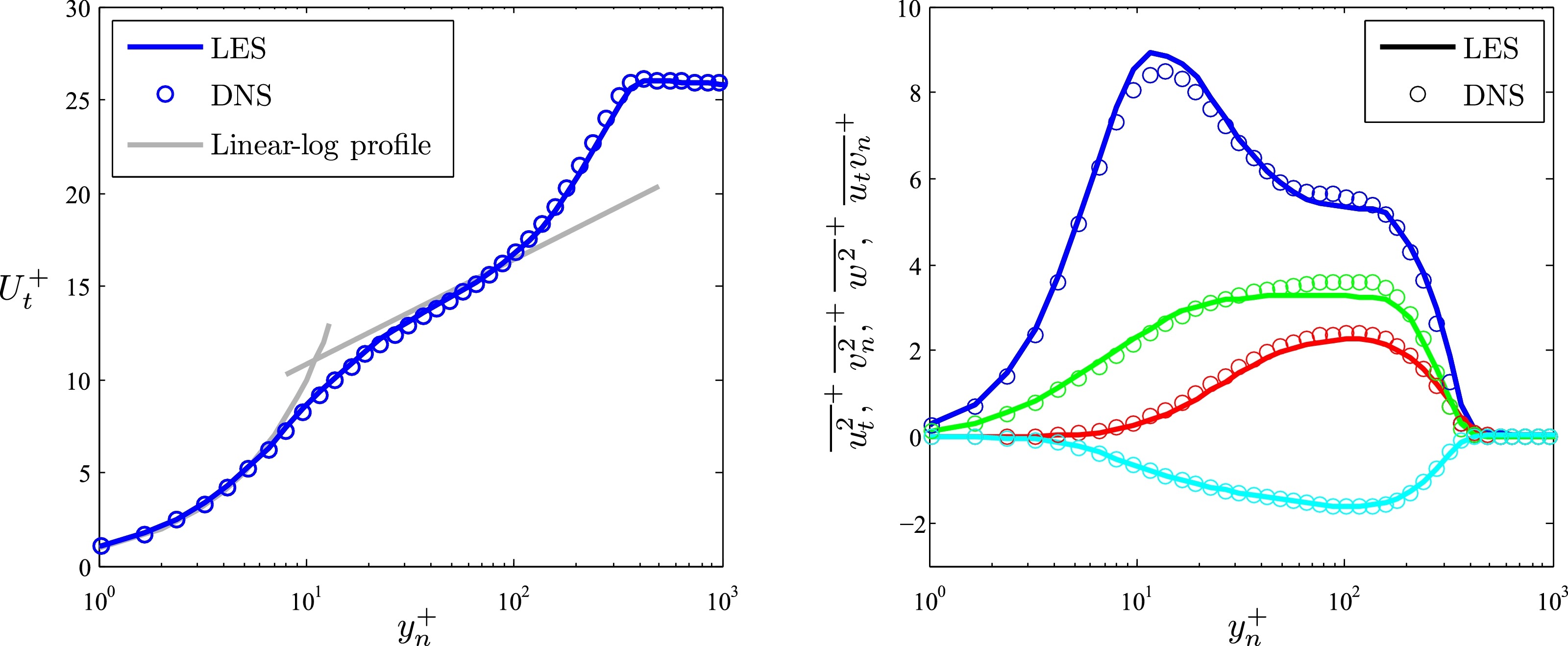}
    \vspace{-.2in}
    \caption{Inner-scaled mean component of wall-tangential velocity and the Reynolds-stress tensors from the 3D NACA4412 simulation at $Re_c=400,000$ with DNS data. {Figure reproduced from Ref~\cite{vinuesa_turbulent_2018} with permission of the publisher (Elsevier)}}
    \label{fig:nek_naca_validation}
\end{figure}

The adequacy of the present mesh for the selected SGS modelling approach has been demonstrated through excellent agreement between DNS and LES data for a NACA4412 airfoil at ${Re}_c = 400{,}000$ and $AoA=5^\circ$ (see figure~\ref{fig:nek_naca_validation}). Specifically, the inner-scaled mean velocity profile and selected components of the Reynolds-stress tensor at $x_{\rm ss}/c = 0.7$ (where \emph{ss} denotes the suction side), expressed in the local tangential and wall-normal reference frame, are compared between both simulations. The comparison reveals excellent agreement in the mean velocity distribution, as well as in the Reynolds shear stress and in the outer-layer behaviour of the tangential velocity fluctuations. These results provide {a} validation of the numerical fidelity of the present configuration.

\paragraph{Environment implementation - Drag reduction setup.}
The control objective is to reduce skin-friction drag using the zero-shot deployment strategy. 
Although the implementation closely follows that adopted for turbulent channel flows described in Sec.~\ref{appendix:Environments:TCF}, several critical modifications are required. First, owing to the spatial development of TBLs, the friction velocity becomes a function of the chordwise coordinate, i.e. $u_\tau = u_\tau (x_{ss}/c)$. This spatial variation is represented using a ninth-order polynomial fit of the $u_\tau$ distribution available in the database reported in~\cite{tanarro_effect_2020}, yielding an exact reconstruction of the reference profile.

As a consequence, (1) the bounds of the actuation amplitude, (2) the wall-normal position of the sensing plane at $y^+ = 15$, and (3) the control update interval ($\Delta t^+ = 0.6$) all become functions of $x_{ss}/c$. The formulation of (1) and (2) follows directly once $u_\tau (x_{ss}/c)$ is known. Ensuring an exact update interval at each streamwise location, however, is considerably more challenging. To address this issue, the update interval is enforced to be uniform within each control block $\Omega_i$, based on a viscous time scale computed using the block-averaged friction velocity, i.e. $t^*_{\Omega_i} = \nu / \langle u\tau \rangle^2_{\Omega_i}$.

Furthermore, a coordinate transformation between the wall-normal and Cartesian reference frames is required. In the wing configuration, actuation is defined in terms {of the} wall-normal velocity, whereas the numerical solver operates in Cartesian coordinates. Consequently, the actuation is imposed through modifications of the horizontal and vertical velocity components such that their projection along the wall-normal direction yields the desired control intensity. Conversely, the observable states are defined in the wall-normal frame. Both horizontal and vertical velocity components therefore contribute to the wall-tangential and wall-normal velocities, with their relative contribution depending on the local curvature of the surface. Owing to the streamwise inhomogeneity of TBLs, the spatial means of both actuation and state variables are obtained by averaging instantaneous quantities exclusively in the spanwise direction.

\paragraph{Environment implementation - Transverse gust setup.}

\begin{figure}[t]
    \centering
    \begin{subfigure}{0.49\textwidth}
        \centering
        \includegraphics[width=\textwidth]{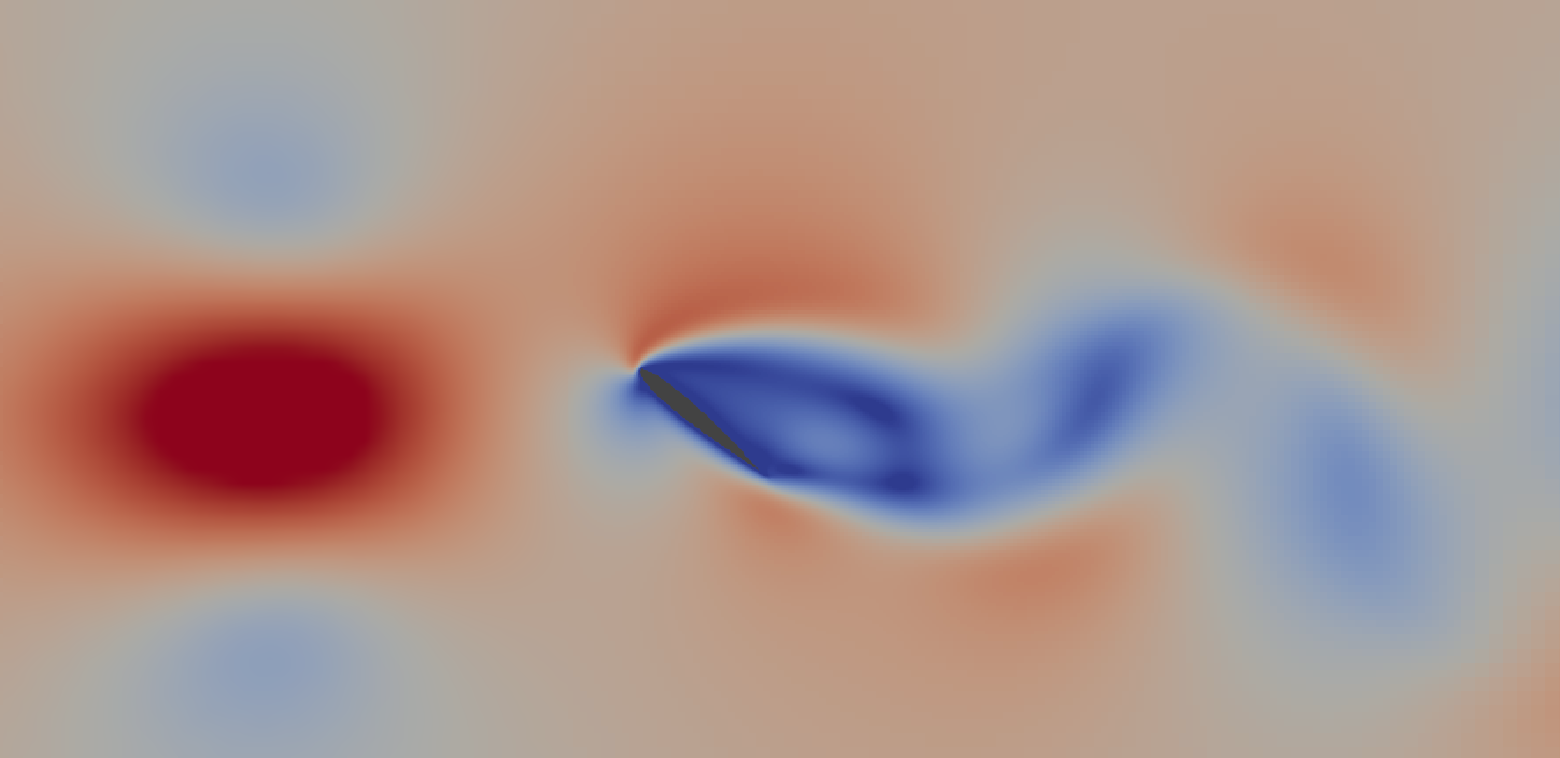}
        \caption{2D environment: $Re=100$ \& $\alpha=40^\circ$}
        \label{appendix:gust_image_2D}
    \end{subfigure}
    \hfill
    \begin{subfigure}{0.49\textwidth}
        \centering
        \includegraphics[width=\textwidth]{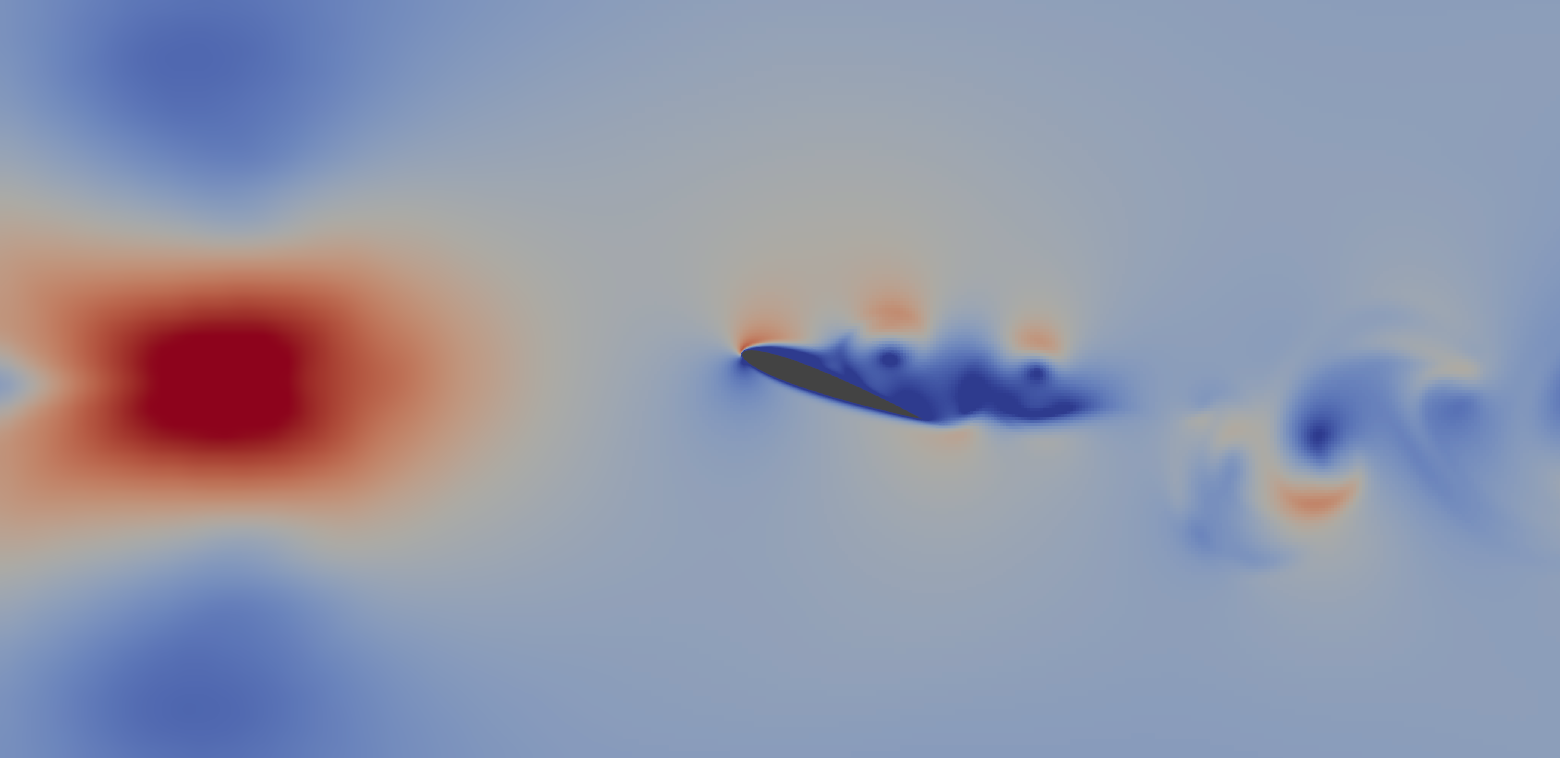}
        \caption{3D environment: $Re=1,000$ \& $\alpha=20^\circ$}
        \label{appendix:gust_image_3D}
    \end{subfigure}
    \caption{Exemplary visualization of the transversal gust scenario for 2D (a) and 3D (b) environments}
    \label{appendix:gust_images}
\end{figure}

\begin{figure}[t]
    \centering
    \includegraphics[width=.95\linewidth]{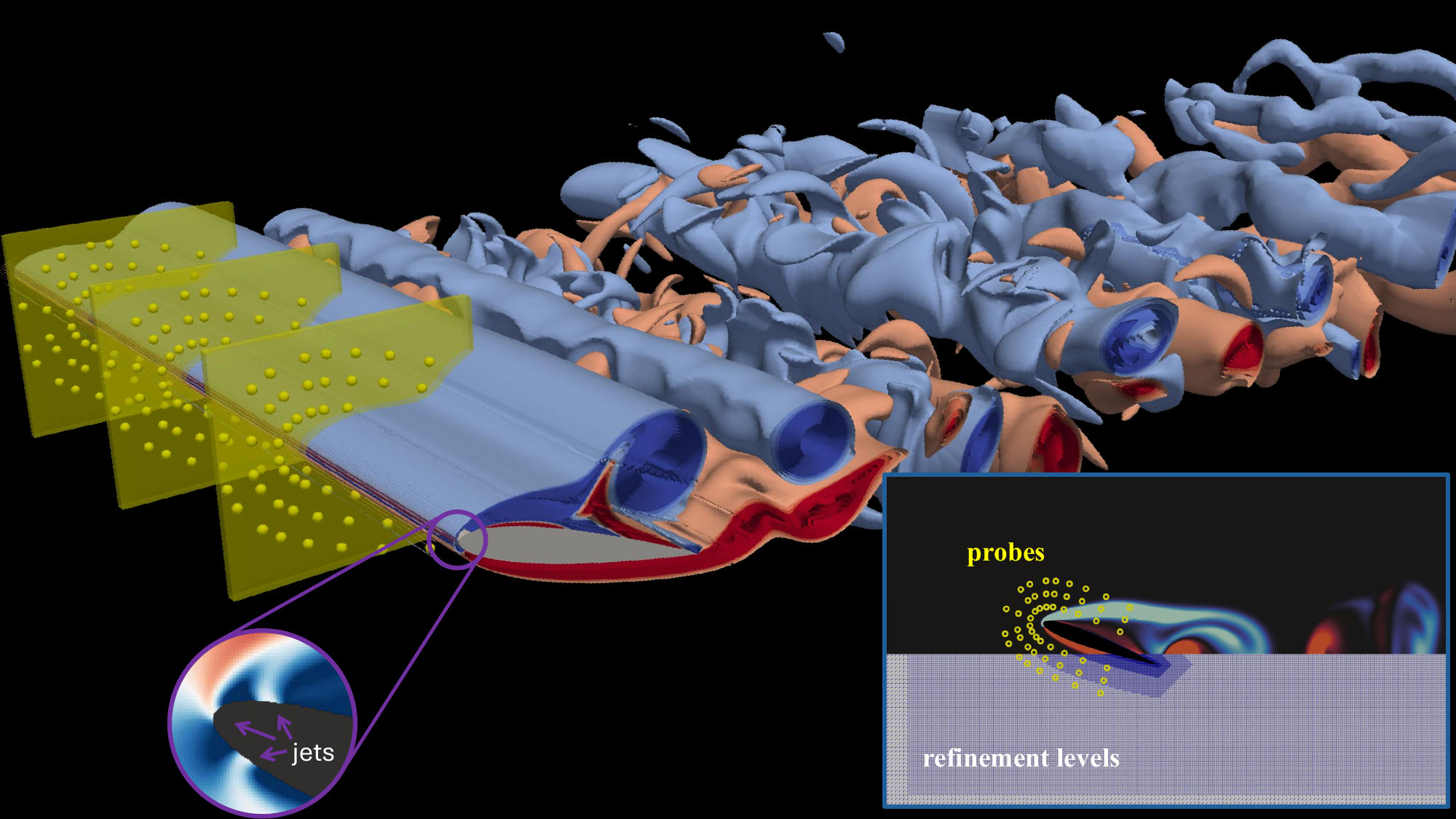}
    \caption{Computational domain setup, sensor probe distribution and actuation strategy for the gust-airfoil interaction environments: Three-dimensional view showing the probe distribution around the airfoil with vortical structures visualized through Q-criterion isosurfaces. The yellow dots in lower right zoom-in highlight sensor probe locations in a single spanwise plane. Active flow control is performed using three independent jet actuators positioned along the leading edge of the NACA 0012 airfoil, each covering $3\%$ of the chord length and extending across the full span (see lower left zoom-in).}
    \label{fig:actuation}
\end{figure}

The transverse gust is implemented through a prescribed velocity perturbation imposed at the inflow boundary. The gust profile follows a 1-cosine approach with a gust factor $G = 2.0$, representing the ratio of peak gust velocity to freestream velocity (see Fig.  \ref{appendix:gust_images}). The gust temporal development is controlled to achieve the desired interaction dynamics with the airfoil. The state space representation consists of user-configurable point probes distributed throughout the computational domain, providing measurements of velocity components, pressure, density, and body forces at specified locations. To facilitate interactions between the RL agent and the flow environment, three jet actuators are distributed across the leading edge of the airfoil (see Fig. \ref{fig:actuation}). Each actuator can be controlled independently, covers $3\%$ of the chord length, and extends over the entire spanwise location. The action spaces are continuous and normalized in the range $[-1.0, 1.0]$.

Depending on the investigated Reynolds number, the number of CFD timesteps varies in the range of $225 - 1,000$, delivering the necessary temporal granularity to represent critical flow physics while preserving computational tractability. The reward function is formulated to minimize gust-induced force fluctuations while maintaining aerodynamic efficiency:
\begin{equation}
R(t) = -|C_L(t)-\overline{C_{L,ref}}| - \omega ~|C_D(t)-\overline{C_{D,ref}}|
\end{equation}
where $C_L(t)$ and $C_D(t)$ are the instantaneous drag and lift coefficients while $\overline{C_{L,ref}}$ and $\overline{C_{D,ref}}$ represent time-averaged coefficients of the unperturbed reference case. The weighting coefficient $\omega=0.25$ is tuned to balance the competing objectives of gust mitigation and drag reduction.


\subsection{Flow around a cube}
\label{appendix:Environments:Cube}

\paragraph{Characteristic physics.}
The flow around a suspended cube represents a canonical problem in fluid dynamics, offering a fundamental case study for three-dimensional bluff body wakes with fixed separation points. Unlike a sphere, where the separation location is Reynolds number dependent, the sharp leading edges of a cube dictate the points of flow separation, leading to a unique and complex evolution of the wake structure as the Reynolds number increases. For Reynolds numbers in the range of 300 to 3,700, the flow transitions through a sequence of distinct regimes, each characterized by specific vortical dynamics and stability properties.

Within the lower end of this Reynolds number range, several key bifurcations govern the wake's topology. The flow is initially steady and symmetric up to a critical Reynolds number of approximately 215 \citep{saha2004three}. As the Reynolds number increases beyond this point, the flow undergoes a first bifurcation, losing its orthogonal symmetry to become steady but asymmetric, a state that persists up to a Reynolds number of about 265. This is followed by a second Hopf bifurcation around a Reynolds number of 270, where the flow becomes unsteady and periodic, characterized by the shedding of hairpin-like vortices \citep{saha2004three, klotz2014experimental}. These hairpin vortices are initially shed in a regular, periodic manner. Experimental investigations using techniques such as Particle Image Velocimetry (PIV) and laser-induced fluorescence have confirmed this sequence of bifurcations, aligning well with numerical simulations \citep{klotz2014experimental}. As the Reynolds number further increases towards 400, the hairpin vortex shedding becomes more complex, eventually transitioning to a chaotic vortex shedding regime around a Reynolds number of 310, where the coherence of the shed structures begins to break down \citep{meng2021wake}.

Modeling the flow around a cube presents significant challenges, particularly as the Reynolds number increases into the turbulent regime. Accurately capturing the precise Reynolds numbers for the initial bifurcations requires high-fidelity numerical methods with sufficient spatial resolution to resolve the thin shear layers separating from the leading edges \citep{saha2004three}. The interaction between these shear layers from the top, bottom, and side faces of the cube creates a highly three-dimensional and inherently unstable wake. As the Reynolds number climbs towards 3,700, the wake becomes fully turbulent, characterized by a broad range of spatial and temporal scales. The breakdown of larger coherent structures into smaller-scale turbulence and the complex interactions between different vortical structures are computationally expensive to resolve. Furthermore, ensuring the numerical schemes accurately capture the sharp-corner-induced separation without introducing excessive numerical diffusion is a persistent challenge. Understanding the physics is complicated by the fully three-dimensional nature of the vortex dislocations and the transition to turbulence in the separated shear layers.

Therefore, the flow around a cube serves as an important test case for the development and benchmarking of flow control strategies, including those based on reinforcement learning. The well-defined sequence of flow regimes offers a varied landscape to test the robustness and adaptability of control algorithms. The objective of such control could be drag reduction, lift suppression, or wake stabilization. The sharp edges provide fixed locations for the placement of actuators, such as synthetic jets or plasma actuators, simplifying one aspect of the control problem. However, the inherent three-dimensionality of the wake presents a significant challenge for flow control. Actuation strategies must be able to influence a complex, three-dimensional flow field, which necessitates a high-dimensional action space for an RL agent. For instance, controlling the interaction and instability of the multiple shear layers simultaneously is a non-trivial task. A key challenge for RL is the development of a low-dimensional state representation from achievable sensor measurements (e.g., surface pressure sensors) that can adequately describe the high-dimensional state of the flow. The agent must learn a control policy that can effectively manipulate the different instabilities present in the various flow regimes, from the steady asymmetry to the chaotic vortex shedding. The time delay between an actuation and its effect on the global flow properties, such as drag, also poses a significant hurdle for the learning process, requiring the RL algorithm to handle delayed rewards and long-term consequences of its actions. The complexity of the cube wake with its multiple interacting shear layers and instabilities therefore provides a challenging and realistic benchmark for advancing the capabilities of reinforcement learning in active flow control.

\paragraph{Numerical setup and validation.}
The cube benchmark follows the same computational setup as the square cylinder case, with the key difference being the three-dimensional bluff body geometry. The cube has a side length $D$ and the Reynolds numbers are $Re \in [300, ~1,000, ~3,700]$. The 3D domain is $(x/D,y/D,z/D) \in [0,32] \times [0,16] \times [0,16]$ with the cube centered at $(x/D,y/D,z/D)=(8,8,8)$ and non-periodic boundary conditions in all directions. All other boundary conditions, grid refinement levels, and resolution criteria follow the square cylinder specifications. Validation results shown in Tab. \ref{tab:3D_cube_comparison} demonstrate comparable accuracy to the square cylinder cases, with good agreement to established literature values where available.

\begin{table}[t!]
\centering
\begin{subtable}[t]{0.49\textwidth}
\centering
\caption{$Re = 300$}
\label{tab:3D_cube_Re300}
\adjustbox{width=0.75\textwidth,center}{%

\begin{tabular}{@{}lcc@{}}
\toprule
Reference & St & $\overline{C_D}$ \\
\midrule
\textbf{Present study} & \textbf{0.102} & \textbf{0.855} \\
Saha et al.\cite{saha2004three} & 0.095 & 0.804 \\
Saha\cite{saha2006three} & 0.097 & 0.833 \\
Haider \& Levenspiel \cite{haider1989drag} & -- & 0.985 \\
Holzer \& Sommerfeld\cite{holzer2008new} & -- & 0.923 \\
Khan et al. \cite{khan2020simulation} & -- & 0.843 \\
\bottomrule
\end{tabular}
}
\end{subtable}
\hfill
\begin{subtable}[t]{0.49\textwidth}
\centering
\caption{$Re = 1,000$}
\label{tab:3D_cube_Re1000}
\adjustbox{width=0.64\textwidth,center}{%
\begin{tabular}{@{}lc@{}}
\toprule
Reference & $\overline{C_D}$ \\
\midrule
\textbf{Present study} & \textbf{0.984} \\
Holzer \& Sommerfeld\cite{holzer2008new} & 0.815 \\
Khan et al.\cite{khan2018flow} & 0.883 \\
Khan et al. \cite{khan2020simulation} &  0.968 \\
Haider \& Levenspiel \cite{haider1989drag} & 1.121 \\
\bottomrule
\end{tabular}
}
\end{subtable}
\vspace{0.5em}
\caption{Comparison of 3D cube flow results at different Reynolds numbers. St: Strouhal number; $\overline{C_D}$: Time-averaged drag coefficient; --: Data not reported or not applicable}
\label{tab:3D_cube_comparison}
\end{table}

\begin{figure}[t!]
    \centering
    \includegraphics[width=.95\linewidth]{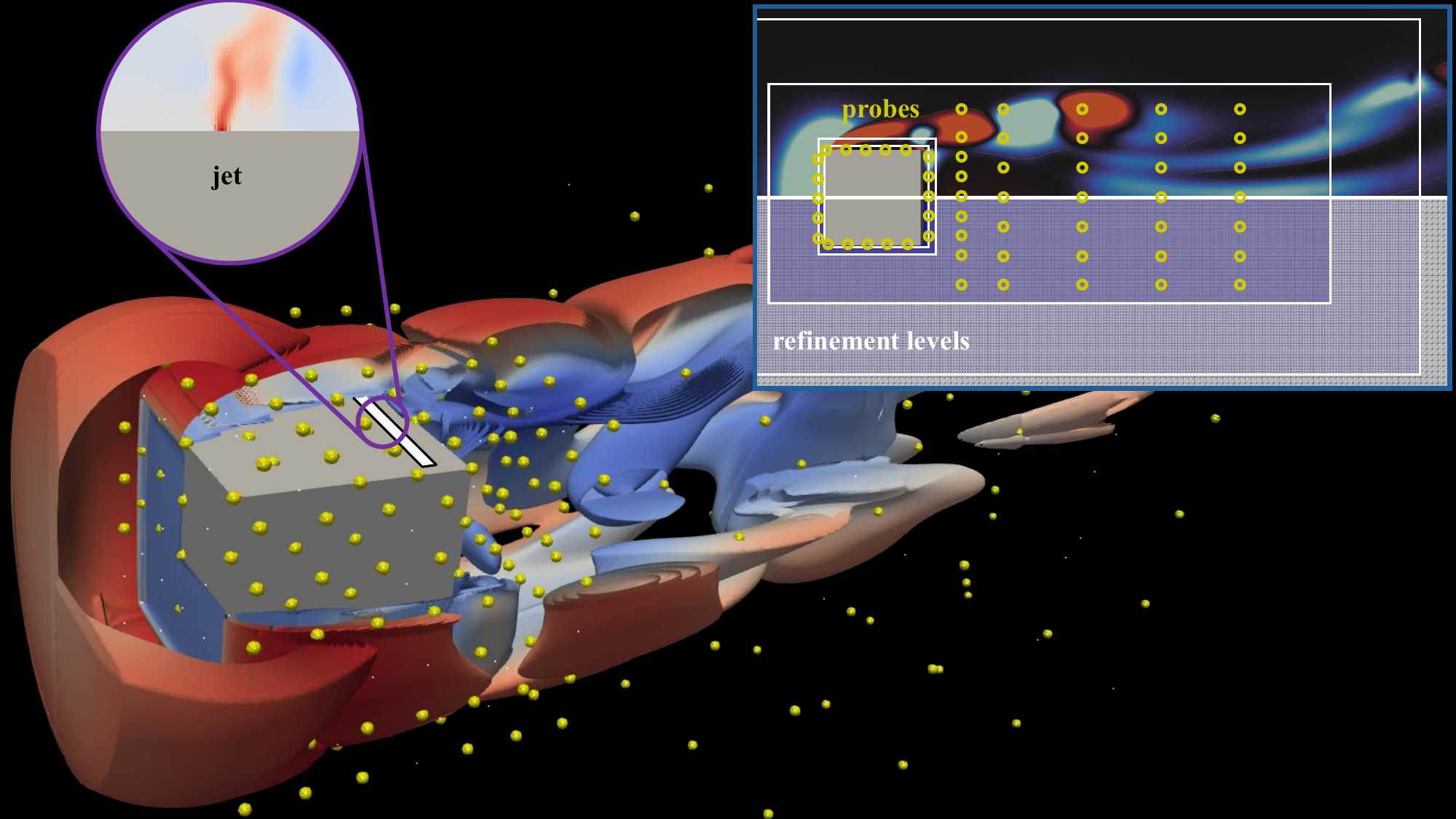}
    \caption{Computational domain setup, sensor probe distribution and actuation strategy for the flow around a cube: Three-dimensional view showing the probe distribution around the cube with vortical structures visualized through Q-criterion isosurfaces. The yellow dots highlight sensor probe locations in a single spanwise plane. Active flow control is performed using independent jet actuators positioned along the leading and trailing edge of the cube.}
    \label{fig:cube_flow_overview}
\end{figure}

\paragraph{Environment implementation.}
The cube environment follows the same fundamental framework as the square cylinder configuration, with identical reward formulation, state space representation, and episode structure. The key distinction is the three-dimensional jet actuation strategy, where jets can be positioned across multiple cube faces: leading/trailing edges with upper, lower, and side face locations (see Fig. \ref{fig:cube_flow_overview}). The RL agent controls each jet's mass flux independently, with actions normalized to $[-1, 1]$.
Temporal coupling follows the square cylinder approach with Reynolds number-specific action frequencies in the range of 1,000 to 1,250 CFD timesteps per control action. This configuration enables the investigation of multi-face actuation strategies for three-dimensional wake control.

\subsection{Flow around a sphere}
\label{appendix:Environments:Sphere}

\paragraph{Characteristic physics.}
Flow past spheres within the Reynolds number regime 300-3,700 demonstrates intricate transitional phenomena marked by sequential symmetry-breaking bifurcations and progressive wake complexity that establishes this configuration as fundamental for understanding three-dimensional bluff body aerodynamics and developing advanced control methodologies. This intermediate Reynolds number range captures the essential transition from organized periodic vortex structures to chaotic subcritical turbulent behavior, presenting distinctive opportunities and obstacles for both theoretical analysis and practical intervention strategies.

The underlying flow mechanics throughout this range are dictated by a systematic bifurcation sequence that progressively destroys flow symmetries and introduces increasingly complex temporal dynamics. The initial critical transition manifests around $Re = 212$ through a supercritical pitchfork bifurcation that eliminates axisymmetry while preserving planar symmetry, establishing steady non-axisymmetric flow patterns~\cite{johnson1999flow}. Subsequently, at approximately $Re = 280$, a supercritical Hopf bifurcation introduces periodic temporal behavior characterized by regular hairpin vortex shedding with Strouhal numbers near 0.18-0.20~\cite{tomboulides2000numerical}. The final primary transition occurs through a Neimark-Sacker bifurcation around $Re = 330$, generating quasiperiodic dynamics with two incommensurate frequencies that create complex modulation patterns while maintaining overall wake organization~\cite{citro2017stability}.

These successive instability mechanisms establish fundamental distinctions between lower Reynolds number steady flows and the complex transitional behavior observed in the $Re=300-3,700$ range. While computational approaches utilizing steady-state assumptions remain viable below $Re = 212$, the emergence of temporal instabilities necessitates fully unsteady three-dimensional simulation methodologies that capture both large-scale wake evolution and fine-scale vortical structures~\cite{rodriguez2011direct, tomboulides2000numerical}. The hairpin vortex morphology characteristic of this regime creates a distinctive flow topology involving interconnected vortex loops and induced secondary structures that significantly influence drag, heat transfer, and mixing properties. Beyond $Re = 800$, the wake transitions toward subcritical turbulent characteristics while retaining coherent large-scale organization, establishing a complex multiscale environment where deterministic and stochastic elements coexist~\cite{poon2014flow}.

Consequently, this rich transitional behavior positions spherical wake flows as exceptionally valuable canonical benchmark for advancing flow control science and validating emerging computational strategies. The geometric simplicity eliminates confounding effects associated with complex shapes while preserving all essential three-dimensional bluff body physics, creating optimal conditions for isolating and understanding control mechanisms. The Reynolds number window of $Re \in [300-3,700]$ provides particularly favorable conditions for these developments because the flow complexity challenges algorithm sophistication while maintaining computational feasibility for extensive training iterations.

In summary, the well-established theoretical foundation surrounding spherical wake physics renders this configuration invaluable as advancing computational methodologies and experimental capabilities continue expanding flow control horizons for purely three-dimensional bluff body aerodynamics. The thorough documentation of fundamental mechanisms enables confident evaluation of whether innovative approaches genuinely capture essential physics or merely exploit numerical artifacts. Consequently, the extensively characterized sphere wake provides an indispensable validation platform for novel developments prior to extension toward more complex geometries and flow physics.

\paragraph{Numerical setup and validation.}
The sphere setup follows the cylinder benchmark scenario with 3D low Mach number flows ($Ma<0.2$) at Reynolds numbers $Re \in [300,~3,700]$ based on the sphere diameter $D$.
The computational domain differs from the cylinder configuration with $(x/D,y/D,z/D) \in [0,32] \times [0,16] \times [0,16]$, featuring an extended z-direction domain. The sphere is centered at $(x/D,y/D,z/D) = (8,8,8)$. Grid spacing and refinement follow the cylinder setup with $\Delta_0^{(3Dsphere)}=32D$ and refinement levels up to $n=14$, leading up to 512  cells per sphere diameter $D$. Inlet and outlet boundary conditions are identical to the cylinder configuration. However, outlet boundary conditions replace the periodic conditions on all lateral boundaries due to the extended z-domain.
Validation results demonstrate similarly good accuracy compared to the cylinder setup, with Strouhal numbers and drag coefficients matching within established literature ranges for sphere flows (see Tab. \ref{tab:3D_sphere_comparison}), where benchmark results are available.

\begin{table}[t!]
\centering
\begin{subtable}[t]{0.49\textwidth}
\centering
\caption{$Re = 300$}
\label{tab:3D_sphere_Re300}
\adjustbox{width=0.6\textwidth,center}{%
\begin{tabular}{@{}lcc@{}}
\toprule
Reference & St & $\overline{C_D}$ \\
\midrule
\textbf{Present study} & \textbf{0.142} & \textbf{0.636} \\
Johnson \& Patel\cite{johnson1999flow} & 0.137 & 0.656 \\
Roos \& Willmarth\cite{roos1971some} & -- & 0.629 \\
Almedeij\cite{almedeij2008drag} & -- & 0.602 \\
Clift et al.\cite{clift2005bubbles} & -- & 0.674 \\
\bottomrule
\end{tabular}
}
\end{subtable}
\hfill
\begin{subtable}[t]{0.49\textwidth}
\centering
\caption{$Re = 3,700$}
\label{tab:3D_sphere_Re3700}
\adjustbox{width=0.68\textwidth,center}{%
\begin{tabular}{@{}lcc@{}}
\toprule
Reference & St & $\overline{C_D}$ \\
\midrule
\textbf{Present study} & \textbf{0.212} & \textbf{0.391} \\
Almedeij\cite{almedeij2008drag} & -- & 0.409 \\
Clift et al.\cite{clift2005bubbles} & -- & 0.378 \\
Haider \& Levenspiel\cite{haider1989drag} & -- & 0.392 \\
Turton \& Levenspiel\cite{turton1986short} & -- & 0.386 \\
Rodriguez et al.\cite{rodriguez2011direct} & 0.215 & 0.394 \\
Yun et al.\cite{yun2006vortical} & 0.21 & 0.355 \\
\bottomrule
\end{tabular}
}
\end{subtable}
\vspace{0.5em}
\caption{Comparison of 3D sphere flow results at different Reynolds numbers. St: Strouhal number; $\overline{C_D}$: Time-averaged drag coefficient; --: Data not reported or not applicable}
\label{tab:3D_sphere_comparison}
\end{table}

\begin{figure}[t!]
    \centering
    \includegraphics[width=.95\linewidth]{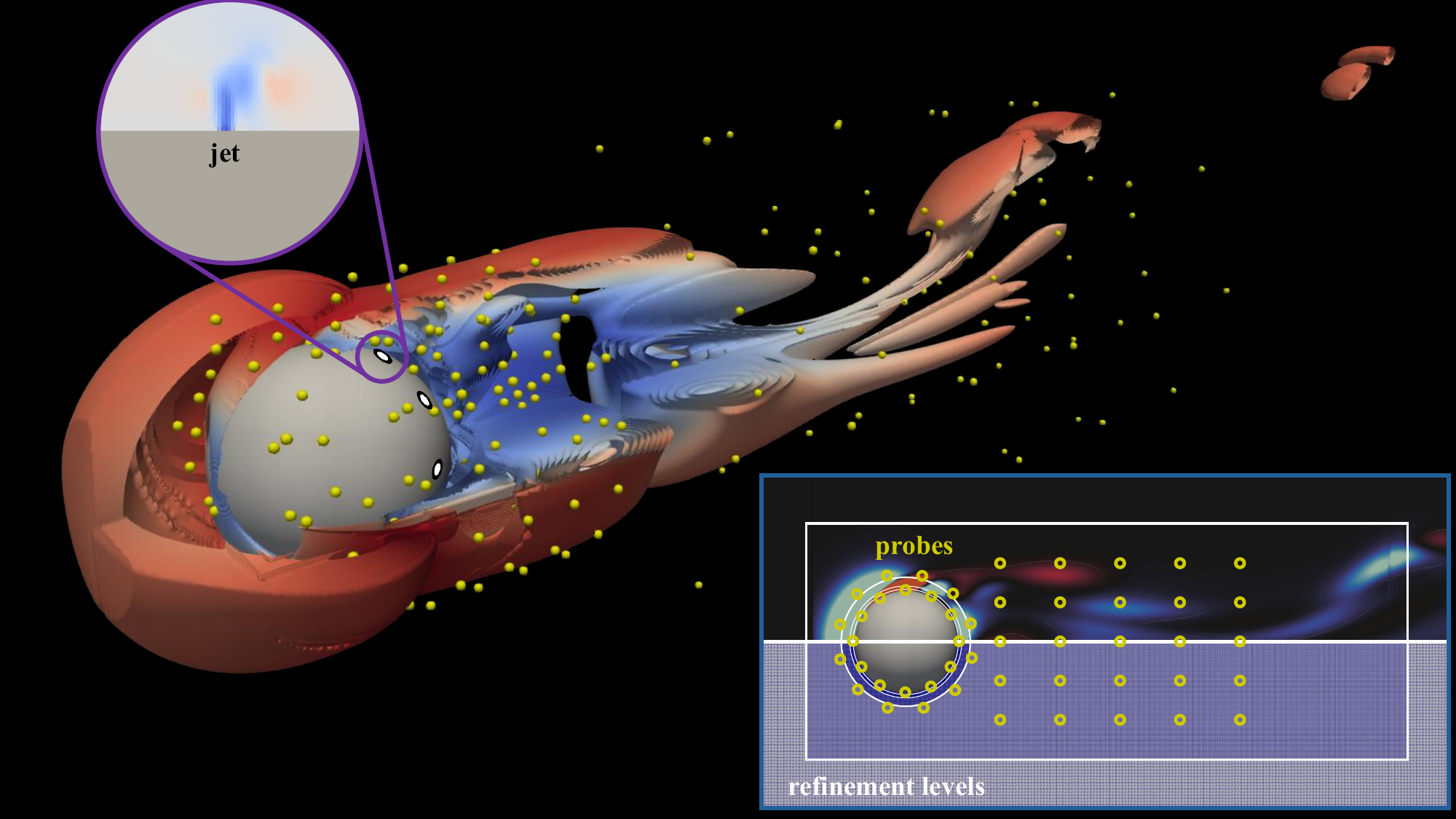}
    \caption{Computational domain setup, sensor probe distribution and actuation strategy for the flow around a sphere: Three-dimensional view showing the probe distribution around the sphere with vortical structures visualized through Q-criterion isosurfaces. The zoom-in highlights sensor probe locations in a single spanwise plane. Independent jet actuators positioned in the sphere's rear half enable active flow control.}
    \label{fig:sphere_flow_overview}
\end{figure}

\paragraph{Environment implementation.}
The sphere flow environment uses the same default reward objective as the cylinder setup, targeting drag minimization and lift force oscillation reduction. The actuation strategy differs from the cylinder configuration through eight cylindrical jets positioned on a circle at the sphere's backside, located at $60^\circ$ around the streamwise x-axis. Unlike the cylinder's slot-like jets, these jets have a circular shape in the z-direction (see Fig. \ref{fig:sphere_flow_overview}) and allow independent mass flux control by the RL agent.

The environment supports the Reynolds numbers $Re = 300, ~3,700$, with an identical state space representation through user-configurable point probes as the cylinder counterpart. Similarly, training episodes and temporal coupling parameters follow the cylinder configuration, featuring 300 and 750 CFD timesteps per control action, respectively.

\subsection{Zero-pressure gradient turbulent boundary layer}
\label{appendix:Environments:ZPGTBL}

\paragraph{Characteristic physics.}
Turbulent boundary layer flows (TBLs) over flat plates represent a canonical flow problem for fundamental turbulence studies and the investigation of viscous drag reduction mechanisms. The flow is dominated by chaotic near-wall coherent structures, such as quasi-streamwise vortices and velocity streaks, which are primarily responsible for high skin-friction drag. Specifically, the buffer layer ($y^+ \approx 10$--$50$) hosts quasi-streamwise vortices with a characteristic diameter of approximately 30 wall units and alternating low- and high-speed velocity streaks with a robust mean spanwise spacing of $\lambda_z^+ \approx 100$~\cite{kline1967structure,jeong1997coherent}. These structures participate in an autonomous self-sustaining cycle comprising streak formation via the lift-up effect, inflectional instability and breakdown of the streaks, and nonlinear regeneration of streamwise vortices~\cite{hamilton1995regeneration,jimenez1999autonomous}. Disrupting any phase of this cycle, for instance through external forcing, can suppress turbulence production and reduce skin friction. While this near-wall cycle is the primary driver of skin-friction drag at moderate Reynolds numbers, its dynamics do not evolve in isolation at higher $Re_\tau$. The inner-scaled peak of the streamwise velocity variance at $y^+ \approx 15$ grows logarithmically with $Re_\tau$, driven by the superposition and amplitude modulation of energy from outer-layer large-scale motions onto the near-wall region~\citep{marusic2010predictive,smits2011high}. At $Re_\tau \gtrsim 1{,}000$, very-large-scale motions (superstructures) with streamwise coherence exceeding $6\delta$ emerge in the logarithmic layer and exert a measurable top-down modulation on the near-wall small-scale activity~\citep{hutchins2007evidence}. Critically, the wall-normal footprint of these outer-layer structures onto the near-wall region intensifies significantly with increasing $Re_\tau$: the superimposed low-frequency signature of the superstructures progressively dominates the near-wall velocity and shear-stress fields, meaning that the inner-layer cycle alone becomes an incomplete picture of the drag-generating dynamics. The Fukagata--Iwamoto--Kasagi identity~\citep{fukagata2002contribution} directly links the skin-friction coefficient to a weighted wall-normal integral of the Reynolds shear stress, providing a quantitative framework for assessing how modifications to the turbulent stress field translate into drag changes. Because the scale separation between inner and outer motions grows linearly with $Re_\tau$, and because the outer-layer footprint on the near-wall region intensifies accordingly, control strategies that are effective at low Reynolds numbers may lose efficacy at higher $Re_\tau$ if they target only the near-wall cycle---motivating actuation methods capable of simultaneously addressing both the inner-layer regeneration mechanism and the outer-layer footprint. Spanning this Reynolds number range with a single benchmark platform therefore enables the systematic investigation of how learned control policies adapt---or fail to adapt---to the evolving multi-scale structure of the turbulent boundary layer.

\begin{figure}[b]
    \centering
    \begin{subfigure}[b]{0.48\textwidth}
        \centering
        \includegraphics[width=\textwidth]{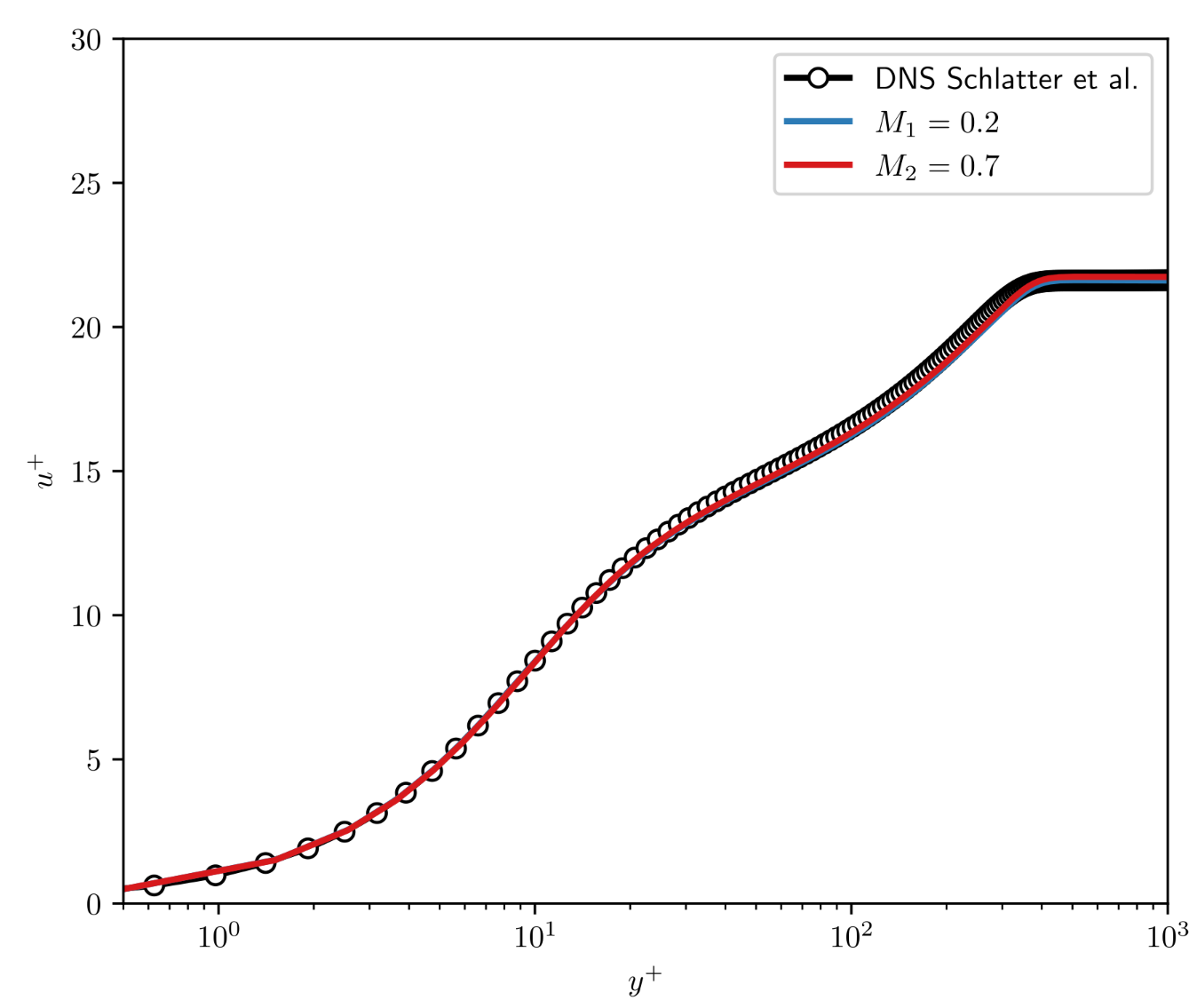} 
        \caption{Streamwise velocity profile}
        \label{fig:velocity}
    \end{subfigure}
    \hfill 
    \begin{subfigure}[b]{0.48\textwidth}
        \centering
        \includegraphics[width=\textwidth]{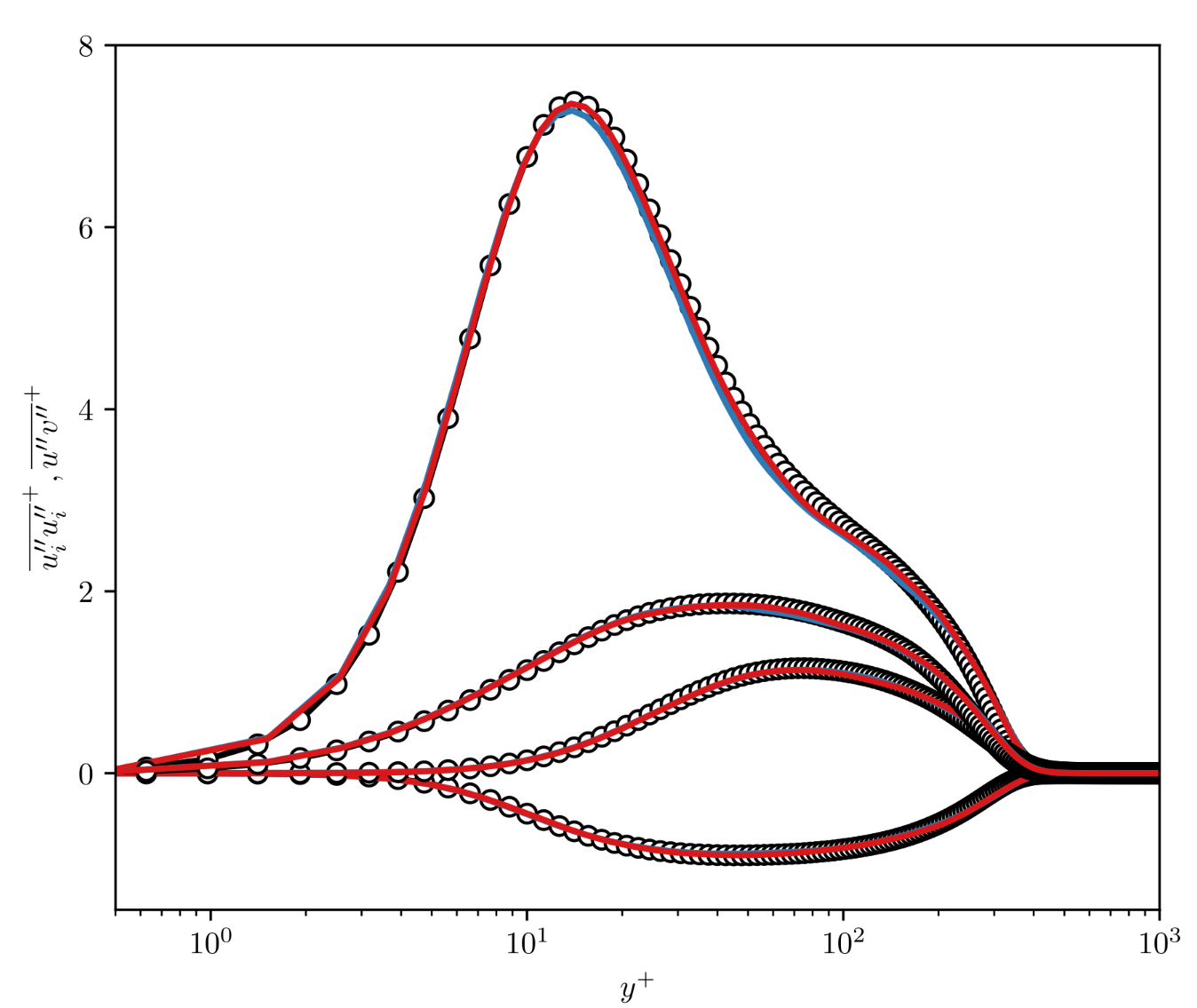}         
        \caption{Reynolds stress profiles}
        \label{fig:reynolds}
    \end{subfigure}
    
    \caption{ZPGTBL Validation for $Re_\theta = 1000$: Wall-normal distributions of (a) the streamwise velocity and (b) the symmetric components of the Reynolds stress tensor and the Reynolds shear stress of the non-actuated reference case at $x = 165$. Results are compared to an incompressible TBL from \cite{schlatter2010assessment}.}
    \label{fig:ZPGTBL_validation}
\end{figure}

\paragraph{Numerical setup and validation.}
The computational domain is a rectangular box spanning $L_x \times L_y \times L_z$ in the streamwise, wall-normal, and spanwise directions, respectively. To validate the baseline numerical setup, an unactuated flat plate boundary layer is simulated at a momentum-thickness Reynolds number of $Re_\theta = 1000$. The grid utilizes a DNS-like near-wall resolution ($\Delta x^+ \approx 12, \Delta y^+_{wall} \approx 1.0, \Delta z^+ \approx 4.0$) to accurately resolve the small-scale turbulent motions. A reformulated synthetic turbulence generation (RSTG) method prescribes the inflow conditions, and a sponge layer is applied at the outflow to prevent the spurious reflection of acoustic waves. Validation of the unactuated baseline demonstrates excellent agreement with reference direct numerical simulation data for both the mean streamwise velocity profile and the Reynolds stress tensor components (see Fig~\ref{fig:ZPGTBL_validation}). To facilitate the study of Reynolds number scaling on control strategies, the HydroGym environment offers pre-configured setups at friction Reynolds numbers $Re_\tau \in[200, 1550, 2200]$.

\paragraph{Environment implementation.}
We investigate two distinct actuation setups: a streamwise traveling surface wave and a jet actuation on the surface. For both setups, the primary reward function for the RL agent is the net power saving ($\Delta P_{net}$), which evaluates the energetic benefit of the drag reduction minus the power required to drive the respective actuation (i.e., the wall deformation or the jets).
\begin{wrapfigure}[25]{r}{0.55\textwidth}
   \centering
    \includegraphics[clip, trim=15.0cm 4.0cm 12.0cm 1.0cm, width=0.55\textwidth]{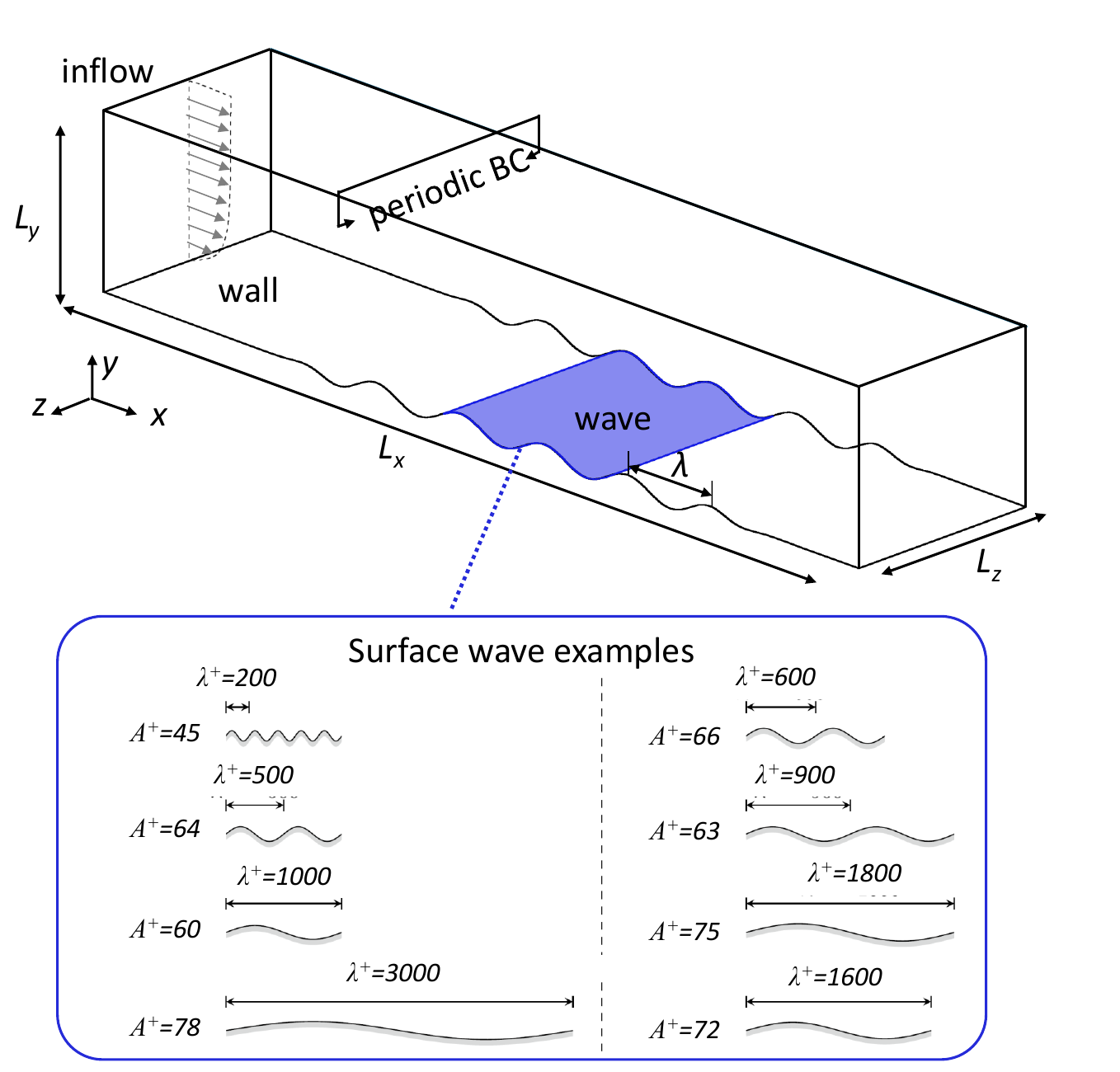}
    \caption{Schematic of the computational setup for zero-pressure gradient turbulent boundary layer environment leveraging a streamwise traveling surface wave.}
    \label{appendix:ZPGTBL_setup}
\end{wrapfigure}
The continuous action spaces are defined according to the setup. For the surface wave, the agent controls the wave geometry and speed, parameterized in outer flow units by the wave amplitude $A \in [0.001, 0.40]$, the wavelength $\lambda \in [1, 8]$, and the phase speed $c \in[0.1, 0.40]$. For the jet actuation, the action space controls the jet velocity, which is bounded within $[-0.1 U_{\infty}, 0.1 U_{\infty}]$. Because macroscopic modifications to the turbulent boundary layer require time to develop, the duration of a control action depends on the setup: one control action comprises $5000$ CFD time steps for the surface wave actuation and $1000$ CFD time steps for the jet actuation. Across both configurations, a single training trajectory consists of $200$ actuation steps. The state observation space is highly customizable; it relies on user-defined probe locations distributed throughout the flow field that can sample instantaneous velocity components ($u, v, w$), static pressure ($p$), density ($\rho$), and local surface forces. Episodes run for the predefined control interactions or terminate early if the solver diverges due to the exploration of non-physical actuation parameters.


\subsection{Supercritical DRA2303 airfoil}
\label{appendix:Environments:DRA2303}

\paragraph{Characteristic physics.}
Moving beyond canonical flat plates, the flow around a realistic lifting surface like the DRA2303 airfoil introduces the complexities of strong streamwise pressure gradients and surface curvature. Operating at a chord-based Reynolds number of $Re_c = 400,000$ and Mach numbers spanning $Ma = [0.2, 0.7]$, the aerodynamic efficiency is dictated by a delicate balance between friction drag and pressure (form) drag. At higher subsonic Mach numbers, compressibility effects become highly pronounced, and the interaction of active flow control mechanisms---such as a dynamically deforming wall or surface jets---with the boundary layer can induce localized supersonic regions or weak shock waves. Applying flow control in this regime poses a distinct control challenge: minimizing skin-friction drag might inadvertently increase pressure drag or penalize the lift coefficient. RL agents are uniquely positioned to discover spatially and temporally varying actuation parameters that navigate these competing aerodynamic forces to optimize global efficiency across different flight regimes.

\begin{figure}[t]
    \centering
    \begin{subfigure}[]{0.3\textwidth}
        \centering
        \includegraphics[width=\textwidth]{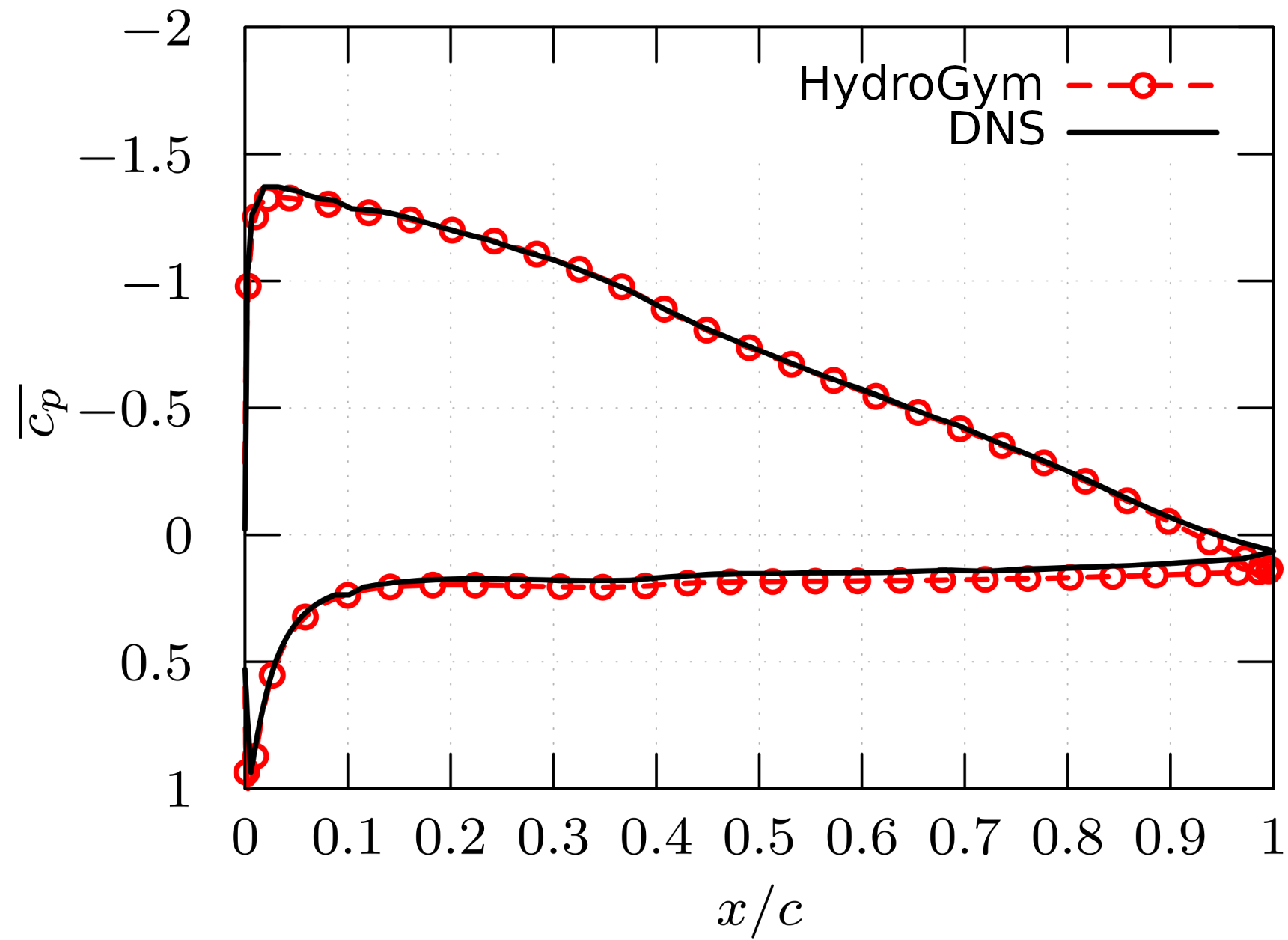} 
        \caption{pressure coefficient $c_p$}
    \end{subfigure}
    \hfill
    \begin{subfigure}[]{0.3\textwidth}
        \centering
        \includegraphics[width=\textwidth]{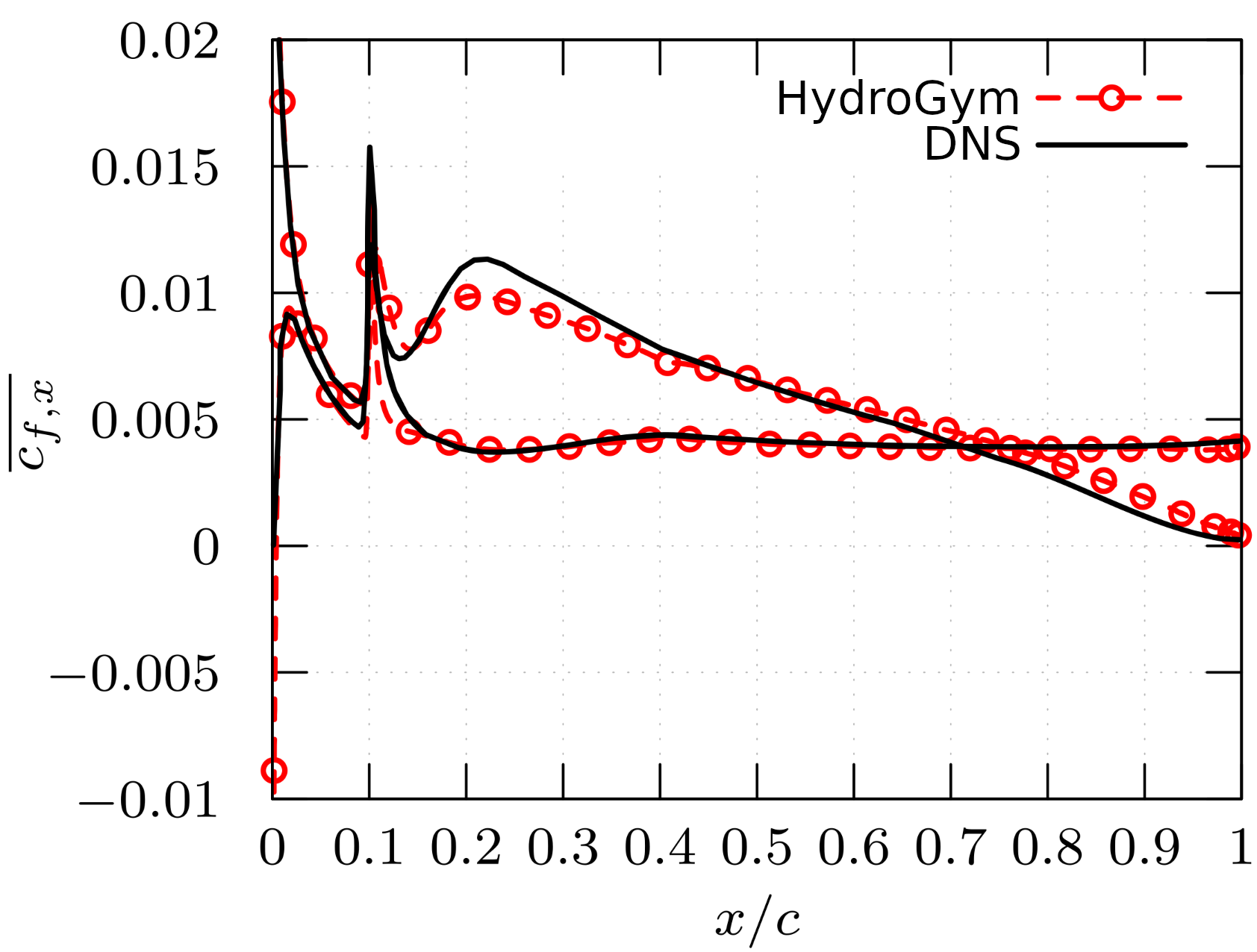} 
        \caption{friction coefficient $c_f$}
    \end{subfigure}
    \hfill
    \begin{subfigure}[]{0.3\textwidth}
        \centering
        \includegraphics[width=\textwidth]{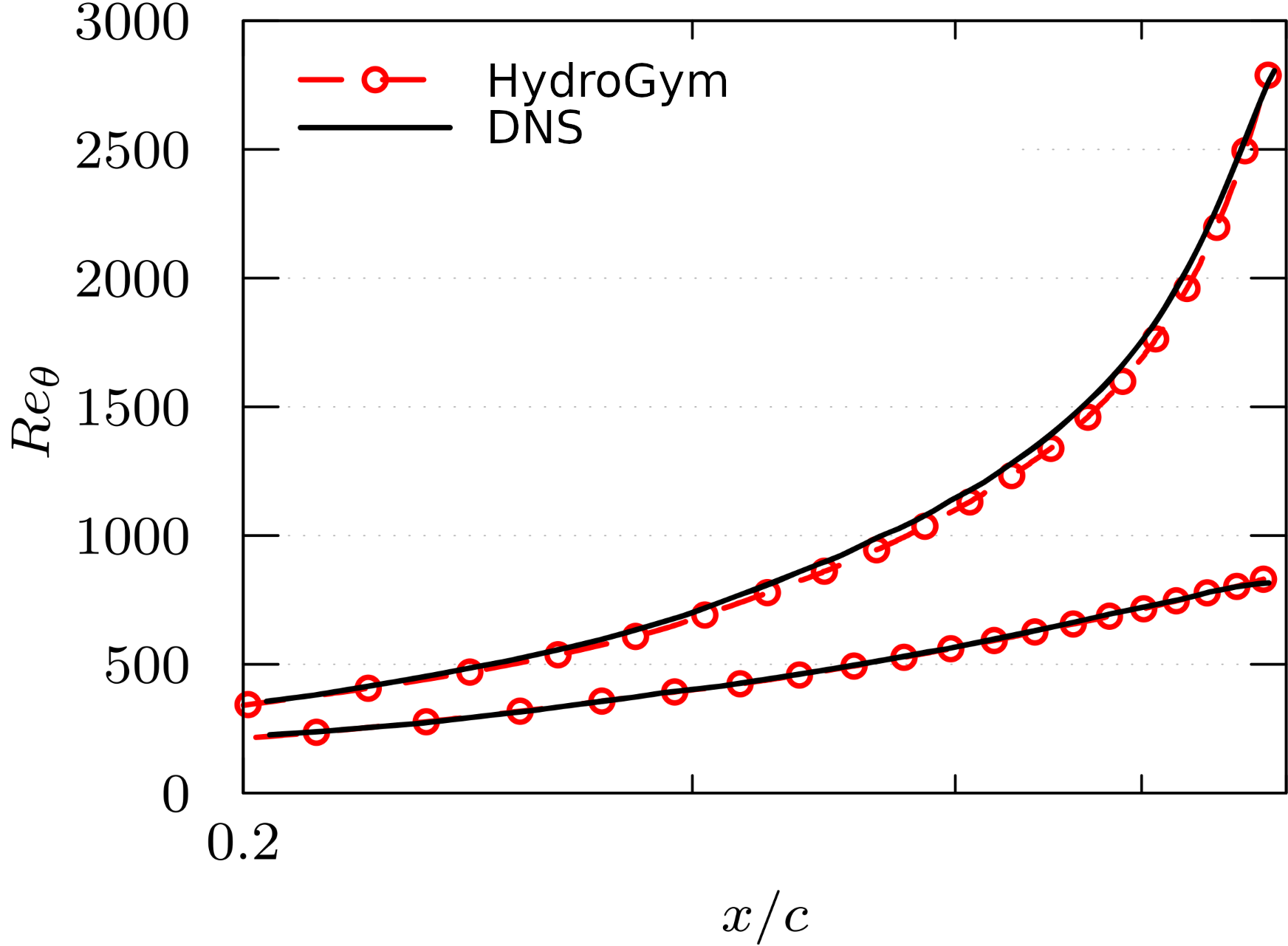} 
        \caption{Reynolds number $Re_\theta$}
    \end{subfigure}
    
    \caption{Validation results of numerical framework for high Re-number airfoil flows: Comparison of the present method applied to the flow around a NACA4412 airfoil with DNS results by \cite{hosseini2016direct} for the (a) pressure coefficient; (b) skin-friction coefficient; and (c) momentum thickness based Reynolds number.}
    \label{fig:validation_dra}
\end{figure}

\paragraph{Numerical setup and validation.}
The flow domain is discretized using a boundary-conforming C-type curvilinear mesh that extends approximately 24 chord lengths upstream and 25 chord lengths in both the downstream and vertical directions. Spanwise periodicity is enforced with a domain width of $L_z = 0.1c$, which is sufficient to ensure the decorrelation of turbulent flow structures along the wingspan. The boundary layer is tripped near the leading edge ($x/c = 0.1$) to ensure stable, fully developed turbulence over the actuated regions of the airfoil. The baseline numerical framework has been rigorously validated against high-resolution large-eddy simulations for similar airfoil geometries under identical flow conditions, showing excellent agreement for chordwise pressure ($C_p$) and skin-friction ($C_f$) distributions (see Fig.~\ref{fig:validation_dra}). The active flow control (traveling wave or jet actuation) is applied over the majority of the suction and pressure sides (typically $0.2 \le x/c \le 08$). For the surface wave setup, a smooth spatial blending function is utilized to transition the surface from static to dynamically wavy.

\paragraph{Environment implementation.}
Similar to the flat plate case, the reward function is formulated to maximize the global net power saving, inherently balancing the reduction in total airfoil drag against the
\begin{wrapfigure}[13]{r}{0.55\textwidth}
   \centering
    \includegraphics[clip, trim=6.0cm 7.5cm 6.0cm 5.5cm, width=0.55\textwidth]{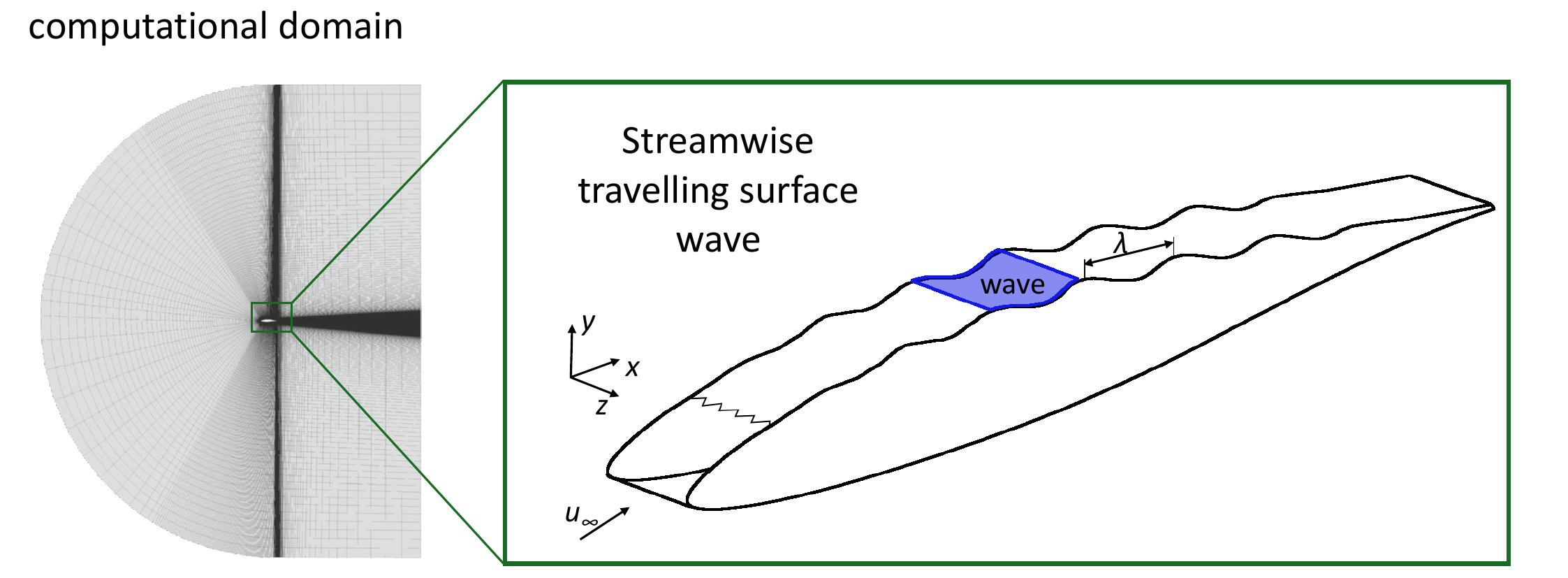}
    \caption{Schematic of the computational setup for zero-pressure gradient turbulent boundary layer environments leveraging a streamwise traveling surface wave.}
    \label{appendix:DRA_setup}
\end{wrapfigure}
energetic cost of the respective actuation (i.e., surface deformation or jet blowing). Both actuation modes expose continuous action spaces tailored to the aerodynamic sensitivities of the curved geometry. For the traveling surface wave, the agent manipulates three parameters in outer units (Fig.~\ref{appendix:DRA_setup}): the amplitude is restricted to $A \in [0.00001, 0.0005]$ to account for the heightened sensitivity of the curved boundary layer, while the wavelength and phase speed are bounded by $\lambda \in [0.005, 0.02]$ and $c \in [0.001, 0.25]$, respectively. The jet actuator exposes a single degree of freedom---the jet exit velocity---constrained to $[-0.1\,U_{\infty},\, 0.1\,U_{\infty}]$. To allow aerodynamic forces and wake dynamics to physically respond to updated wall boundary conditions, the action interval is set to $5000$ CFD iterations for the surface wave and $1000$ iterations for the jet. Each training trajectory comprises $200$ actuation steps across both configurations. Flow observations are recorded through user-defined probe arrays positioned within the boundary layer or downstream wake, capturing instantaneous velocities, pressure, density, and local aerodynamic forces. The environment accommodates a range of freestream Mach numbers ($Ma = 0.2$--$0.7$), supporting the development of control policies that generalise across compressible flow regimes. Episodes conclude either upon completion of the prescribed actuation sequence or upon early termination triggered by solver divergence under non-physical actuation parameters.


\subsection{Flow through a stenotic pipe}
\label{appendix:Environments:Stenotic_pipe}

\paragraph{Characteristic physics.}
Flow through a stenotic pipe has been extensively studied as a canonical configuration to investigate complex hemodynamic and biofluidic phenomena. 
The presence of a localized narrowing introduces strong acceleration of the fluid jet, 
followed by an adverse pressure gradient downstream that promotes flow separation, shear layer instabilities, and, at sufficiently high Reynolds numbers, transition to turbulence~\cite{Varghese2007}. 
These dynamics create characteristic features such as recirculation zones, vortex shedding, and highly fluctuating wall shear stresses, which are of particular relevance in cardiovascular applications, for instance, in the progression of atherosclerosis and thrombosis. 
The flow field is highly sensitive to the severity and geometry of the stenosis as well as the Reynolds number. 
As such, stenotic pipe flow serves as a benchmark problem for studying transitional shear flows as well as fluid–structure interaction in arterial geometries.

Beyond its relevance to vascular hemodynamics, the stenotic pipe is also a valuable generic case for studying respiratory flows under obstructed conditions.
The localized constriction mimics the airflow limitations encountered in the human nasal cavity during pathological states such as septal deviation or turbinate hypertrophy.
In this context, a level-set approach has been developed for monitoring the change in pressure and temperature distributions when iteratively modifying the surface from a stenosed to an unconstricted pipe~\cite{Waldmann2022}.
After successfully testing this on the stenotic pipe, the authors have extended this framework to capture surface modifications from pre- to post-surgical nasal geometries.
Building on this methodological foundation, subsequent studies integrated RL with CFD by letting an agent control interpolation factors that modify the geometry via the same level-set representation~\cite{Ruettgers2021}.
The agent's goal was to find a compromise between reducing pressure loss and improving the heating capability.
In the context of nasal obstruction surgery, this approach has been extended to nasal cavity geometries to design optimal surgical intervention strategies~\cite{Ruettgers2024}.
In this sense, the stenotic pipe serves as a bridge between canonical fluid mechanics problems and clinically relevant respiratory applications, making it a suitable benchmark for developing and validating RL–CFD platforms.

Beyond steady inflow conditions, stenotic pipe flow has also been investigated under pulsatile forcing, which more closely resembles the physiological environment in cardiovascular and respiratory systems. 
The periodic acceleration and deceleration of the fluid introduce additional instabilities that interact with the stenosis-induced shear layers, often amplifying vortex shedding and enhancing the complexity of the downstream flow structures~\cite{Varghese2007a}. 
These effects can result in intermittent turbulence, increased wall shear stress fluctuations, and significant cycle-to-cycle variability in recirculation zones. 

In addition to pulsatile inflow, oscillatory flow through stenotic geometries has been examined as a canonical setup for studying shear reversal and resonance phenomena under zero-mean periodic forcing. 
Unlike pulsatile flow, which superimposes forward and reverse phases on a mean flow, purely oscillatory forcing accentuates the interaction between unsteady shear layers and flow separation in both directions, leading to enhanced mixing and strongly asymmetric vortex dynamics. Under such conditions it has been demonstrated how the frequency of oscillation relative to the stenosis geometry governs vortex strength and flow reversal patterns~\cite{Jain2020}.

Building on this canonical framework, the present study explores a novel RL–CFD coupling 
in which an RL agent interacts with oscillatory stenotic pipe flow by controlling the inlet temperature. 
The agent’s objective is to maintain the temperature field downstream of the stenosis as close as possible to a prescribed target value. 
This setup not only leverages the complex unsteady dynamics of oscillatory stenotic flows as a challenging benchmark for flow control, but also provides a direct link to biomedical applications. 
Specifically, the stenotic pipe serves as a generic surrogate for obstructed respiratory passages, where maintaining controlled temperature and flow conditions is critical. 
In this context, the methodology holds promise for extension to clinically relevant scenarios such as the pre-calibration of mechanical ventilation machines, where RL-driven control strategies could help adapt inlet conditions to patient-specific airway geometries and pathologies.

\paragraph{Implementation validation.}
The computational domain used for validating the stenotic pipe flow and its coordinate system are illustrated in Fig.~\ref{fig:domain_sten_pipe}.
Details about the curvature of the stenosis can be found in~\cite{Varghese2007}.
The Reynolds number based on the pipe diameter at the inlet $D$ is set to $Re=(U_{ref} \cdot D)/\nu=500$ to match the conditions in~\cite{Varghese2007},
where $D=1m$,
$\nu=1.63\cdot10^5~m^2/s$ is the kinematic viscosity of air,
and $U_{ref}$ the spatially averaged velocity at the inlet.
The domain length is $L=38D$.

\begin{figure*}[t!]
\centering
\include{figures/SI/domain_sten_pipe}
\vspace{-2\baselineskip}
\captionsetup{format=hang}
\caption{Computational domain for validating the stenotic pipe flow.}
\label{fig:domain_sten_pipe}
\end{figure*}

At the inlet, the velocity profile of a fully developed laminar pipe flow is prescribed
and the density of the fluid is linearly extrapolated from the neighboring inner cells.
At the outlet, a constant pressure $p_{out}$ is set, 
and the velocity is extrapolated from the inner cells.
To satisfy the no-slip condition at the pipe walls,
an interpolated bounce-back scheme is used~\cite{Bouzidi2001}.
These boundary conditions match the boundary conditions of the reference simulation in~\cite{Varghese2007}.

\begin{figure}
\captionsetup[subfigure]{aboveskip=8pt,belowskip=8pt}
\begin{subfigure}{.33\linewidth}
\centering
\begin{tikzpicture}
  \scriptsize
  \begin{axis}[
    width=1.0\textwidth,
    height=0.95\textwidth,
    scaled ticks=false,
    axis y line=left,
    axis x line=bottom,
    ymin=0, ymax=0.52, xmin=-0.5, xmax=5.5,
    ytick={0, 0.1, 0.2, 0.3, 0.4, 0.5},
    yticklabels={{0}, {0.1}, {0.2}, {0.3}, {0.4}, {0.5}},
    xtick={0, 1, 2, 3, 4, 5},
    xticklabels={{0}, {1}, {2}, {3}, {4}, {5}},
    ylabel={\scriptsize $r/D$},
    ylabel near ticks,
    xlabel={\scriptsize $u/U_{ref}$},
    xlabel near ticks,
    axis on top,
    line width = 0.8pt,
    legend pos=north east,  
    legend cell align={left},  
    legend style={draw=black, fill=white, font=\scriptsize}  
    ]

    \addplot[color=teal, thick] table[x index=1,y index=0,col sep=space] {sten_pipe_grid_ref_200.dat};
    \addlegendentry{$\Delta x=D/200$};

    \addplot[color=blue, thick] table[x index=1,y index=0,col sep=space] {sten_pipe_grid_ref_100.dat};
    \addlegendentry{$\Delta x=D/100$};

    \addplot[color=red, thick] table[x index=1,y index=0,col sep=space] {sten_pipe_grid_ref_50.dat};
    \addlegendentry{$\Delta x=D/50$};

    \addplot[color=black, thick] table[x index=1,y index=0,col sep=space] {sten_pipe_grid_ref_varg.dat};
    \addlegendentry{DNS~\cite{Varghese2007}};

  \end{axis}  

\draw[-, dashed, line width=0.25mm, draw=black] (0.35,0) -- (0.35,4.0);

\draw[-, line width=0.25mm, draw=black] (0.25,4.0) -- (0.45,4.0);
 
\end{tikzpicture}
\vspace{-2\baselineskip}
\caption{Mesh refinement study at $x=5D$.} \label{fig:sten_pipe_grid_ref}
\end{subfigure}
\begin{subfigure}{.33\linewidth}
\centering
\begin{tikzpicture}
  \scriptsize
  \begin{axis}[
    width=1.0\textwidth,
    height=0.95\textwidth,
    scaled ticks=false,
    axis y line=left,
    axis x line=bottom,
    ymin=0, ymax=0.52, xmin=-0.5, xmax=5.5,
    ytick={0, 0.1, 0.2, 0.3, 0.4, 0.5},
    yticklabels={{0}, {0.1}, {0.2}, {0.3}, {0.4}, {0.5}},
    xtick={0, 1, 2, 3, 4, 5},
    xticklabels={{0}, {1}, {2}, {3}, {4}, {5}},
    ylabel={\scriptsize $r/D$},
    ylabel near ticks,
    xlabel={\scriptsize $u/U_{ref}$},
    xlabel near ticks,
    axis on top,
    line width = 0.8pt,
    legend pos=north east,  
    legend cell align={left},  
    legend style={draw=black, fill=white, font=\scriptsize}  
    ]

    \addplot[color=teal, thick] table[x index=1,y index=0,col sep=space] {sten_pipe_prof_1000.dat};
    \addlegendentry{$\Delta x=D/200$};

    \addplot[color=black, thick] table[x index=1,y index=0,col sep=space] {sten_pipe_prof_varg.dat};
    \addlegendentry{DNS~\cite{Varghese2007}};

  \end{axis}  

\draw[-, dashed, line width=0.25mm, draw=black] (0.35,0) -- (0.35,4.0);

\draw[-, line width=0.25mm, draw=black] (0.25,4.0) -- (0.45,4.0);
 
\end{tikzpicture}
\vspace{-2\baselineskip}
\caption{Inlet profiles with $\Delta x=D/200$.} \label{fig:inl_prof}
\end{subfigure}
\begin{subfigure}{.33\linewidth}
\centering
\begin{tikzpicture}
  \scriptsize
  \begin{axis}[
    width=1.0\textwidth,
    height=0.95\textwidth,
    scaled ticks=false,
    axis y line=left,
    axis x line=bottom,
    ymin=0, ymax=0.52, xmin=-0.5, xmax=5.5,
    ytick={0, 0.1, 0.2, 0.3, 0.4, 0.5},
    yticklabels={{0}, {0.1}, {0.2}, {0.3}, {0.4}, {0.5}},
    xtick={0, 1, 2, 3, 4, 5},
    xticklabels={{0}, {1}, {2}, {3}, {4}, {5}},
    ylabel={\scriptsize $r/D$},
    ylabel near ticks,
    xlabel={\scriptsize $u/U_{ref}$},
    xlabel near ticks,
    axis on top,
    line width = 0.8pt,
    legend pos=north east,  
    legend cell align={left},  
    legend style={draw=black, fill=white, font=\scriptsize}  
    ]

    \addplot[color=teal, thick] table[x index=1,y index=0,col sep=space] {sten_pipe_prof_2_5_1000.dat};
    \addlegendentry{$\Delta x=D/200$};

    \addplot[color=black, thick] table[x index=1,y index=0,col sep=space] {sten_pipe_prof_2_5_varg.dat};
    \addlegendentry{DNS~\cite{Varghese2007}};

  \end{axis}  

\draw[-, dashed, line width=0.25mm, draw=black] (0.35,0) -- (0.35,4.0);

\draw[-, line width=0.25mm, draw=black] (0.25,4.0) -- (0.45,4.0);
 
\end{tikzpicture}
\vspace{-2\baselineskip}
\caption{Profiles at $x=2.5D$ with $\Delta x=D/200$.} \label{fig:prof_at_2_5D}
\end{subfigure}
\caption{Profiles of $u/U_{ref}$ for different mesh resolutions (a) or at different locations with the highest resolution (b, c) compared to the reference case from the DNS in~\cite{Varghese2007}.}
\label{fig:vel_profiles}
\end{figure}

Figure~\ref{fig:sten_pipe_grid_ref} shows results of a mesh refinement study.
Profiles of the normalized streamwise velocity component $u/U_{ref}$ at $x=5D$ with various mesh resolutions are compared to the reference solution from~\cite{Varghese2007} (black).
The coarse mesh (blue) has a resolution of $\Delta x=D/50$ and $2.5 \cdot 10^6$ cells.
Results with the coarse mesh indicate a smaller recirculation zone between $r/D=0.4$ and $r/D=0.5$ 
and an underprediction of the maximum velocity at $r/D=0$ by $5.2\%$.
The solution of the medium mesh (red) with a resolution of $\Delta x=D/100$ and $20 \cdot 10^6$ cells agrees better with the reference solution near the recirculation zone,
but still shows a slight underprediction of the maximum velocity by $2.5\%$.
Finally, the fine grid (green) has a resolution of $\Delta x=D/200$ and features $160 \cdot 10^6$ cells. The corresponding velocity profile matches well with the reference solution and has only a deviation in the maximum velocity of $1.4\%$.

Figure~\ref{fig:inl_prof} shows that the inlet profile of the simulation with the LB method matches the inlet profile in~\cite{Varghese2007}.
The velocity profiles at $x=2.5D$ in Fig.~\ref{fig:prof_at_2_5D} reveal that the LB simulation agrees well with the reference solution in the near-wall region and recirculation zone.
Deviations below $5\%$ are observed near $r/D=0$.

The comparisons of velocity profiles and recirculation dynamics across different grid resolutions demonstrate that the fine mesh solution agrees closely with the established DNS benchmark~\cite{Varghese2007}.
Together with the previously verified thermal model~\cite{Ruettgers2022}, these results confirm that the present LB solver reliably reproduces the key flow features of the stenotic pipe.

\paragraph{Environment implementation.}
The stenotic pipe control environment is likewise formulated as a discrete-time Markov Decision Process. 
The underlying flow is driven by an oscillatory inflow with a Womersley number of $\alpha=\tfrac{1}{2}D\sqrt{\tfrac{\omega}{\nu}}=4$, where $\omega$ denotes the angular frequency. 
To introduce variability and mimic physiological fluctuations, a Gaussian noise term with mean $0$, standard deviation $1$, and amplitude $0.05$ is superimposed on the sinusoidal signal. 
The agent interacts with the system by modulating the inlet temperature. 
Actions are continuous and normalized, with the initial condition corresponding to a normalized inlet-to-wall temperature ratio of $T_{\mathrm{inlet}}/T_{\mathrm{wall}}=0.9$. 
At each time step, the agent can adjust the inlet temperature within the bound $|\Delta T_{\mathrm{inlet}}/T_{\mathrm{wall}}| \leq 0.05$.

The environment advances in intervals of 4,000 simulation time steps between two consecutive control decisions. 
This duration ensures that thermal perturbations introduced at the inlet have sufficient time to advect downstream and alter the velocity and temperature fields in the vicinity of the stenotic region. 
The state space is defined by 10 velocity, pressure, and temperature probes placed along the pipe centerline, distributed between $x=-3.5D$ upstream of the stenosis and $x=D$ downstream (see Fig.~\ref{fig:uncontrolled_sten_pipe}). 
These probes capture the temporal development of the flow response to the applied control, providing the agent with the necessary feedback for learning effective strategies.
The reward function is formulated as $R(t)=-100\cdot|(T_{target}(t)/T_{wall})-0.95|$,
where $T_{target}$ is the temperature on the centerline at $x=D$.

\begin{figure*}[t!]
\vspace{-.1in}
\centering
\include{figures/SI/sten_pipe_flow}
\captionsetup{format=hang}
\vspace{-.35in}
\caption{Normalized velocity and temperature fields of states I-IV of the uncontrolled 3D stenotic pipe flow.}
\label{fig:uncontrolled_sten_pipe}
\end{figure*}

Figure~\ref{fig:uncontrolled_sten_pipe} presents the normalized velocity and temperature fields of the uncontrolled 3D stenotic pipe flow for four characteristic states.
In state I, the inflow velocity reaches its peak and cold fluid enters the pipe.
In state II, the inflow velocity transitions from positive to negative, resulting in near-zero velocities throughout most of the domain. Under these conditions, wall-to-fluid heat transfer is enhanced, leading to a rise in fluid temperature.
In state III, the reverse flow attains its maximum, driving warm fluid into the pipe from the outlet.
Finally, state IV marks the transition from negative back to positive inflow velocity, again associated with low velocities and elevated heat transfer.

\subsection{Turbulent channel flow}
\label{appendix:Environments:TCF}
\paragraph{Characteristic physics.}
Turbulent channel flow (TCF) represents a canonical configuration for investigating wall-bounded turbulence. 
Owing to its periodicity in two spatial directions, TCF provides an ideal training environment for multi-agent reinforcement learning (MARL), as this feature enables the generation of numerous pseudo-environments that can be exploited for parallel training. 
Consequently, global invariance is preserved while learning efficiency is maximized. 
This distinctive advantage is absent in turbulent boundary layers (TBLs), where inherent spatial development substantially increases computational cost. Nevertheless, at comparable Reynolds numbers, TCF exhibits near-wall dynamics and coherent structures that closely resemble those observed in TBLs.

These considerations motivate the zero-shot MARL methodology, whereby policies trained in a channel configuration are subsequently deployed on TBLs without additional training. 
The TCF setup is therefore designed to reproduce the mean characteristics of the TBL within the targeted control region, thereby serving as an analogue of the flow environment encountered in the boundary layer. In the present study, the reference TBL corresponds to the suction side of a NACA0012 wing at $Re_c=200{,}000$ and $AoA=0^\circ$, as described in Sec.~\ref{appendix:Environments:NACA0012}, representing a boundary layer subjected to a mild adverse pressure gradient (APG). This strategy facilitates high-Reynolds-number deep reinforcement learning (DRL) control in complex flow configurations while substantially reducing computational expense.

Following the zero-shot framework, the principal objective is to match the momentum-thickness-based Reynolds number ($Re_\theta$) as the target flow characteristic. 
This quantity is defined as $Re_\theta = U_0 \theta/\nu$, where {momentum thickness ($\theta$) is defined as
\[ 
\theta=\int^{\infty}_0 ({U}/{U_0}) \left (1- (U/U_0) \right){\rm d}y.
\]}
Here, $U_0$ corresponds to the boundary-layer edge velocity ($U_e$) in TBLs and to the centerline velocity ($U_{cl}$) in TCFs. The resulting configuration yields $Re_\theta=365$, which aligns closely with the chordwise-averaged value $\langle Re_\theta \rangle_x = 382$ on the wing, corresponding to a deviation of only $4.5\%$ from the target. This setup consequently produces a friction Reynolds number of $Re_\tau = 206$.

\paragraph{Numerical setup and validation.}
The three-dimensional open channel flow is simulated using direct numerical simulation (DNS)
based on the spectral-element method (SEM).
The computational domain has dimensions $(L_x, L_y, L_z) = (2\pi, 1, \pi)h$ in the
streamwise, wall-normal, and spanwise directions, respectively, and is discretized into
$(N_x, N_y, N_z) = (10, 8, 9)$ spectral elements.
A polynomial order of $N = 7$ is adopted, yielding $(80, 64, 72)$ grid points in the
streamwise, wall-normal, and spanwise directions, respectively.
The resulting inner-scaled spatial resolutions are $\Delta x^+ < 18.8$,
$\Delta y^+|_{{\rm min}, {\rm max}} < (0.5, 11.3)$, and $\Delta z^+ < 10.1$ in the
streamwise, wall-normal, and spanwise directions, respectively.
This resolution is selected to ensure that the effective size of the control actuators
closely matches that employed in the wing simulations expressed in viscous units, thereby
providing the grid-level consistency required to justify zero-shot transfer.
A no-slip boundary condition is imposed at the lower wall, whereas a symmetry condition
is applied at the upper boundary to avoid additional complications associated with
wall actuation.
Periodic boundary conditions are enforced in both the streamwise and spanwise directions,
and the flow is driven by a prescribed constant bulk velocity to maintain statistically
stationary turbulence.

\begin{figure}[t!]
    \centering
    \includegraphics[width=\linewidth]{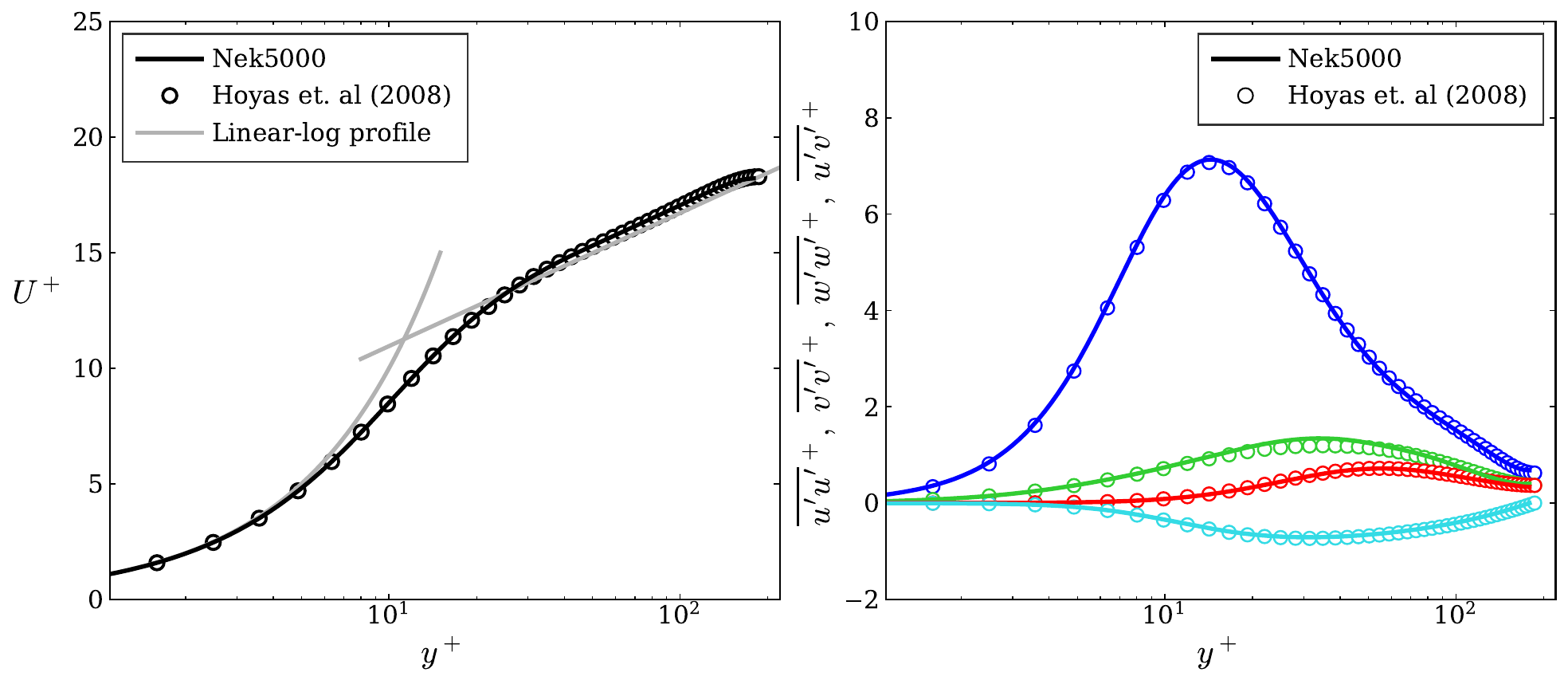}
    \vspace{-.2in}
    \caption{Inner-scaled (left) mean component of streamwise velocity and (right) Reynolds stresses as a function of inner-scaled wall-normal distance $y^+$ compared to {Hoyas et al.} DNS data of a full channel at $Re_\tau = 180$.}
    \label{fig:nek_channel_valid}
\end{figure}

To validate the numerical framework, a DNS of a full channel flow (i.e. with no-slip conditions
imposed at both walls) is performed in Nek5000 at $Re_\tau = 180$, following the
configuration detailed in~\cite{wang_opposition_2025}.
The inner-scaled mean streamwise velocity profile and Reynolds-stress components are compared
against the reference DNS database of~\cite{Hoyas_tcf180_2008} at $Re_\tau = 180$ as
functions of the inner-scaled wall-normal distance $y^+$, presented in logarithmic scale
in figure~\ref{fig:nek_channel_valid}.
The excellent agreement observed across all quantities confirms the accuracy of the present
numerical setup. 

\paragraph{Environment implementation.}
The control objective is to reduce the wall-shear stress $\tau_w = \rho \nu \left({{\rm d}U}/{{\rm d}y}\right)_{y=0}$ in the channel flow. Accordingly, the reward ($R$) is defined as the relative reduction in $\tau_w$:
\begin{equation}
R = 1 - \tau^{\rm ctrl}_w/\tau^{\rm ref}_w
\end{equation}
\noindent where \emph{\rm ctrl} and \emph{\rm ref} denote the controlled and uncontrolled flow states, respectively. For a fully developed TCF, this quantity is spatially uniform at the wall, thereby constituting a global invariant well suited for MARL. The actuation consists of zero-mean wall-normal blowing and suction at the wall, as detailed in Sec.~\ref{appendix:Implementation:nek5000}, with amplitudes constrained within the range $[-u_\tau, u_\tau]$, where $u_\tau = \sqrt{\tau_w/\rho}$.
Control actions are updated at intervals of $\Delta t^+ = 0.6$, scaled using the viscous time unit $t^* = \nu/u_\tau^2$. 
Each training episode spans a duration of $t^+ = 1{,}500$. 
The observable states available to each agent consist of the streamwise and wall-normal velocity components sampled at $y^+=15$ (with $y^+ = y u_\tau/\nu$). 
For each component, the spatial mean computed as the ensemble average over all homogeneous directions is removed. In fully developed TCF, this mean value is identically zero for all velocity components. 
Further implementation details can be found in \cite{Guastoni2023}, as the present setup closely follows the methodology established in that work.

\subsection{Differentiable turbulent channel flow}
\label{appendix:Environments:Channel}

\paragraph{Characteristic physics.}
The turbulent channel flow at $Re_{\tau} = 180$ is an ideal example of a fully developed turbulent flow. It exhibits the phenomena of wall-bounded flows in general, serving as an ideal case for understanding flow over surfaces. In HydroGym, the channel flow is solved on a 3D domain of size $(2 \pi \times 2 \times
\pi )$ discretized into $(72 \times 72 \times 72)$  cells. The streamwise and spanwise directions contain uniform spacing, while a tunable hyperbolic tangent function is applied to the spacing in the wall-normal direction resulting in a finer grid close to the wall. A no-slip boundary condition is implemented at the walls, and periodic boundary conditions are applied in the streamwise and spanwise directions. A Neumann boundary condition is applied on the walls for pressure, $\frac{\partial p}{\partial z}_{wall} = 0$. The channel flow is driven by a constant, external body force in the x-direction, $f = (fx, 0, 0)$, to ensure a statistically stationary turbulent flow~\cite{nelson2017}. The flow is initialized with a laminar flow plus a random perturbation to generate turbulence. The forcing term sets the friction Reynolds number, $Re_{\tau}$, through the following relation: 

\begin{equation}
  f_x = \frac{8 \nu^2}{H^3} Re_\tau^2,
\label{eq:ReynoldsFriction}
\end{equation}
where $H$ is the height of the channel. The velocity and pressure are both solved at the cell-centroids of the grid. The timestep, $\Delta t$, of the simulation is dynamically updated to ensure a maximum Courant number of 0.9~\cite{vuorinen2016dnslab}.

\begin{figure}[t!]
    \centering
    \includegraphics[width=\linewidth]{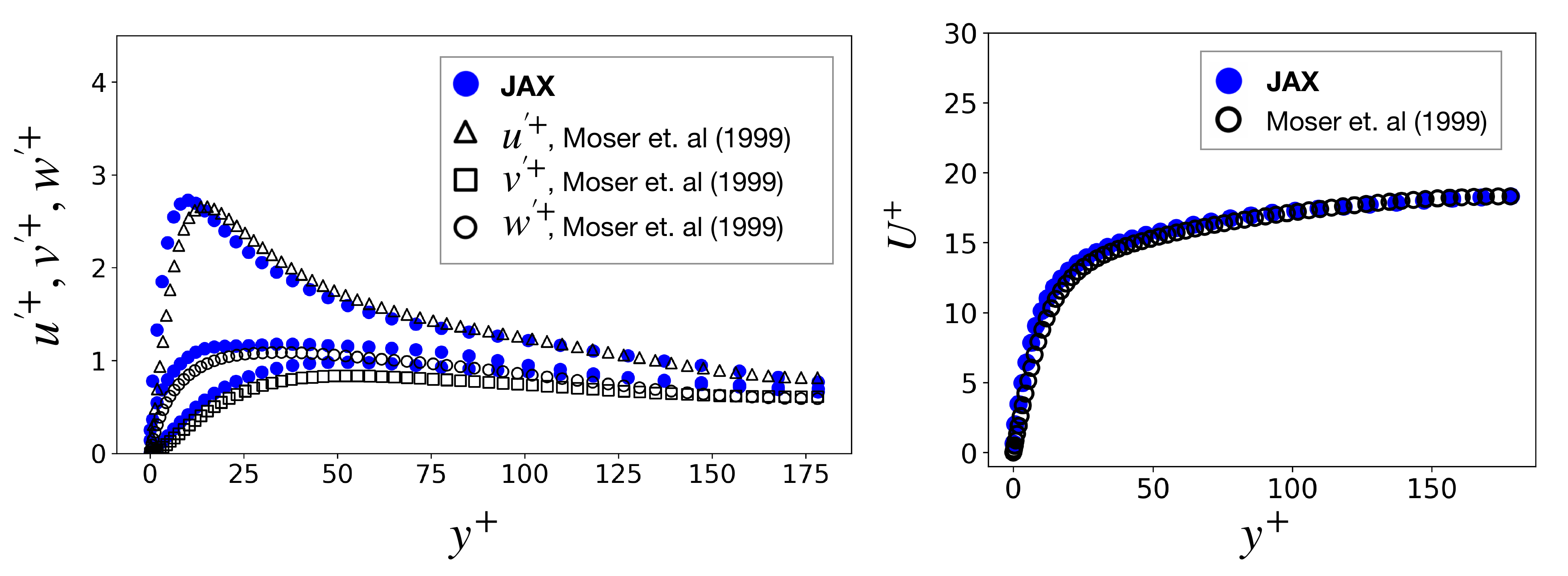}
    \vspace{-.2in}
    \caption{Dimensionless velocity fluctuations and mean velocity from the 3D channel flow compared to Moser et al. DNS data.}
    \label{fig:channel_valid}
\end{figure}

\paragraph{Numerical setup and validation.}
The 3D channel flow problem is solved with a finite-difference direct numerical solver (DNS) inspired by work of Vuorinen et al.~\cite{vuorinen2016dnslab}. A compact way to numerically solve the derivatives in equation \ref{eq:navier_stokes} is with differentiation matrices.  Consider a derivative operator $\frac{\partial}{\partial x}$ and the velocity $u$ and pressure $p$ at the cell-centroid points. 
For a grid of size $(N, N, N)$, the velocity and pressure can be expressed as column vectors of size ($N^3, 1$), and the differentiation operators, say $\frac{\partial }{\partial x}$, as a sparse differentiation matrix $\mathbf{D_x}$ of size ($N^3, N^3$). All together, there are six differentiation matrices for velocity and six for pressure, ($\mathbf{D_x}, \mathbf{D_{xx}}, \mathbf{D_y}, \mathbf{D_{yy}}, \mathbf{D_z}, \mathbf{D_{zz}}$). The derivatives in the Navier-Stokes equations can then be computed with basic linear algebra $\frac{\partial u}{\partial x} \approx \mathbf{D_x} u $. For instance, the diffusive term in $x$ would be computed as: $\nu (\mathbf{D_{xx}}u + \mathbf{D_{yy}}u + \mathbf{D_{zz}}u )$. The Poisson equation is used to solve for the pressure, using the formula:  $\mathbf{M}p = \mathbf{D_x} u + \mathbf{D_y} u + \mathbf{D_z} u $ where $\mathbf{M}$ is the Poisson operator. This is a linear system, which is solved using the Biconjugate gradient solver from lineax \texttt{BiCGStab}~\cite{rader2023lineaxunifiedlinearsolves}. Once the pressure is solved for, the velocity is corrected using the pressure gradient: $u^* = u - \mathbf{D_{x,p}} p$. To validate the solver, the dimensionless velocity fluctuations and mean velocity profile were plotted against the well-known Moser et al. 1999 DNS study~\cite{moser1999direct}, as shown in Figure \ref{fig:channel_valid}.

\paragraph{Environment implementation.}
The control task for the 3D channel flow is to locally minimize the wall-shear stress with synthetic jets. Specifically, zero-mass-flux jets with periodic blowing and suction are implemented at the wall of the channel. This is essentially a wall-normal velocity distribution with a parabolic velocity profile, ensuring it is maximal at the centerline and smoothly decays to zero at the jet’s edges. The actuation time is determined by the travel time of the velocity streaks, and ranges from 50$\Delta t$ to 150$\Delta t$ depending on the amount of jets implemented and the friction Reynolds number. The environment observation is the point value of the velocity in the x-direction, $u$, sampled from an off-wall location parallel to the jet locations. The sampling height was chosen as $y^+_s \approx 15$, where there exists a strong correlation between wall-normal velocity and near-wall vortical structures~\cite{hammond1998}. Similar works have also used this sampling plane to ensure the best results~\cite{stroh2015}. The objective was to reduce the skin friction, or wall-shear stress,  of the channel. The wall shear stress was computed as $\tau_w = \nu u(y) / y$, where $y$ was chosen within the viscous sub-layer. Mathematically, this is equivalent to computing the velocity gradient with respect to the wall, since u(y=0) = 0~\cite{lagemann2024uncovering,lagemann2025deep}. To decrease $\tau_w$,  reward signal was calculated as the ratio between the magnitude of the uncontrolled wall shear stress and the controlled wall shear stress. Both GPPO and PPO used the same observations and rewards, with the main distinction being that the GPPO training uses the reward signal in the loss function to differentiate through the environment dynamics when updating the policy.  


\subsection{Kolmogorov flow}
\label{appendix:Environments:Kolmogorov}

\begin{figure}[t!]
    \centering
    \includegraphics[width=\linewidth]{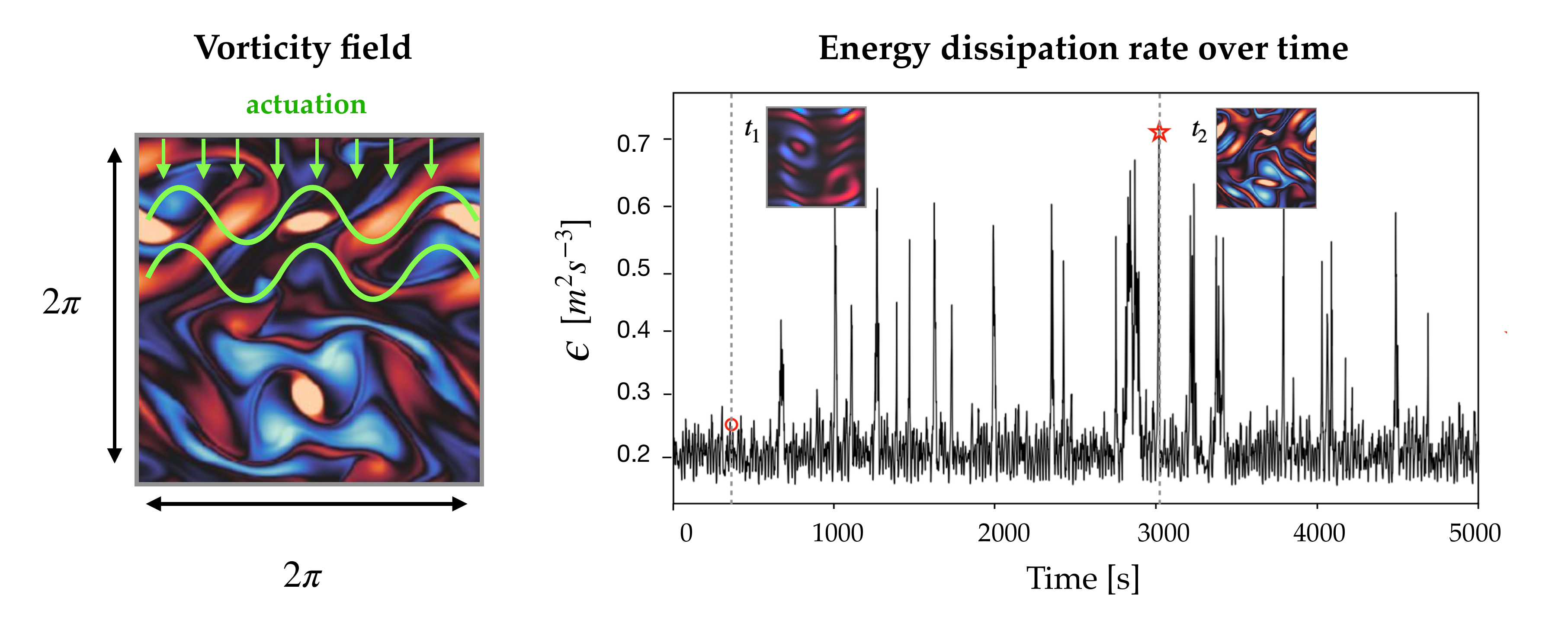}
    \vspace{-.25in}
    \caption{(a) A snapshot of the vorticity field for a Re$\approx$100 case of the Kolmogorov flow, and 
    (b) the energy dissipation rate over time for Re$\approx$40. Highlighted with the red circle is the stable state, and the red star indicates one of the few extreme energy dissipation bursts. }
    \label{fig:kolmogorovvalidation}
\end{figure}

\paragraph{Characteristic physics.} Chaotic dynamical systems often exhibit extreme events in which the system significantly deviates from its expected behavior. These extreme events exist in both nature and engineered systems, such as oceanic rogue waves, earthquakes, and shocks in power grids, often leading to adverse financial and societal impacts~\cite{Farazmand2019}. As such, it is important to be able to predict and control these events. In turbulence, extreme events manifest as energy bursts, which are random and intermittent in both space and time~\cite{Moffatt2021}. To demonstrate this phenomena, a canonical two-dimensional turbulent flow driven by a sinusoidal forcing term is developed. This flow is known as the Kolmogorov flow, and in a certain parameter regime, it  exhibits sporadic energy dissipation bursts due to a non-linear energy transfer between scales~\cite{Blanchard2019}. This often happens between Reynolds numbers 40 to 80~\cite{kumar2023stabilizing}, as depicted in Figure \ref{fig:kolmogorovvalidation} (b). For this implementation, it was found that the energy transfer that occurs is between a larger fluid structure with wavenumber (1,0), which transfers  energy to smaller structures with wavenumber (0,$k_f$)~\cite{Blanchard2019}. This rapidly increases the energy dissipation, as turbulent energy dissipates at smaller scales.

\paragraph{Numerical implementation.} Earlier in this section, the vector form of the incompressible Navier-Stokes equation \ref{eq:navier_stokes} was introduced. This can be reformulated in vorticity space by taking the curl of the momentum equation, which can then be solved with Fourier methods. To demonstrate this, recall that $\nabla \times \mathbf{u} = \omega$,  
$\nabla \cdot \mathbf{u} = 0$ for an incompressible fluid, and $\nabla \times \nabla  \cdot q = 0 $ where $q$ is a scalar quantity.  Equation \ref{eq:navier_stokes} can then be simplified to:
\begin{equation}
  \frac{\partial \mathbf{\omega}}{\partial t} + \mathbf{u} \cdot \nabla \omega = \nu \nabla ^2 \omega + \mathbf{f}, \quad \omega = - \nabla ^2 \Psi;
\label{eq:vorticitykol}
\end{equation}
 where the stream function $\Psi$, defined as $\mathbf{u} = \nabla \times \Psi$, ensures continuity. Equation \ref{eq:vorticitykol} is also known as the vorticity equation~\cite{Boffetta2012}, and is approximated numerically on a doubly-periodic domain of $[ (0, 2 \pi) \times (0, 2 \pi) ]$ with a sinusoidal forcing term of $\mathbf{f} = (\sin{\text{k}_f  y}, 0)$, where $k_f$ is the forcing wavenumber. A pseudo-spectral method with 2/3 aliasing was developed to solve the equation~\cite{YIN2004509}. For the time integration, a fourth order Runge Kutta and Crank Nicolson method from~\cite{Dresdner2022-Spectral-ML} was implemented.

\paragraph{Environment implementation.} The Kolmogorov flow contained two control objectives: (1) mitigating energy dissipation, $\epsilon$, and (2) increasing turbulent kinetic energy (TKE). While the obvious control objective is to predict and mitigate the energy burst, it can instead be leveraged to drive the system into a more turbulent regime, thereby increasing TKE and mixing. For the first objective, the energy dissipation rate is computed as  
$\epsilon = 2 \nu <S_{ij} S_{ij}>,$
where $S_{ij}$ is the strain rate tensor~\cite{bird_transport_phenomena}. The reward function penalizes both energy dissipation magnitude and events characterized as extreme. In this case, any energy dissipation value that is greater than three standard deviations is penalized. For the second objective, the average turbulent kinetic energy (TKE) is computed as: $\frac{1}{n^2} \sum_{i=1}^{n} \sum_{j=1}^{n} [\frac{1}{2} (u_i - U)^2 + (v_i - V)^2],$ where $(u,v)$ are the $x$ and $y$ velocity components with mean fields of $U$ and $V$, respectively. In this case, the reward is proportional to the TKE. The observations are the same for both objectives - equally spaced velocity point values across the domain. The actuation is formulated as amplitudes of a forcing wavenumber, where the amplitudes are the actions of the agent. Four wavenumbers are chosen to manipulate the system, and the control function took the form $c = a_1 f(k_{f_1}) + a_2 f(k_{f_2})  + a_3 f(k_{f_3}) + a_4 f(k_{f_4})$, where $f$ is the sinusoidal forcing function with wavenumber $k_f$. The wavenumbers selected are greater than the forcing wavenumber, as any additional energy in the large-scale structures  dominated the dynamics. In the case presented in this work, the specific forcing wavenumbers of $k_1, k_2, k_3, k_4 = 4, 5, 6, 7$ are chosen. For both objectives, the actions are added as a term to the reward function to penalize large actions and promote more efficient controllers. 

\subsection{Shock vector control in a single divergent nozzle}
\label{appendix:Environments:TVC}

\paragraph{Characteristic physics.}
Shock vector control (SVC) is a fluidic thrust vectoring technique in which a secondary jet is injected into the divergent section of a nozzle, deflecting the primary supersonic flow and generating a controllable lateral force. This approach offers a lightweight and mechanically simple alternative to conventional systems based on movable flaps or gimbaled nozzles, thereby reducing structural complexity, weight, and maintenance requirements~\cite{Das2025}. The underlying physics are governed by compressible flow interactions: the injected secondary jet generates a bow shock that induces an adverse pressure gradient, forcing the primary flow to turn.
This results in an asymmetric wall pressure distribution, which is the primary source of the lateral force and the resulting thrust vector deflection. 
The effectiveness of SVC is highly sensitive to operating conditions such as the pressure ratios of the primary and the secondary flow. Developing control strategies is therefore essential to ensure stable and predictable performance across a wide operating envelope.

The nozzle geometry and flow conditions considered in this study are based on the experimental work of Waithe et al.~\cite{Waithe2003}, who examined the flow in a single divergent planar nozzle. A schematic of the geometry is shown in Fig.~\ref{fig:tvc_schematic}.
In addition to the planar (2D) configuration, we also investigate the corresponding three-dimensional bell nozzle. The 3D geometry is constructed by revolving the 2D nozzle contour about the $x$-axis.
 The nozzle contour consists of linear and circular segments, with a linear convergent section. The throat diameter is $D_t=27.48\mathrm{mm}$ and the injector width is $2.032\mathrm{mm}$. The injection direction is aligned with the planar injector normal.
 The flow conditions are parameterised by the nozzle pressure ratio $NPR=P_0/P_{\infty}$ and the secondary pressure ratio $SPR=P_s/P_0$. Here, $P_0$ and $P_s$ denote the total pressure of the primary and secondary flow, respectively. The ambient pressure is denoted by $P_{\infty}$. Key performance metrics are the thrust vector angles $\delta_0$ and $\delta_1$. Given the thrust vector $\mathbf{F}=[F_x,F_y,F_z]$, they are defined as
\begin{equation}
    \delta_{0} = \tan^{-1}(F_y/F_x), \quad \delta_{1} = \tan^{-1}(F_z/F_x).
\end{equation}

\paragraph{Numerical setup and validation.} The simulations are performed using the JAX-Fluids CFD solver~\cite{bezgin2023jax, bezgin2025jax}. JAX-Fluids solves the compressible Navier-Stokes equations on a Cartesian mesh using a high-order Godunov-type method. In particular, a fifth-order WENO-Z scheme~\cite{Borges2008} is employed for cell-face reconstruction. The numerical cell-face flux is then computed with the HLLC approximate Riemann solver~\cite{Toro1994}. Temporal integration is performed using the third-order total-variation-diminishing Runge-Kutta scheme~\cite{Gottlieb1998a}. Solids are modeled with a conservative cut-cell immersed boundary method~\cite{Hu2006}, where the interface is represented as the zero level-set of a signed-distance function.

Figure~\ref{fig:tvc_schematic} shows a schematic of the computational domain. The domain extends over $(x/D_t,\, y/D_t,\, z/D_t) \in [0,250] \times [-75,75] \times [-75,75]$, where the nozzle centerline coincides with the $x$-axis. The mesh consists of a highly resolved uniform region around the nozzle and a stretched region, increasing cell size towards the domain boundaries to reduce boundary effects. Two grid resolutions are considered: a coarse grid with $D_t / \Delta x = 50$, resulting in $(N_x, N_y, N_z) = (500,300,300)$ cells, and a fine grid with $D_t / \Delta x = 100$, resulting in $(N_x, N_y, N_z) = (900,500,500)$ cells.

At the primary and secondary inlets, stagnation pressure and a total temperature of $T_0 = 300\,\mathrm{K}$ are prescribed. At the secondary inlet, no assumption is made a priori regarding choking. Instead, the local static pressure at the boundary is compared to the injector total pressure $P_s$. If the corresponding pressure ratio indicates choked conditions, the boundary state is imposed assuming isentropic expansion from $P_s$ to sonic conditions ($M=1$). Otherwise, the local static pressure is imposed, and the remaining flow variables are obtained from isentropic relations between $P_s$ and the local pressure. The ambient pressure is set to $P_\infty = 1\,\mathrm{bar}$. The fluid is an ideal gas with specific heat ratio $\gamma=1.4$ and specific gas constant $R = 287.14\,\mathrm{J kg^{-1} K^{-1}}$.

In the present work, viscous effects are neglected, eliminating the need to resolve the boundary layer and thereby significantly reducing computational cost. As a consequence, boundary layer separation is not captured. Figure~\ref{fig:tvc_validation} shows the resulting Mach number field for the planar nozzle at $NPR=4.6$ and $SPR=0.7$. The injected secondary jet generates a bow shock that penetrates into the supersonic core flow, inducing a strong flow deflection and forming an asymmetric shock structure responsible for thrust vectoring.
In the absence of viscous effects, the adverse pressure gradient does not lead to boundary layer separation. In viscous flows, the resulting separation bubble modifies the shock system such that, instead of a strong bow shock located close to the injector, a comparatively weaker oblique (separation) shock forms further upstream~\cite{Waithe2003,Das2025}. This mechanism is absent in the inviscid case, which is also reflected in the wall pressure distributions in Fig.~\ref{fig:tvc_validation}. Here, the overall trend is well captured, but the inviscid solution exhibits a more pronounced pressure peak closer to the injector. Nevertheless, the predicted thrust vector angles show good agreement with RANS and experimental data, indicating that the essential flow dynamics governing the shock-vectoring mechanism are well captured despite the absence of boundary layer effects. Future work will extend the present approach to wall-modeled large-eddy simulation to account for these near-wall effects.

\begin{figure}[!t]
\centering
\begin{tikzpicture}[>={Latex}]

\begin{scope}[scale=50, yshift=0.02cm]
    \coordinate (A) at (0.0, -0.01559);
    \coordinate (O) at (0.0, 0.0);
    \coordinate (B) at (0.0, 0.0352);
    \coordinate (C) at (0.02329, 0.02954);
    \coordinate (D) at (0.05049, 0.01552);
    \coordinate (E) at (0.0608076, 0.0140462);
    \coordinate (G) at (0.104, 0.0224412);
    \coordinate (G1) at (0.102, 0.022052);
    \coordinate (G2) at (0.106, 0.0228299);
    \coordinate (F) at (0.05779, 0.02962);
    \coordinate (H) at (0.11557, 0.0246888);
    \coordinate (EDGE) at (0.11557, 0.045);

    \coordinate (Am) at (0.0, 0.01559);
    \coordinate (Cm) at (0.02329, -0.02954);
    \coordinate (Dm) at (0.05049, -0.01552);
    \coordinate (Em) at (0.0608076, -0.0140462);
    \coordinate (Fm) at (0.05779, -0.02962);
    \coordinate (Hm) at (0.11557, -0.0246888);
    \coordinate (EDGEm) at (0.11557, -0.045);

    \draw[thick]
    (A) ++(90:0.05079) 
    arc[start angle=90, end angle=62.69, radius=0.05079];
        
    \draw[thick]
    (F) ++(-76.5413:0.0158657)
    arc[start angle=-76.5413, end angle=-117.2884, radius=0.0158657];

    \draw[thick, <->] (0.05779,-0.0137543) -- (0.05779,0.0137543) node[at start, above, xshift=0.3cm] {$D_{t}$};

    \draw[thick]
    (Am) ++(-90:0.05079)
    arc[start angle=-90, end angle=-62.69, radius=0.05079];

    \draw[thick]
    (Fm) ++(76.5413:0.0158657)
    arc[start angle=76.5413, end angle=117.2884, radius=0.0158657];

    \draw[thick] (Cm) -- (Dm);
    \draw[thick] (Em) -- (Hm);
    
    \node[below] at (A) {$A$};
    \node[above] at (B) {$B$};
    \node[above] at (C) {$C$};
    \node[below left] at (D) {$D$};
    \node[below right] at (E) {$E$};
    \node[above] at (F) {$F$};
    \node[below, red] at (G) {$G$};
    \node[above] at (H) {$H$};

    \draw[thin, ->] (A) -- (0.015, 0.033);
    \draw[thin, ->] (F) -- (0.054, 0.014);
    
    \draw[thick] (C) -- (D);
    \draw[thick] (E) -- (G1);
    \draw[thick, red] (G1) -- (G2);
    \draw[thick] (G2) -- (H);

    \end{scope}
    
    \fill (A) circle (2pt);
    \fill (B) circle (2pt);
    \fill (C) circle (2pt);
    \fill (D) circle (2pt);
    \fill (E) circle (2pt);
    \fill (F) circle (2pt);
    \fill[red] (G) circle (2pt);
    \fill (H) circle (2pt);

    \draw[thick, red, ->] ([xshift=-0.1cm,yshift=0.514cm]G) -- (G) node[at start, above] {$P_s$};
    
    \draw[thick, dash dot] (O) -- (H|-O);

    \draw[thick, ->] (O) -- ([yshift=1cm]O) node[at end, right] {$y$};
    \draw[thick, ->] (O) -- ([xshift=1cm]O) node[at end, below] {$x$};


    \draw[thick, ->] ([xshift=-1cm,yshift=1cm]O) -- ([xshift=-0.2cm,yshift=1cm]O) node[midway, above] {$P_0$};

    \draw[thick, ->] (H|-O) -- ([xshift=1.5cm]H|-O) node[midway, above] {$F_x$};;
    \draw[thick, ->] ([xshift=1.5cm]H|-O) -- ([xshift=1.5cm,yshift=-0.5cm]H|-O) node[midway, right] {$F_y$};;
    \draw[thick, ->] (H|-O) -- ([xshift=1.5cm,yshift=-0.5cm]H|-O) node[midway, below] {$F_r$};
    
    \node at ([xshift=1cm, yshift=1cm]H|-O) {$P_{\infty}$};

    \node[anchor=west,inner sep=0, scale=0.75] at (-1.2cm,-2cm) {
    \begin{tabular}{c cccccccc}
        \toprule
         & A & B & C & D & E & F & G & H \\
        \midrule
        $x$ [mm] & 0.00 & 0.00 & 23.29 & 50.49 & 60.81 & 57.79 & 104.00 & 115.57 \\
        $y$ [mm] & -15.59 & 35.20 & 29.54 & 15.52 & 14.05 & 29.62 & 22.44 & 24.69 \\
        \bottomrule
    \end{tabular}
    };
    
    \begin{scope}[xshift=9.5cm, scale=15]

    \coordinate (A) at (0.0, -0.01559);
    \coordinate (O) at (0.0, 0.0);
    \coordinate (B) at (0.0, 0.0352);
    \coordinate (C) at (0.02329, 0.02954);
    \coordinate (D) at (0.05049, 0.01552);
    \coordinate (E) at (0.0608076, 0.0140462);
    \coordinate (F) at (0.05779, 0.02962);
    \coordinate (H) at (0.11557, 0.0246888);
    \coordinate (EDGE) at (0.11557, 0.045);

    \coordinate (Am) at (0.0, 0.01559);
    \coordinate (Bm) at (0.0, -0.0352);
    \coordinate (Cm) at (0.02329, -0.02954);
    \coordinate (Dm) at (0.05049, -0.01552);
    \coordinate (Em) at (0.0608076, -0.0140462);
    \coordinate (Fm) at (0.05779, -0.02962);
    \coordinate (Hm) at (0.11557, -0.0246888);
    \coordinate (EDGEm) at (0.11557, -0.045);

    \coordinate (LR) at (0.0, 0.2);
    \coordinate (RR) at (0.5, 0.2);
    \coordinate (RL) at (0.5, -0.2);
    \coordinate (LL) at (0.0, -0.2);

    \coordinate (LR1) at (0.0, 0.06);
    \coordinate (RR1) at (0.2, 0.06);
    \coordinate (RL1) at (0.2, -0.06);
    \coordinate (LL1) at (0.0, -0.06);
        
    \draw[thick, blue]
    (A) ++(90:0.05079)
    arc[start angle=90, end angle=62.69, radius=0.05079];

    \draw[thick, blue]
    (F) ++(-76.5413:0.0158657)
    arc[start angle=-76.5413, end angle=-117.2884, radius=0.0158657];

    \draw[thick, blue] (C) -- (D);
    \draw[thick, blue] (E) -- (H) -- (EDGE) -- (EDGE-|O);

    \draw[thick, blue]
    (Am) ++(-90:0.05079)
    arc[start angle=-90, end angle=-62.69, radius=0.05079];

    \draw[thick, blue]
    (Fm) ++(76.5413:0.0158657)
    arc[start angle=76.5413, end angle=117.2884, radius=0.0158657];

    \draw[thick, blue] (Cm) -- (Dm);
    \draw[thick, blue] (Em) -- (Hm) -- (EDGEm) -- (EDGEm-|O);

    \draw[thick] (B) -- (LR) -- (RR) -- (RL) -- (LL) -- (Bm);

    \end{scope}
    
    \draw[thick] (B) -- (LR) node[midway, rotate=90, above] {Outflow};
    \draw[thick] (LL) -- (Bm) node[midway, rotate=90, above] {Outflow};
    \draw[thick] (LR) -- (RR) node[midway, above] {Constant extrapolation};
    \draw[thick] (RR) -- (RL) node[midway, rotate=90, below] {Constant extrapolation};
    \draw[thick] (RL) -- (LL) node[midway, below] {Constant extrapolation};

    \draw[thick, red] (B) -- (Bm) node[midway, rotate=90, above] {Inflow};
    \draw[thin, dashed] (RR1) -- (RL1) -- (LL1);
    \draw[thin, dashed] (LR1) -- (RR1) node[above, midway] {uniform region};
    
    \draw[thin, <->] ([yshift=0.2cm]LL) -- ([yshift=0.2cm]RL) node[midway, above] {$250D_t$};
    \draw[thin, <->] ([xshift=-0.2cm]RR) -- ([xshift=-0.2cm]RL) node[midway, rotate=90, above] {$150D_t$};

    \draw[thin, <->] ([yshift=-0.2cm]LL1) -- ([yshift=-0.2cm]RL1) node[midway, below] {$8D_t$};
    \draw[thin, <->] ([xshift=0.2cm]RR1) -- ([xshift=0.2cm]RL1) node[midway, rotate=90, below] {$4D_t$};

\end{tikzpicture}
\caption{Schematic of the nozzle geometry and the computational domain.}
\label{fig:tvc_schematic}
\end{figure}

\begin{figure}[t!]
\centering
\input{figures/SI/tvc_validation}
\caption{Results for the planar nozzle with $SPR=0.7$. Visualized is an instantaneous snapshot of the Mach number field for $NPR=4.6$ and the corresponding mean wall pressure $p_w/P_0$ over the centerline $x/x_t$, where $x_t$ denotes the throat location. Additionally, the mean thrust vector angle $\delta_0$ over a variation of the nozzle pressure ratio $NPR$ is shown. 
}
\label{fig:tvc_validation}
\end{figure}

\paragraph{Environment implementation.}
Two environments are considered: a planar (2D) and an axisymmetric (3D) nozzle configuration, each simulated at two grid resolutions as described above. The flow is characterized by a fixed nozzle pressure ratio $NPR=4.6$, while the secondary pressure ratio is user-defined within $SPR \in [0.7,0.9]$.
In the 2D configuration, two injectors are present, one located on the upper and one on the lower wall, resulting in two control inputs. In the 3D configuration, the number of injectors $N_{\text{inj}}$ is user-defined, and the injectors are uniformly distributed along the nozzle circumference, yielding $N_{\text{inj}}$ independent control inputs, as illustrated in Fig.~\ref{fig:tvc_env}.
The control input is defined through normalized actuator signals $a_i \in [0,1]$, $i=1,\dots,N_{\text{inj}}$, for each injector. Each actuator controls the total pressure of the corresponding secondary jet according to
\begin{equation}
    P_{s,i} = p_{\text{local},i} + a_i \left(P_{s,\max} - p_{\text{local},i}\right),
\end{equation}
where $p_{\text{local},i}$ denotes the instantaneous local static pressure at injector $i$, and $P_{s,\max} = P_{\infty} \cdot NPR \cdot SPR$.

The observations are the instantaneous thrust vector angles ($\delta_0$ and $\delta_1$), the instantaneous target thrust vector angles ($\delta_{0}^{\text{ref}}$ and $\delta_{1}^{\text{ref}}$), and wall pressure measurements downstream of the injectors.
In particular, two pressure probes are placed along the wall between each injection location and the nozzle exit, resulting in a total of $2N_{\text{inj}}$ pressure probes.
The reward $r$ is defined based on tracking a reference trajectory of the thrust vector angles. Using the instantaneous thrust vector angles and their corresponding reference values, the reward is formulated as
\begin{equation}
    r = - \sqrt{\left(\delta_0 - \delta_{0}^{\text{ref}}\right)^2 + \left(\delta_1 - \delta_{1}^{\text{ref}}\right)^2},
\end{equation}
which penalizes deviations from the desired thrust vector direction.
Note that the environment admits time-dependent reference values for the thrust vector direction.

\begin{figure}[t!]
\centering
\input{figures/SI/tvc_3d.tex}
\caption{Illustration of the 3D bell nozzle environment with $N_{\text{inj}}=8$. The injector locations are indicated by white patches on the nozzle surface, which is colored by the static pressure. A 2D $xy$-slice at $z=0$ shows the corresponding Mach number field. The computational grid is also depicted, highlighting the highly resolved uniform region and the stretched region towards the domain boundaries.}
\label{fig:tvc_env}
\end{figure}
\end{appendices}

\end{bibunit}
\end{document}